\documentclass[11pt]{article}
\usepackage{jheppub}
\usepackage{graphicx}
\usepackage{float} 
\usepackage{slashed}
\usepackage[dvipsnames,table]{xcolor}
\usepackage{hyperref} 
\usepackage{xspace}
\usepackage[tight]{subfigure}
\usepackage{amssymb}
\usepackage{amsfonts}
\usepackage{mathrsfs}
\usepackage{comment}
\usepackage{amsmath}
\DeclareMathOperator{\Tr}{Tr}
\usepackage{afterpage}
\usepackage{tikz}
\usetikzlibrary{quantikz2}
\usepackage[compat=1.1.0]{tikz-feynman}
\usepackage[T1]{fontenc}
\usepackage{verbatim}
\usepackage{booktabs}
\usepackage{array}
\usepackage[title,titletoc]{appendix}
\usepackage{tabularx}
\usepackage{orcidlink}
\newcolumntype{C}[1]{>{\centering\arraybackslash}p{#1}}

\newcommand{\newc}{\newcommand}
\newc{\beq}{\begin{equation}}
\newc{\eeq}{\end{equation}}
\newc{\bit}{\begin{itemize}}
\newc{\eit}{\end{itemize}}
\newc{\ben}{\begin{enumerate}}
\newc{\een}{\end{enumerate}}
\newc{\bce}{\begin{center}}
\newc{\ece}{\end{center}}
\newc{\bfi}{\begin{figure}}
\newc{\efi}{\end{figure}}

\newcolumntype{.}{D{.}{.}{-1}}
\newcolumntype{d}[1]{D{.}{.}{#1}}
\colorlet{tableoverheadcolor}{gray!37.5}
\colorlet{tableheadcolor}{gray!25}
\colorlet{tablerowcolor}{gray!12.5}

\def\draftdate{\relax}
\def\mda{\relax}
\def\mua{\relax}
\def\mla{\relax}
\def\draft{
\def\thtystars{******************************}
\def\sixtystars{\thtystars\thtystars}
\typeout{}
\typeout{\sixtystars**}
\typeout{* Draft mode!
         For final version remove \protect\draft\space in source file *}
\typeout{\sixtystars**}
\typeout{}
\def\draftdate{\today}
\def\mua{\marginpar[\boldmath\hfil$\uparrow$]%
                   {\boldmath$\uparrow$\hfil}\color{black}%
                    \typeout{marginpar: $\uparrow$}\ignorespaces}
\def\mda{\color{red}\marginpar[\boldmath\hfil$\downarrow$]%
                   {\boldmath$\downarrow$\hfil}%
                    \typeout{marginpar: $\downarrow$}\ignorespaces}
\def\mla{\marginpar[\boldmath\hfil$\rightarrow$]%
                   {\boldmath$\leftarrow $\hfil}%
                    \typeout{marginpar: $\leftrightarrow$}\ignorespaces}
\def\Mua{\marginpar[\boldmath\hfil$\Uparrow$]%
                   {\boldmath$\Uparrow$\hfil}\color{black}%
                    \typeout{marginpar: $\uparrow$}\ignorespaces}
\def\Mda{\color{red}\marginpar[\boldmath\hfil$\Downarrow$]%
                   {\boldmath$\Downarrow$\hfil}%
                    \typeout{marginpar: $\downarrow$}\ignorespaces}
\def\Mla{\marginpar[\boldmath\hfil\textcolor{red}{$\Rightarrow$}]%
                   {\boldmath\textcolor{red}{$\Leftarrow $}\hfil}%
                    \typeout{marginpar: $\leftrightarrow$}\ignorespaces}
\overfullrule 5pt
\oddsidemargin 15mm
\marginparwidth 29mm
}

\begin{document}


\title{Physics inspired quantum algorithm for QCD splitting functions } 
\author{Gabriel Rouxinol $\,^{a,b}$,}\emailAdd{G.Rouxinol@lmu.de}
\author{Yacine Haddad$\,^{c,d}$,}\emailAdd{yacine.haddad@cern.ch}
\author{Cenk T\"uys\"uz$\,^{c}$,}
\author{Sofia Vallecorsa$\,^{c}$}
\author{and Michele Grossi$^{c}$}\emailAdd{michele.grossi@cern.ch}

\affiliation{$\,^{a}$Department of Physics and Arnold Sommerfeld Center for Theoretical Physics (ASC), Ludwig Maximilian University of Munich, 80333 Munich, Germany}
\affiliation{$\,^{b}$Munich Center for Quantum Science and Technology (MCQST), 80799 Munich, Germany}
\affiliation{$\,^{c}$European Organisation for Nuclear Research (CERN), 1211 Geneva, Switzerland}
\affiliation{$\,^{d}$Albert Einstein Center for Fundamental Physics, Laboratory for High Energy Physics, University of Bern, Sidlerstrasse 5, CH-3012 Bern, Switzerland}

\abstract{
We introduce a modular quantum circuit primitive to model entanglement dynamics in QCD parton splitting and use it as a composable building block for data-driven, physics-consistent modelling of parton-shower kinematics. For the pure-gluon channel, we derive an analytic expression for the helicity entanglement generated at the splitting vertex, quantified via the concurrence, and construct a two-qubit circuit whose $\sigma_3$ expectation values encode the momentum shared between outgoing gluons. The circuit reproduces the QCD-predicted concurrence by construction, which fixes a two-qubit correlator that the momentum fractions alone do not determine. Iterating the primitive, the concurrence obeys a product rule across successive splittings. Calibrating the circuit parameters to LHC jet substructure data relates the reconstructed momentum-sharing fractions directly to circuit rotation angles. Composing multiple splitting primitives yields multi-prong momentum-fraction distributions; we compare the three- and four-prong cases against experimental data, reproducing the core of the distributions. For the three-prong configuration, we execute the circuit on superconducting quantum hardware, recover the momentum-fraction distributions after a device-bias correction calibrated on a held-out subset, and study how noise alters the entanglement-dependent observable. This work provides a concrete framework for quantum-native parton-shower modules that encode quantum correlations at the level of splitting dynamics, and offers physics-informed ansätze for future quantum algorithms for QCD.
}

\maketitle

\section{Introduction}\label{sec:intro}

Quantum computers offer an opportunity to transform computing in the basic sciences and their applications~\cite{Preskill_2018}. In high-energy physics (HEP), the scale of current and future collider programs motivates exploring quantum approaches to the most computationally demanding tasks in simulation and analysis~\cite{PRXQuantum.5.037001}. Recent work has investigated quantum algorithms for parton showers~\cite{PhysRevLett.126.062001, Bepari2021Towards, Bepari2022QuantumWalk, Bauer2024QuantumPartonKinematics, Chawdhry2023QuantumColour}, quantum simulation of quantum field theories~\cite{Jordan_2012, Garc_a_lvarez_2015, halimehUniversalFrameworkQuantum2026} and lattice gauge theories~\cite{Tagliacozzo2013Simulation, Silvi2017FiniteDensity, Klco2020SU2, Fromm:2024caq}, as well as applications to collider data analysis~\cite{Schuhmacher_2023, tuysuztracking, Funcke_2023, luxetracking, Wei2020QuantumJet, Cheng_2025, Magano2022QuantumSpeedup}.

Event generation is a growing contributor to LHC computing cost, already contributing to $20\%$ of the ATLAS CPU resources in 2017~\cite{valassiChallengesMonteCarlo2021}, while the full Monte Carlo pipeline, including reconstruction and detector simulation, takes around half of all computational resources~\cite{Albrecht2019Roadmap}. As Next-to-Next-to-Leading Order (NNLO) calculations become necessary for the High Luminosity LHC (HL-LHC) precision requirements, the computational cost of event generation is expected to grow significantly, driven by the higher jet multiplicities involved~\cite{azziStandardModelPhysics2019}. A simulated event combines a perturbative hard interaction with a subsequent parton shower, followed by non-perturbative hadronization once the evolution reaches the QCD scale $\Lambda_{\mathrm{QCD}}$~\cite{buckley2021practical}. The non-perturbative regime remains model-driven~\cite{ANDERSSON198331, FIELD198365}, and recent quantum-simulation efforts highlight the longer-term potential of quantum devices for strongly coupled dynamics~\cite{mildenbergerConfinement$$mathbbZ_2$$Lattice2025, alexandrouRealizingStringBreaking2025}. Looking ahead to the HL-LHC, improved algorithms to sample QCD splittings that constitute the parton shower would reduce the computational demands of event generation.

State-of-the-art generators such as \textsc{Pythia}~\cite{Sjostrand:2014zea}, \textsc{Herwig}~\cite{B_hr_2008}, and \textsc{Sherpa}~\cite{Gleisberg_2004} implement parton showers through Markov Chain Monte Carlo algorithms that repeatedly sample QCD splitting functions~\cite{Ellis_Stirling_Webber_1996}. While highly successful, these methods treat each branching step in a purely classical manner, discarding quantum coherence between emissions. A quantum algorithm for parton showers can naturally capture quantum correlations between particles in the shower, with particular focus on quantum entanglement, which has been linked to spin-flavour symmetries and has motivated both theoretical and experimental investigations~\cite{PhysRevD.104.074014, PhysRevLett.122.102001, ATLAS:2024entanglement, yazganMeasurementsTopQuark2026, florioQuantumRealtimeEvolution2024, florioThermalizationQuantumEntanglement2025}. The study of scattering processes in HEP using quantum information tools is a rapidly growing field~\cite{afikQuantumInformationMeets2025}, with work done on treating particle splitting using Kraus operators~\cite{aoudeDecoherenceEffectsEntangled2026}, the entanglement of Non-Abelian scattering processes ~\cite{mcginnisSymmetryEntanglementSmatrix2025} and their realization using quantum gates~\cite{mcginnisQuantumComputationalStructure2025}. The present work extends this research direction by linking these studies with real collider data.

In this work, we address this challenge by introducing a quantum algorithm for the splitting function of parton showers, whose core component is a quantum-native, modular splitting unit for the $g \to gg$ channel. The unit implements parton splitting functions, enforces kinematic consistency via total momentum conservation and momentum fraction ordering, and encodes the quantum correlations intrinsic to the underlying QCD interaction. 
Using helicity states as a two-qubit basis, we compute the entanglement produced in the splitting and obtain an analytic concurrence $C_{\mathrm{QCD}}(z)$. We then design a two-qubit circuit whose expectation values encode the splitting fractions $z$ and $1-z$ and whose entanglement matches $C_{\mathrm{QCD}}(z)$. Finally, we calibrate the circuit to LHC jet substructure by extracting momentum-sharing fractions from the AspenOpenJets dataset~\cite{Amram:2024fjg} and use them to determine circuit parameters that replicate the two-prong structure. By applying the splitting module with parameters fitted to the two-prong structure, we find the three-prong and four-prong momentum-fraction distributions are reproduced across the centre of the range, with quantified departures at large momentum fraction. We further demonstrate the feasibility of this approach on superconducting quantum hardware for the three-prong case.

By iterating this unit, we obtain a composable route to shower-like multi-prong structures and a data-driven framework for physics-consistent modelling of parton-shower kinematics validated against LHC data. Classical machine learning has shown strong performance in HEP simulation tasks~\cite{baldi_searching_2014, guest_deep_2018, deoliveira_jet-images_2016, Qu_2020, komiske_energy_2019}, including parton showers~\cite{deepshower_2018, lai_explainable_2020} and hadronization~\cite{ghosh_towards_2022, chan_flavor_2023, chan_fitting_2023}. By basing the quantum circuit on the splitting function, we take an approach compatible with variational quantum circuit training as explored in the emerging field of Quantum Machine Learning (QML)~\cite{lloyd_quantum_2013, cai_entanglement-based_2015, Schuld_2016, schuld_supervised_2018, gibbs_dynamical_2024, Belis_2024, Tuysuz:2025mrv}. Since our building block mirrors the physical entanglement properties of the QCD process, it provides a structured starting point for physics-informed ansätze, an approach which has been shown to improve QML performance~\cite{meyer_exploiting_2023, park_hamiltonian_2024, west_provably_2024, marrero_expressivity_2021, wang2023symmetryenhancedvariationalquantum}. Although our circuit is derived in perturbation theory, the data-driven calibration lays the groundwork for future extensions that may capture non-perturbative aspects of QCD. For instance, instead of fixing the parameters of the circuit to match the entanglement of the perturbative QCD process, the parameters could be trained on jet substructure observables, which inherently include hadronization, encoding non-perturbative corrections into the entanglement structure of the physics-motivated splitting circuit.

The paper is organised as follows. In Sec.~\ref{sec:Section2}, we present a quantum-information description of the splitting function, highlight the role of entanglement generated in three-gluon scattering, and construct a quantum circuit that reproduces this entanglement and the associated momentum fractions while conserving total momentum, before showing how the module can be composed to model shower-like evolution. In Sec.~\ref{sec:Section3}, we use data on the LHC jet substructure from the AspenOpenJets dataset to obtain the circuit parameters. In Sec.~\ref{sec:Section4}, we compose the splitting module to obtain the three-prong and four-prong structures, and for the three-prong case, we run the circuit on IBM hardware. Finally, Sec.~\ref{sec:Section5} summarises our findings and outlines directions for future work.

\section{Splitting Function}\label{sec:Section2}

\subsection{The Role of Entanglement in a QCD splitting}

In this section, we use quantum information tools to study the splitting function $P_{gg}(z)$ for the $g \to gg$ interaction, a central object in the QCD sector of the Standard Model (SM). The argument $z$ is the momentum fraction of the harder daughter obtained in the splitting process, such that $z\in[1/2, 1]$. We focus on the pure-gluon channel given its uniqueness in QCD relative to Quantum Electrodynamics (QED). This is due to the non-abelian nature of the $\mathrm{SU}(3)$ gauge group and its dominant role in high transverse momentum parton shower evolution at the LHC, especially via the $gg\to ggg$ interaction~\cite{campbellEventGeneratorsHighenergy2024, khalekUltimatePartonDistributions2018,abreuMeasurementTriplegluonVertex1993, campbellHardInteractionsQuarks2006}. A deep understanding of splitting functions is fundamental in the design of parton shower algorithms. Collinear factorisation~\cite{Collins:1988ig} allows one to decompose the full shower into a sequence of elementary splitting blocks $a\to b+c$, each weighted by the corresponding splitting function. A more complete treatment of the underlying concept of factorisation and the study of the $g\to gg$ process using Quantum Field Theory (QFT)~\cite{peskin1995introduction,schwartz2014quantum} tools is presented in Appendix~\ref{app:MatrixElements}.

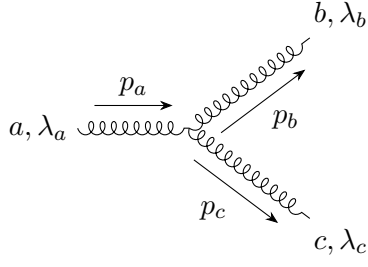
\begin{figure}[h]
    \centering
    \begin{tikzpicture}
      \begin{feynman}
        \vertex (a) at (-2,0) {\(a, \lambda_a\)};
        \vertex (o) at (0,0);
        \vertex (b) at (2,1.5) {\(b, \lambda_b\)};
        \vertex (c) at (2,-1.5) {\(c, \lambda_c\)};
    
        \diagram* {
          (a) -- [gluon, momentum=\(p_a\)] (o),
          (o) -- [gluon, momentum'=\(p_b\)] (b),
          (o) -- [gluon, momentum'=\(p_c\)] (c),
        };
      \end{feynman}
    \end{tikzpicture}
    \caption{Feynman diagram for the interaction between an incoming gluon of momentum $p_a$, color $a$ and helicity $\lambda_a$ with outgoing gluons of momenta $p_b, p_c$, colors $b, c$ and helicities $\lambda_b, \lambda_c$}
    \label{fig:ThreePointGluonDiagram}
\end{figure}

Consider now that the incoming gluon has colour $a$, helicity $\lambda_a$, momentum $p_a$ and polarization $\varepsilon_a$, while the outgoing ones have colours $b$ and $c$, helicities $\lambda_b$ and $\lambda_c$, momenta $p_b$ and $p_c$ and polarizations $\varepsilon_b$ and $\varepsilon_c$. Their interaction can be represented in the Feynman diagram of Fig.~\ref{fig:ThreePointGluonDiagram}.

We now define the state of the incoming gluon in the helicity basis as $\ket{p_a, a, \lambda_a}$, where $p_a$ denotes the particle's momentum, $a$ its color, and $\lambda_a$ its helicity, while the final state is \begin{equation}
    \ket{p_b,b,\lambda_b; p_c,c,\lambda_c}=\ket{p_b,b,\lambda_b}\otimes\ket{p_c,c,\lambda_c}.
\end{equation}
The operator responsible for the scattering is the S-Matrix $S=\mathbb{I}+iT$, where the matrix elements of $iT$, $ \bra{p_b,b,\lambda_b; p_c,c,\lambda_c} iT\ket{p_a, a, \lambda_a}$, are the scattering amplitudes $i\mathcal{M}(\lambda_a\to \lambda_b\lambda_c)$. The computation of these amplitudes is presented in Appendix~\ref{app:MatrixElements}.

We can now look into the entanglement produced in the process, by following a procedure similar to the one presented in Refs.~\cite{afik2022quantum, cheng2024quantum}, where they define the matrix $R^{\lambda_a}_{\lambda_b\lambda_c, \lambda_{b'}\lambda_{c'}}$, known as the R-Matrix, as
\begin{equation}
    R^{\lambda_a}_{\lambda_b\lambda_c, \lambda_{b'}\lambda_{c'}}=\sum_{\text{colours}}\bra{p_b,b,\lambda_b; p_c,c,\lambda_c} T\ket{p_a, a, \lambda_a} \bra{p_a, a, \lambda_a} T^\dagger \ket{p_{b},b,\lambda_{b'}; p_{c},c,\lambda_{c'}},
    \label{eqn:RMatrixDefinition}
\end{equation}
where, for a definite final momentum, the two helicity eigenstates L (left-handed) and R (right-handed), which label the rows and columns of the R-matrix, are identified with the qubit states $\ket{0}$ and $\ket{1}$, respectively. To construct the two-qubit density matrix $\hat{\rho}_{sc}$, also referred to as the spin density matrix, we need to sum over the initial helicities and normalise, resulting in 
\begin{equation}
    \hat{\rho}_{sc}=\frac{1}{N}\sum_{\lambda_a} R^{\lambda_a}_{\lambda_b\lambda_c, \lambda_{b'}\lambda_{c'}},
    \label{eqn:SpinDensityDefinition}
\end{equation}
where $N=\Tr(\sum_{\lambda_a} R^{\lambda_a}_{\lambda_b\lambda_c, \lambda_{b'}\lambda_{c'}})$, which ensures a unity trace. The spin density matrix encodes the final quantum state, with the four helicity combinations of left and right helicities of the two outgoing gluons mapped to the four two-qubit states. Using this formalism, we recast the scattering process as a two-qubit problem, where each qubit encodes the helicity of one outgoing gluon after tracing out the colour degrees of freedom. The entanglement of the system is then quantified via the concurrence~\cite{PhysRevLett.80.2245}, a standard measure for general two-qubit states. The spin density matrix for the tree-level $g \to gg$ process is obtained in Appendix~\ref{app:MatrixElements}, from which we find that the generated entanglement is  
\begin{equation}
    \mathcal{C}_\text{QCD}(z)= \left( \frac{z(1-z)}{1-z(1-z)}\right) ^2.
    \label{eqn:ConcurrenceQCDProcess}
\end{equation}
The resulting expression exhibits three key features. Firstly, it peaks at $z = 0.5$, corresponding to equal momentum sharing between the two gluons. Secondly, the entanglement vanishes as $z\to 0$ or $z\to 1$, where one gluon carries negligible momentum, and the two-particle state approaches a product state. Finally, the function is symmetric under $z\to 1-z$, which is consistent with the symmetry of the scattering process: gluon exchange does not alter the process.

In this section, we derived an analytical expression for the entanglement generated in a parton splitting, which will be reproduced using a quantum circuit in the next section. The pure-gluon channel is chosen for its simplicity; the same approach applies to other splitting channels, even when the resulting entanglement function is more complex.

\subsection{A quantum circuit for QCD splitting}\label{subsec:SplittingQuantumCircuit}

In this section, we construct a two-qubit quantum circuit of which the measurement outcomes encode the momentum fractions $z$ and $1-z$ for a $g\to gg$ splitting. The circuit is designed to enforce the entanglement computed in the previous section while having measurement outcome probabilities match those given by the splitting function. A general overview of the Quantum Computing basics is presented in~\ref{app:QCBasics}.

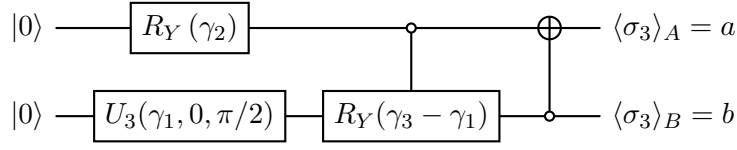
\begin{figure}[H]
\begin{center} 
    \begin{quantikz}
    \lstick{$\ket{0}$}&\gate[1]{R_Y\left(\gamma_2\right)}&\ctrl[open]{1}&\targ{}&\rstick{$\langle\sigma_3\rangle_A =a$}\\
    \lstick{$\ket{0}$}&\gate[1]{U_3(\gamma_1,0, \pi/2)}&\gate[1]{R_Y(\gamma_3-\gamma_1)}&\ctrl[open]{-1}&\rstick{$\langle\sigma_3\rangle_B =b$}
    
    \end{quantikz}
    \caption{The quantum circuit building block for parton showers, with free parameters $\{\gamma_1, \gamma_2, \gamma_3\}$ and expected values of the Pauli matrix $\sigma_3$ for qubit $A$ and $B$ are denoted as $a$ and $b$. The parameters are constrained such that $a+b=1$ and $a>b$, ensuring that qubit A always carries the higher momentum fraction.}
    \label{fig:TwoQubitBuildingBlock}
\end{center}

\end{figure}

To achieve this, we design the quantum circuit of Fig.~\ref{fig:TwoQubitBuildingBlock}, encoding it in a way that the splitting fraction of the interaction, $z$, is the expected value of the third Pauli matrix $\sigma_3$ of the first qubit, meaning $a=z$, and consequently,  the expected value of  $\sigma_3$ of the second qubit is $b=1-z$. To ensure $a+b=1$, the circuit parameters must satisfy the condition $2\cos(\gamma_1)\sin^2(\gamma_2/2)=1$. This condition is solvable for $\gamma_1 \in [-\pi/3, \pi/3]$ and $\gamma_2\in [\pi/2, 3\pi/2]$, which can be restricted to the intervals $\gamma_1 \in [0,\pi/3]$ and $\gamma_2\in [\pi/2, \pi]$, since these ranges will be sufficient to cover all physically allowed values of $z \in [0,1]$ and the entanglement of the process, as verified in Appendix~\ref{app:QCParameters}, which also shows that $a=z(\gamma_1, \gamma_3)$ and $b=1-z(\gamma_1, \gamma_3)$, with
\begin{equation}
    z(\gamma_1, \gamma_3)=\frac{1}{2}\left(1+\cos(\gamma_3)(\sec(\gamma_1)-2)\right),
    \label{eqn:valueofz}
\end{equation}

As the total system is in a pure state, we can compute the concurrence using only the reduced states~\cite{PhysRevLett.80.2245} and obtain  
\begin{equation}
\begin{gathered}
    \mathcal{C}_\text{circuit}(\gamma_1, \gamma_3)=\sqrt{\frac{3}{4}-\frac{1}{4}(\sec (\gamma_1)-2)^2 \cos ^2(\gamma_3)+\frac{1}{2} \frac{\sec (\gamma_1)-2}{\cos(\gamma_1)}(\sin
   (\gamma_1) \sin (\gamma_3)+1)} =\\
   =\sqrt{\frac{3}{4}-\left(z-\frac{1}{2}\right)^2+\left(z-\frac{1}{2}\right)(\tan
   (\gamma_1) \tan (\gamma_3)+ \sec (\gamma_1)  \sec (\gamma_3) )},
   \label{eqn:ConcurrenceCircuitAnal}
\end{gathered}
\end{equation}
where we now seek the combination of values $(\gamma_1, \gamma_3)$ such that the circuit with output $z$ in the first qubit satisfies $\mathcal{C}_\text{circuit}(\gamma_1, \gamma_3)=\mathcal{C}_\text{QCD}(z(\gamma_1, \gamma_3))$. This construction allows the two qubits of the splitting block to carry the momentum fraction of the produced parton in the expectation values of $\sigma_3$, while imposing the concurrence from the QCD process as a design constraint. The resulting state is therefore distinct from the spin density matrix $\hat{\rho}_{sc}$ in Eq.~\eqref{eqn:SpinDensityDefinition}, in which the momentum fraction enters only through a nonlinear function of the helicity correlations. Encoding it linearly in a single qubit observable allows a composite structure where successive splitting blocks will return the product of splitting fractions directly. A single splitting block produces a pure state, whereas $\hat{\rho}_{sc}$ is mixed. When blocks are composed, tracing out the ancillary qubits gives a mixed state for the produced partons. With this encoding, kinematic observables are ensemble averages over many circuit executions. Generating a separate event from each execution would require a different circuit structure. 

Because the momentum fractions are single-qubit expectation values, they only depend on the combination of parameters that determines $z$. Nevertheless, at fixed $z$ the remaining freedom in $\gamma_1$ changes the concurrence without altering any $\sigma_3$ expectation value. Matching $\mathcal{C}_\text{QCD}$ removes this freedom, leaving a signature on the two-qubit correlator of the block $\langle\sigma_3\otimes\sigma_3\rangle = \sec\gamma_1 - 1$, independent of $\gamma_3$ and of $z$. Together with $\langle\sigma_3\rangle$, the correlator determines the concurrence. The correlator is measured in the same basis as the momentum fractions and requires no additional settings.

From the functional form of $ \mathcal{C}_\text{circuit}$ it is not immediately evident that it is invariant under the transformation of $z$ to $1-z$ and zero for $z=0,1$, as required to match $\mathcal{C}_\text{QCD}(z)$. The presence of these properties is discussed in Appendix~\ref{app:QCDiscussion}. It is also defined that the lower-momentum gluon is always mapped to the lower qubit and the higher-momentum gluon to the upper qubit. This ordering ensures we always track which qubits correspond to the highest-momentum particle, essential information for applying the next splitting circuit. Due to the running of the coupling in QCD, these higher-momentum particles are more likely to split again, which means that by tracking the qubit carrying the highest momentum fraction, we know where to apply the next splitting circuit.

Up to this point, the circuit implements a single splitting function, yet it already encompasses the entanglement structure of QCD. This circuit receives as input the state $\ket{00}$, which, via the mapping between momentum fraction and the expected value of $\sigma_3$, corresponds to an incoming particle with momentum fraction $z=1$, the initial state, with the second qubit free to store the new particle created in the process. However, in the parton shower, multiple splittings occur in succession, so our circuit needs to now be able to have a state different from $\ket{00}$ as input, where the splitting fraction is mapped onto one of the qubits, while the other will still serve as a way to store the momentum of the newly produced particle. In addition, the Sudakov factors governing the splitting probability~\cite{buckley2021practical} tell us that higher-momentum particles are more likely to split again. Given our circuit design, in which the upper qubit has the most momentum, we aim to implement a sequence of two building blocks, with the momentum fraction of the particle with the highest momentum serving as the input to the next splitting circuit. 

\begin{figure}[H]
\begin{center} 
    \begin{quantikz}
    &\gate[1]{R_Y\left(\gamma_2\right)}&\ctrl[open]{1}&\targ{}&\\
    &\gate[1]{U_3(\gamma_1,0, \pi/2)}&\gate[1]{R_Y(\gamma_3-\gamma_1)}&\ctrl[open]{-1}&
    \end{quantikz}
    $\equiv$
    \begin{quantikz}
    &\gate[2]{U_S(z(\gamma_1, \gamma_3))}&\\
    &&
    \end{quantikz}
\end{center}
    \caption{The two-qubit splitting circuit $U_S(z(\gamma_1, \gamma_3))$), used as the modular building block of the parton shower. Its explicit gate decomposition in terms of one- and two- qubit gates is shown on the left. Its compact representation on the right will be used throughout this paper.}
    \label{fig:DefinitionBuildingBlock}
\end{figure}
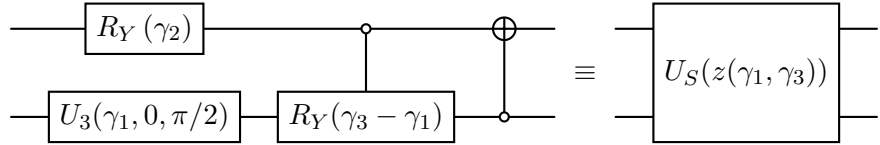

To achieve this goal, we provide a general procedure for scaling the problem, starting from a single splitting block. To do this, let us define our previous splitting circuit as $U_S(z(\gamma_1, \gamma_3))$, as presented in Fig.~\ref{fig:DefinitionBuildingBlock}. Consider as well that from the study of our original two-qubit quantum circuit we know the pairs of parameters $(\gamma_1, \gamma_3)$ and $(\gamma'_1,\gamma_3')$ that allow our two-qubit quantum circuit to output momentum fraction $z^{(1)}$ and $z^{(2)}$, respectively, while producing the same entanglement as the equivalent process in QCD. The circuit $U_S(z)$ acts as a splitting primitive: given a parent parton encoded in one qubit and an ancilla initialized to $|0\rangle$, it outputs two qubits whose $\sigma_3$-expectation values encode the daughter momentum fractions and whose entanglement is consistent with the QCD prediction for splitting fraction $z$. Multi-prong structures arise by iterating this primitive on the parton's qubit with the highest momentum fraction, implemented by CNOT gates that propagate the parent's momentum fraction into the next splitting stage. In this construction, the measured expectation values propagate multiplicatively, so successive emissions yield products of splitting fractions that can be directly compared to momentum fractions of reconstructed prongs. The design also provides access to internal splitting information by measuring intermediate wires (e.g., the first splitting fraction in Fig.~\ref{fig:FourParticleCircuit}), which is useful for comparison with declustering trees in data.

\begin{figure}[H]
\begin{center}
\scalebox{0.9}{
    \begin{quantikz}
        \lstick{$\ket{0}$}&&&\gate[2]{U_S(z(\gamma'_1, \gamma'_3)=z^{(2)})}&\meter{}\rstick{$\langle\sigma_3\rangle_A =z^{(1)}z^{(2)}$} \\
        \lstick{$\ket{0}$}&&\targ{}&&\meter{}\rstick{$\langle\sigma_3\rangle_B=z^{(1)}(1-z^{(2)})$}\\
        \lstick{$\ket{0}$}&\gate[2]{U_S(z(\gamma_1, \gamma_3)=z^{(1)})}&\ctrl{-1}&& \\
        \lstick{$\ket{0}$}&&&&\meter{}\rstick{$\langle\sigma_3\rangle_D =1-z^{(1)}$}
    \end{quantikz}}
\end{center}
    \caption{Three-particle splitting circuit constructed by iterating the $U_S$ primitive. The first block with parameters $(\gamma_1, \gamma_3)$, associated with the splitting fraction $z^{(1)}$, is applied to two qubits initialized to $\ket{00}$. A CNOT gate connects the qubit with the highest momentum to a second block with parameters $(\gamma_1', \gamma_3')$, which encodes a splitting with fraction $z^{(2)}$. The expected value of $\sigma_3$ of qubits $A, B$, and $D$ encodes the momentum fractions $z^{(1)}z^{(2)}$, $z^{(1)}(1-z^{(2)})$, and $1-z^{(1)}$ of the three final-state particles. The unmeasured qubit carries no momentum fraction of a final-state particle, but gives access to intermediate splitting information.}
    \label{fig:ThreeParticleCircuit}
\end{figure}
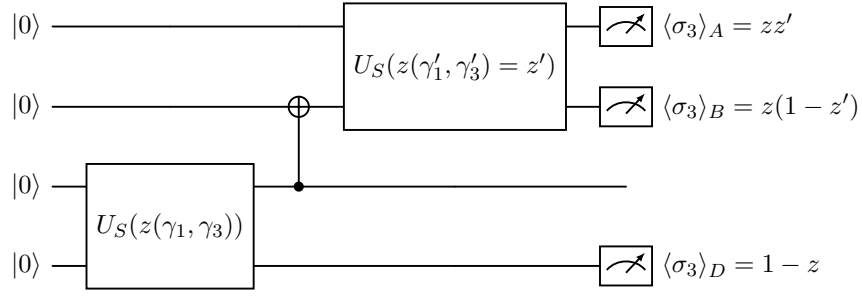

We now consider the case introduced above, in which a first splitting produces a particle with momentum fraction $z^{(1)}$, after which the produced particle with the highest momentum, mapped to the top qubit, undergoes a new splitting, now with splitting fraction $z^{(2)}$. To ensure physical consistency, we need to verify that stacking another $U_S$ block outputs the expected values of $\sigma_3$ for the three qubits to be the momentum fractions of each one of the final-state particles: $z^{(1)}z^{(2)}, z^{(1)}(1-z^{(2)})$ and $1-z^{(1)}$. To accomplish this, consider the quantum circuit of Fig.~\ref{fig:ThreeParticleCircuit} in which the measurement $\sigma_3$ expected value of the bottom qubit remains the same, $1-z^{(1)}$, which is equivalent to saying that the particle mapped to the lowest qubit does not split again. Furthermore, the CNOT gate feeds the momentum fraction of the first splitting circuit into the second circuit. Labelling the qubits from top to bottom as A to D, we obtain the reduced density matrix of the first two qubits (A and B), given in Appendix~\ref{app:QCParameters}, which allows us to verify that $$\Tr(Z\hat{\rho}_A)=\langle Z \rangle_A=\frac{z^{(1)}}{2}[1+(\sec (\gamma'_1)-2) \cos (\gamma'_3)]=z^{(1)}z^{(2)}$$ and that
\begin{equation}
    \langle Z\rangle_B=\frac{z^{(1)}}{2}[1-(\sec (\gamma'_1)-2) \cos (\gamma'_3)] =  z^{(1)}-\frac{z^{(1)}}{2}[1+(\sec (\gamma'_1)-2) \cos (\gamma'_3)] = z^{(1)}(1-z^{(2)}),
\end{equation}
as expected.

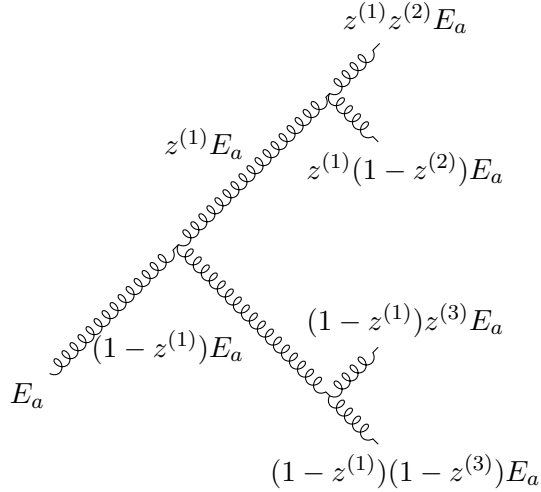
\begin{figure}[H]
    \centering
    \begin{tikzpicture}
      \begin{feynman}
        \vertex (a) at (-2,-2){\(p_T\)};
        \vertex (o) at (0,0) ;
        \vertex (b) at (2,2) ;
        \vertex (c) at (2,-2) ;
        \vertex (d) at (3,3) {\(z^{(1)}z^{(2)}p_T\)};
        \vertex (e) at (3,1) {\(z^{(1)}(1-z^{(2)})p_T\)};
        \vertex (f) at (3,-1) {\((1-z^{(1)})z^{(3)}p_T\)};
        \vertex (g) at (3,-3) {\((1-z^{(1)})(1-z^{(3)})p_T\)} ;
    
        \diagram* {
          (a) -- [gluon] (o),
          (o) -- [gluon, edge label=\(z^{(1)}p_T\)] (b),
          (o) -- [gluon, edge label'=\((1-z^{(1)})p_T\)] (c),
          (b) -- [gluon](d),
          (b) -- [gluon] (e),
          (c) -- [gluon](f),
          (c) -- [gluon] (g),
        };
      \end{feynman} 
    \end{tikzpicture}
    \caption{Feynman diagram for a four-particle final state parton shower. The momentum fraction $z^{(i)}$ labels each vertex, where $i$ indexes the splitting order. An initial particle with  transverse momentum $p_T$ undergoes a splitting with momentum fraction $z^{(1)}$, producing daughters with momenta $z^{(1)}p_T$ and $(1-z^{(1)})p_T$. Both particles produced in this first splitting interact again with momentum fractions $z^{(2)}$ and $z^{(3)}$, producing the four final particles.}
    \label{fig:4particleFeynman}
\end{figure}
\begin{center}

\begin{figure}[H]
\begin{center}
    \begin{quantikz}
        \lstick{$\ket{0}$}&&&\gate[2]{U_S(z^{(2)})}&\meter{}\rstick{$\langle\sigma_3\rangle_A =z^{(1)}z^{(2)}$} \\
        \lstick{$\ket{0}$}&&\targ{}&&\meter{}\rstick{$\langle\sigma_3\rangle_B=z^{(1)}(1-z^{(2)})$}\\
        \lstick{$\ket{0}$}&\gate[2]{U_S(z^{(1)})}&\ctrl{-1}&& \\
        \lstick{$\ket{0}$}&&\ctrl{2}&& \\
        \lstick{$\ket{0}$}&&&\gate[2]{U_S(z^{(3)})}&\meter{}\rstick{$\langle\sigma_3\rangle_E =(1-z^{(1)})z^{(3)}$}\\        \lstick{$\ket{0}$}&&\targ{}&&\meter{}\rstick{$\langle\sigma_3\rangle_F =(1-z^{(1)})(1-z^{(3)})$}
    \end{quantikz}
\end{center}
    \caption{Four-particle splitting circuit obtained by iterating the $U_S$ primitive twice. The first block acts on two qubits initialized to $\ket{00}$, producing two particles with momentum fractions $z^{(1)}$ and $1-z^{(1)}$. Its outputs are each fed via a CNOT gate to two independent splitting blocks with fractions $z^{(2)}$ and $z^{(3)}$, which encode the splitting of the daughter partons of the first splitting. The expected value of $\sigma_3$ of the four outer qubits, $A, B, E$, and $F$ is measured, to obtain the same momentum fraction of the final-state particles of the Feynman diagram in Fig.~\ref{fig:4particleFeynman}. The two unmeasured intermediate qubits carry information about the first splitting fraction $z^{(1)}$ and can be read out to access the internal splitting structure.}
    \label{fig:FourParticleCircuit}
\end{figure}
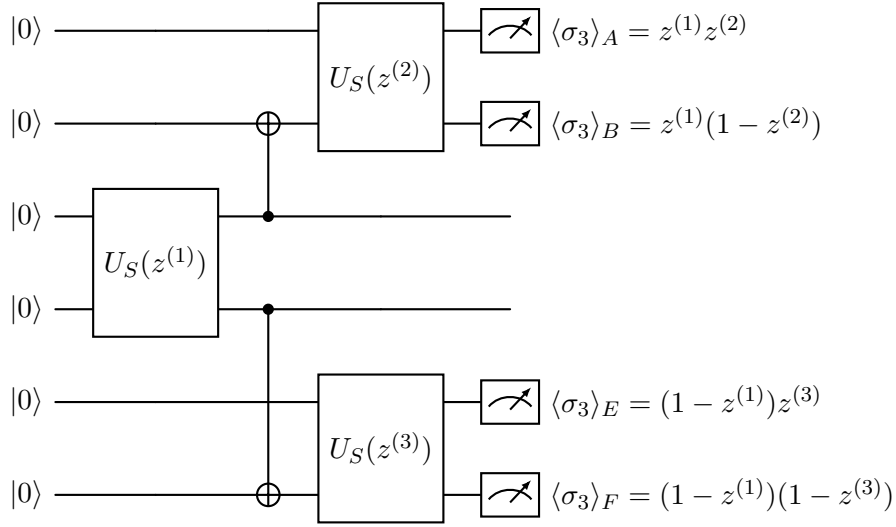

\end{center}
We can observe that although we only measure the momentum fraction of the three-particle final state, the circuit requires four qubits, where one of them serves as an ancillary qubit for the controlled operation. Now consider the  Feynman diagram in Fig.~\ref{fig:4particleFeynman}, where $z^{(i)}$ are the splitting fractions of the $i^{\text{th}}$ splitting and $p_T$ the transverse momentum of the initial particle. The quantum circuit that would allow us to measure the final momentum fractions of all four particles is presented in Fig.~\ref{fig:FourParticleCircuit}, whose four measured qubits encode the momentum fractions of the four final-state particles. On the other hand, if we choose to measure the two qubits in the middle (qubits C and D), we can access information about the first splitting fraction $z^{(1)}$, which means this circuit allows us to access information about intermediate splitting fractions. Moreover, the circuit exhibits a structure distinct from the ladder-like form of the dominant emission circuit. This structural difference indicates that incorporating secondary processes could provide a pathway beyond classical methods, as any ladder-structured circuit can be mapped to a matrix product state tensor network~\cite{vidalEfficientClassicalSimulation2003}. 

\begin{figure}[H]
\begin{minipage}[t]{0.48\textwidth}
    \centering
    \includegraphics[width=\linewidth]{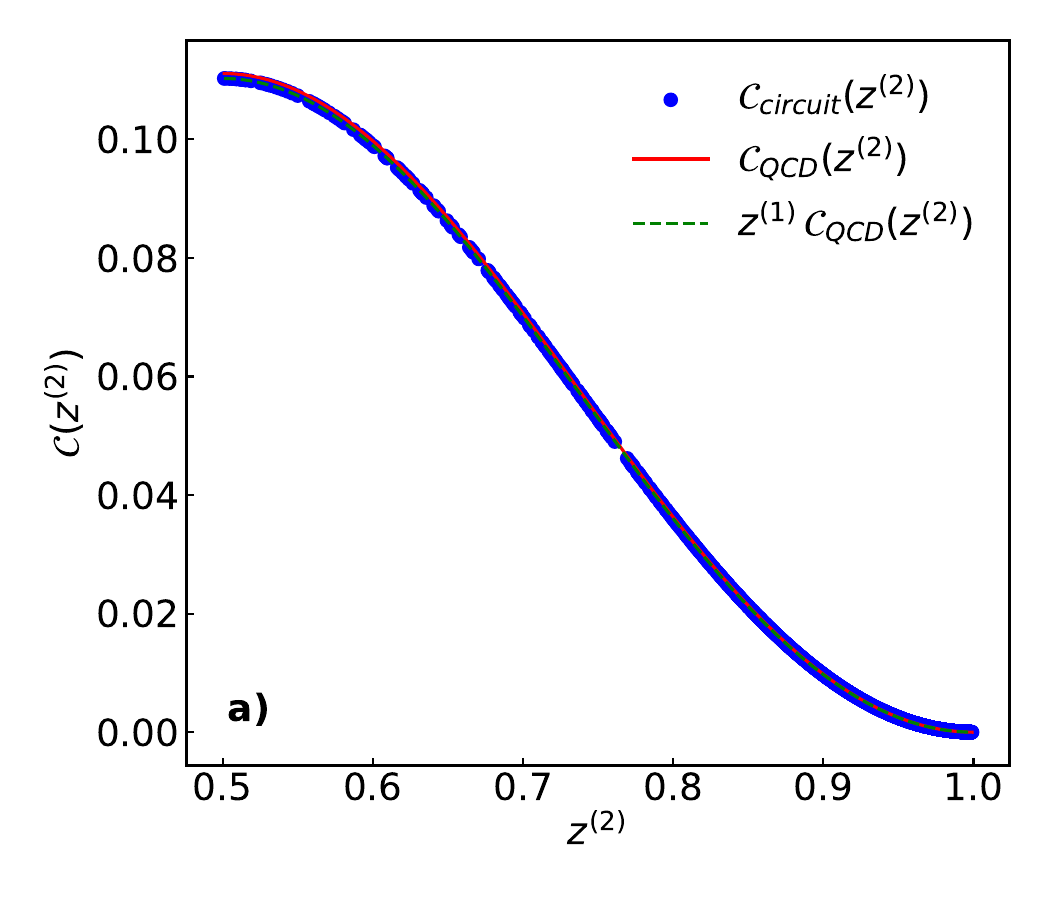}
    \vspace{-10mm}
\end{minipage}\hfill
\begin{minipage}[t]{0.48\textwidth}
    \centering
    \includegraphics[width=\linewidth]{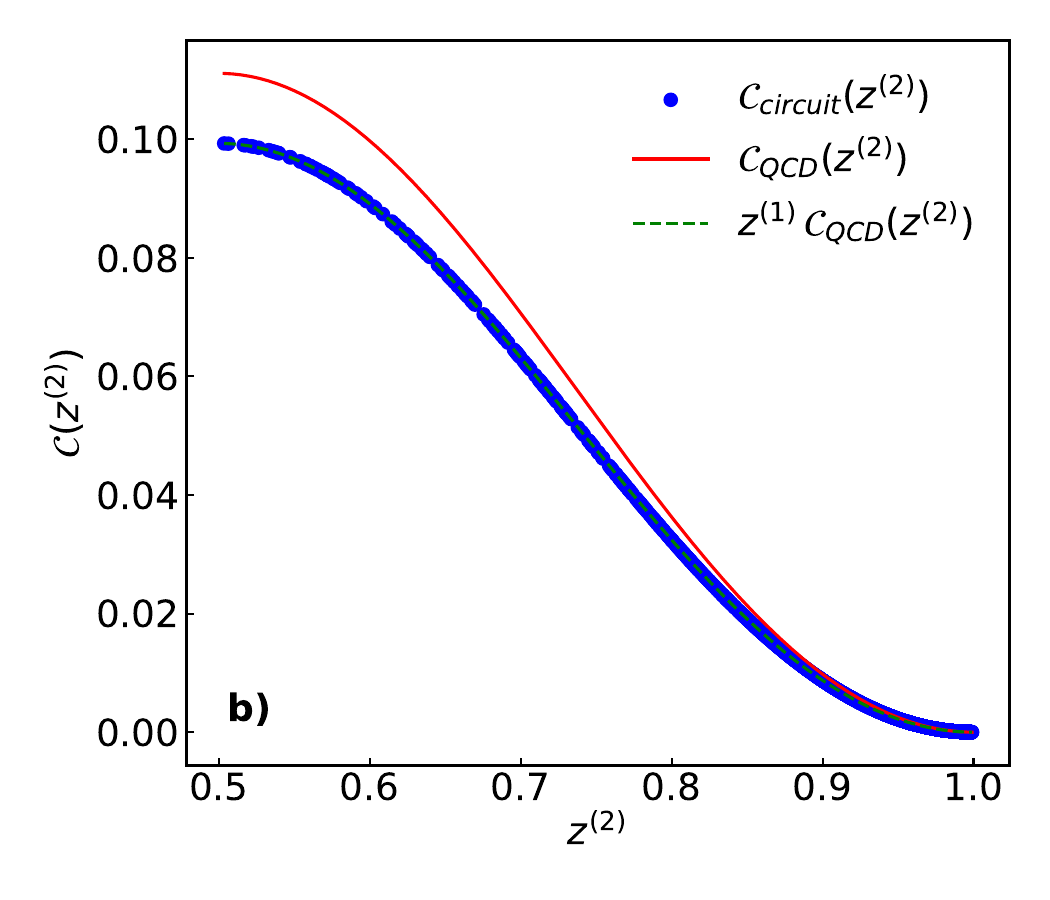}
    \vspace{-10mm}
\end{minipage}
\caption{Concurrence of the two-qubit reduced density matrix as a function of the second splitting fraction $z^{(2)}$, for fixed first splitting a) $z^{(1)}=0.9924$ and b) $z^{(1)}=0.8937$. Blue dots show the circuit results $\mathcal{C}_\text{circuit}(z^{(2)})$; the red curve shows the QCD prediction $\mathcal{C}_\text{QCD}(z^{(2)})$ from Eq.~\eqref{eqn:ConcurrenceQCDProcess}. The green dashed curve shows $z^{(1)}\mathcal{C}_\text{QCD}(z^{(2)})$, the scaling relation derived in Sec.~\ref{subsec:SplittingQuantumCircuit}. The near-perfect agreement in a) is a consequence of the limit $z^{(1)} \to 1$ restoring an input state of $\ket{00}$ to the second block. In b), the input state for the second block is now away from $\ket{00}$, and the difference between the two curves increases.}
\label{fig:ConcurrencePlots}
\end{figure}

In the preceding analysis, we have shown that the successive application of the splitting quantum circuit yields the product of the momentum fractions. Despite this, it is not obvious that the entanglement between the two qubits after the second splitting matches the QCD prediction, as the input is no longer the state $\ket{00}$, but a mixed state that depends on the first momentum fraction $z^{(1)}$. To assess this, we fix the first splitting fraction $z^{(1)}$ and compute, for each value of the second splitting fraction $z^{(2)}$, the concurrence of the reduced density matrix of the first two qubits, and compare it to the QCD value as expressed in Eq.~\eqref{eqn:ConcurrenceQCDProcess}. The result is presented in Fig.~\ref{fig:ConcurrencePlots} for initial splitting $z^{(1)}=0.9924$ and $z^{(1)}=0.8937$, where the red line is the curve given by Eq.~\eqref{eqn:ConcurrenceQCDProcess} while the blue circles represent the result from the quantum circuit. These plots show that the circuit concurrence for the same $z^{(2)}$ can vary with the parameters of the first splitting, and can deviate from the QCD prediction. When the initial splitting is close to $z^{(1)}=1$, the input to the second block approaches $\ket{00}$, which can be seen in Eq.~\eqref{eqn:FirstReducedState}, recovering the ideal condition under which the circuit was derived, and hence the agreement with QCD improves. 
The first splitting fraction is transferred using a CNOT, so the input state of the second block is a diagonal density matrix. The circuit concurrence is $\mathcal{C}_{\mathrm{circuit}}(z^{(2)}) = z^{(1)}\mathcal{C}_{\mathrm{QCD}}(z^{(2)})$. This relation is verified in the case of $z^{(1)}=0.9924$, where the relative deviation of the circuit concurrence from the QCD prediction is $1-0.9924=0.76\%$, while for $z^{(1)}=0.8937$ it is $1-0.8937=10.63\%$, for the whole range of $z^{(2)}$. This means that after $k$ splittings with fractions $\{z^{(i)}\}$, with $i=1,...,k$, the concurrence between the qubits produced at splitting $k$ follows $\mathcal{C}_{\mathrm{circuit}}(z^{(k)})=\prod_i^{k-1}z^{(i)}\mathcal{C}_{\mathrm{QCD}}(z^{(k)})$. Measurements of the intermediate wires give access to the earlier splitting fractions, from which the QCD-predicted concurrence can be recovered. Each new block acts only on the qubits descending from the harder daughter, so the entanglement produced at earlier splittings is left unchanged. The final state therefore retains the contribution of every vertex, each following the relation above. The two-qubit correlator introduced in Sec.~\ref{subsec:SplittingQuantumCircuit} behaves differently. At any depth it equals $\sec\gamma_1 - 1$ for the parameters of that block alone, unaffected by the preceding splittings. The entanglement imposed at each vertex therefore leaves a signature that composition does not degrade, even though the concurrence itself is reduced by the accumulated momentum fraction. The CNOT hands each new block an incoming qubit that is in $\ket{0}$ or $\ket{1}$ with probabilities set by the earlier splittings, and since the correlator takes the same value for either input, it does not depend on those probabilities.

\section{Splitting function on experimental data} \label{sec:Section3}

\subsection{Data preparation}\label{sec:DataPrep}
The splitting circuit introduced in Sec.~\ref{sec:Section2} is defined for a fixed momentum fraction $z$. To assess whether such a physics-informed quantum primitive can reproduce realistic splitting patterns, we calibrate and validate it on LHC data containing jet substructure. Concretely, we use reconstructed jets to extract momentum fractions that act as proxies for the earliest QCD branchings, and we use the resulting empirical $z$-distribution to determine circuit-parameter distributions for the two-qubit splitting module and to benchmark the distributions obtained after composing multiple splitting modules into three- and four-prong topologies.

We rely on the AspenOpenJets dataset~\cite{Amram:2024fjg}, which is derived from the 2016 CMS Open Data~\cite{cmscollaborationJetHTRun2016GUL2016_MiniAODv2v2MINIAOD2024, cmscollaborationJetHTRun2016HUL2016_MiniAODv2v2MINIAOD2024}. This dataset provides a large collection of reconstructed jets in a format optimised for physics and machine learning analyses. Our jet selection and declustering procedure closely follows the procedure established in~\cite{Larkoski:2017Exposing}.
 
Jets are initially reconstructed with the anti-$k_T$ algorithm~\cite{Cacciari:2008AntiKt} using a radius parameter $R = 0.8$. We select jets with transverse momentum $p_T > 300~\mathrm{GeV}$, pseudo-rapidity $|\eta| < 2.4$, and a tagging variable consistent with a high-quality reconstruction. From the selected events, we retrieve the associated particle-flow candidates and construct their four-momenta. To remove very soft radiation, we apply a constituent-level transverse momentum cut of $p_T > 1~\mathrm{GeV}$. This fixed threshold effectively suppresses soft wide-angle components. The filtered constituents are reclustered using the Cambridge/Aachen (C/A) algorithm~\cite{Dokshitzer:1997CA, Wobisch:1998CA} with a radius parameter $R = 0.4$. The resulting clustering tree is de-clustered to obtain a desired number of prongs. From these prongs, we compute the momentum fractions $\zeta_i = p_{Ti} / \sum_j p_{Tj}$, which serve as proxies for the momentum sharing at early stages of the parton shower. Throughout, $z$ and $z^{(k)}$ denote splitting fractions at a vertex, while $\zeta_i$ denotes the momentum fraction of a final prong. All momentum-fraction distributions shown in this work are normalised to unit area. For the two-prong sample used to calibrate the circuit, $z$ is the larger of the two momentum fractions, as defined in Appendix~\ref{app:MatrixElements}. This procedure provides a clean, reproducible reconstruction of angular-ordered prongs without relying on additional grooming parameters, thereby offering direct access to the kinematics of the first few QCD branchings.

\subsection{Quantum circuit parameters}\label{sec:QCParameters}
\begin{figure}[H]
    \centering
    \includegraphics[width=\linewidth]{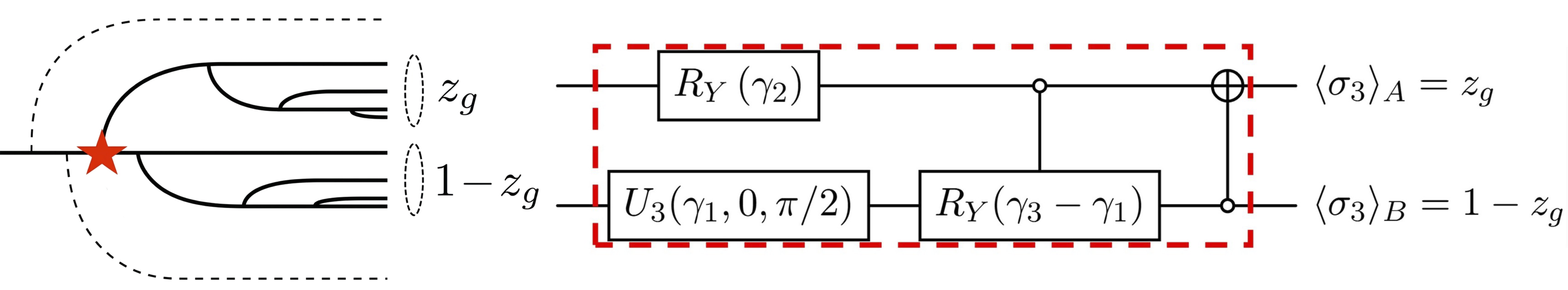}
    \caption{Mapping between the two-prong structure and the corresponding quantum circuit. The point where the two prongs split is represented by a red star (left), which is mapped to the quantum circuit with the red dashed square (right).}
    \label{fig:OneSplittingScheme}
\end{figure}

We now determine the circuit parameters that reproduce the measured momentum-fraction distribution while matching the QCD entanglement. From the two-prong sample extracted in Sec.~\ref{sec:DataPrep}, we obtain a dataset $\{z_1, ..., z_N\}$ of the momentum fractions for each event, one per jet. From this dataset, we determine the pairs of values $(\gamma_{1,i}, \gamma_{3,i})$, such that $z_i=z(\gamma_{1,i}, \gamma_{3,i})$ and $\mathcal{C}_\text{circuit}(\gamma_{1,i}, \gamma_{3,i})=\mathcal{C}_\text{QCD}(z_i)$. This constitutes a two-equation system with two variables, which can be solved numerically. Once $\gamma_{1,i}$ is determined, $\gamma_{2,i}$ follows from $\sin^2(\gamma_{2,i}/2)=1/(2\cos(\gamma_{1,i}))$. To achieve this, given a certain $z_i$, we define 
\begin{equation}
    \gamma_{3,i}(\gamma_{1,i}, z_i)=\arccos\left(2\frac{z_i-0.5}{\sec(\gamma_{1,i})-2}\right),
\end{equation}
which expresses $\gamma_{3,i}$ as a function of $\gamma_{1,i}$ and $z_i$. It can then be substituted into the concurrence-matching condition, which gives
\begin{equation}
    \mathcal{C}_\text{circuit}(\gamma_{1,i},\gamma_{3,i}(\gamma_{1,i}, z_i) )=\mathcal{C}_{QCD}(z_i),\label{eqn:entanglementmatching}
\end{equation}
where we numerically find the value of $\gamma_{1,i}$ that satisfies it. This uniquely determines all three parameters $(\gamma_{1,i}, \gamma_{2,i},\gamma_{3,i})$ for each $z_i$. After obtaining the parameter distribution, we obtain a mapping between the physical process and a quantum circuit, illustrated in Fig.~\ref{fig:OneSplittingScheme}. The red star marks the splitting vertex (with momentum fraction $z_g$), which is represented by the quantum circuit on the right. The distributions of $(\gamma_{1,i},\gamma_{3,i})$ are shown in Appendix~\ref{app:QCParameters}.

After determining the circuit's parameter distribution, we implement it with \textsc{Pennylane}~\cite{bergholm2022pennylane} and study its properties. Specifically, by computing the concurrence of the circuit's output density matrix, we validate the agreement with the concurrence of the scattering described in Eq.~\eqref{eqn:ConcurrenceQCDProcess}.  This validation is presented in Appendix~\ref{app:QCParameters}. We confirm that the circuit reproduces the empirical $z$-distribution of the two-prong jets and that its concurrence agrees with the prediction from QCD. The construction of a circuit that encodes entanglement structure and reproduces jet reconstruction data enables future studies of quantum information properties of jets, such as Von Neumann entropy, on quantum hardware, extending analyses that have so far been restricted to classical approaches~\cite{neillEntropyJet2019}.

\section{Quantum Computer Execution} \label{sec:Section4}
\begin{figure}[H]
\begin{minipage}[t]{0.33\textwidth}
    \centering
    \includegraphics[width=1.0\linewidth]{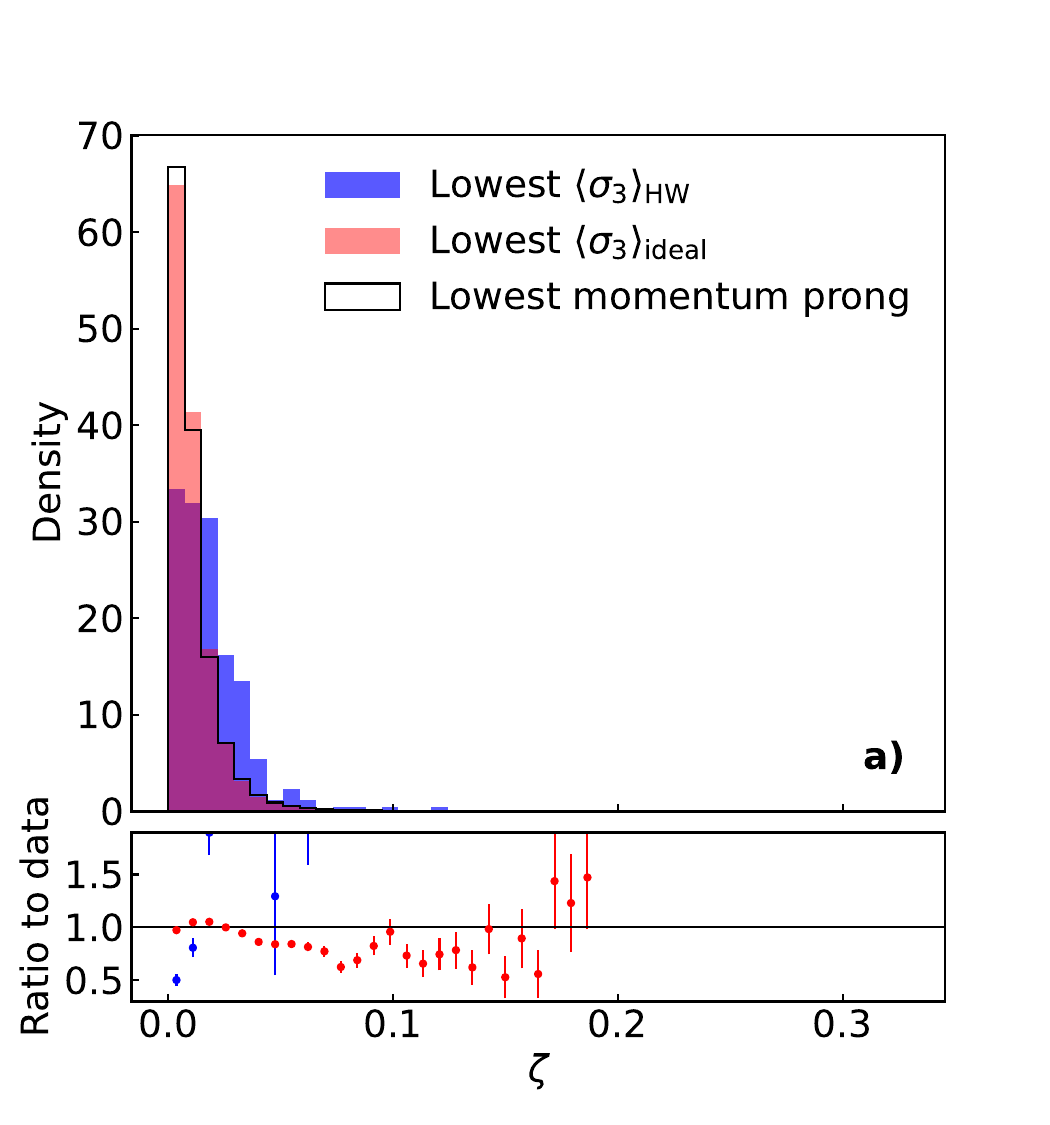}
\end{minipage}\hfill
\begin{minipage}[t]{0.33\textwidth}
    \centering
    \includegraphics[width=1.0\linewidth]{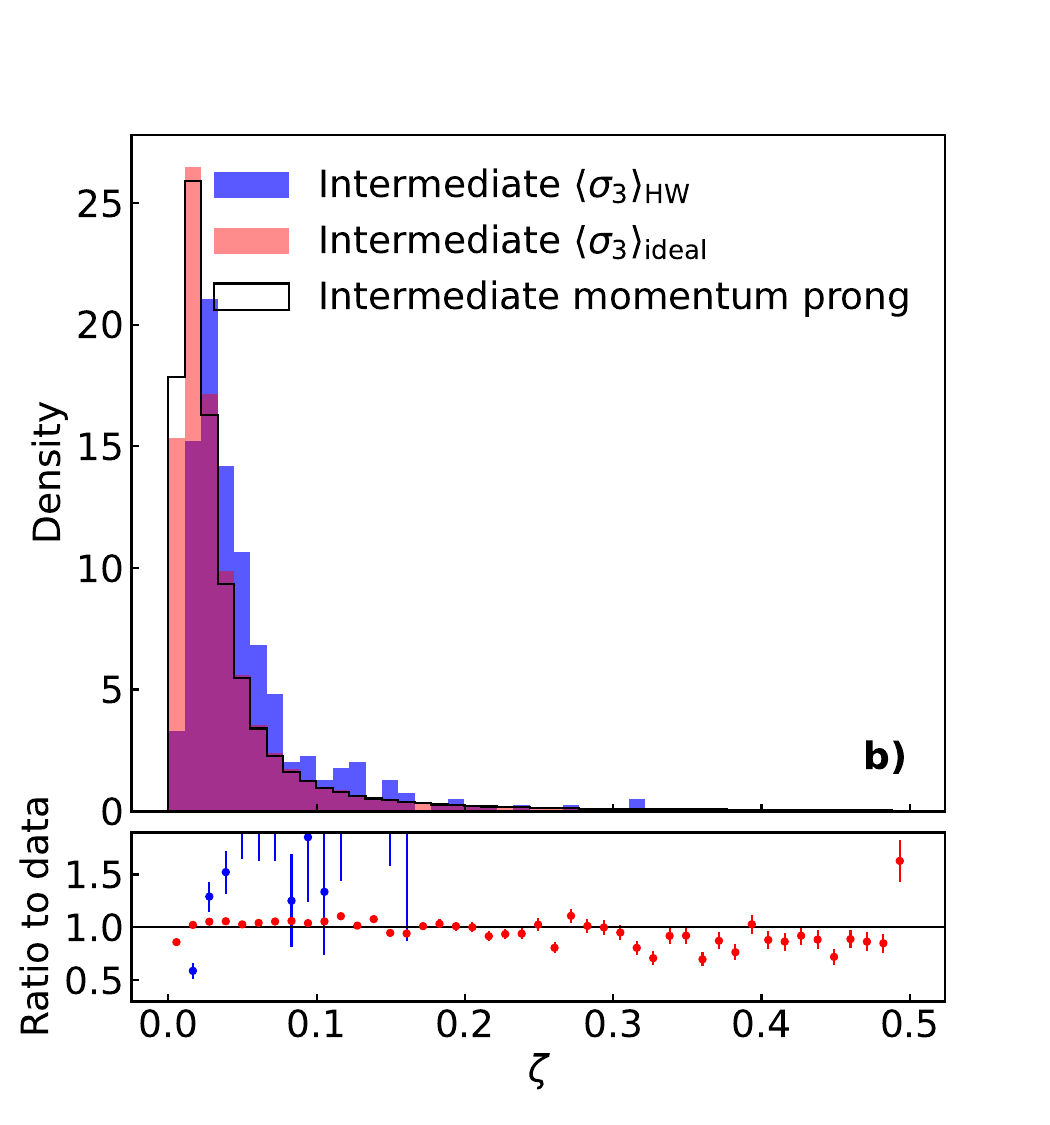}
\end{minipage}\hfill
\begin{minipage}[t]{0.33\textwidth}
    \centering
    \includegraphics[width=1.0\linewidth]{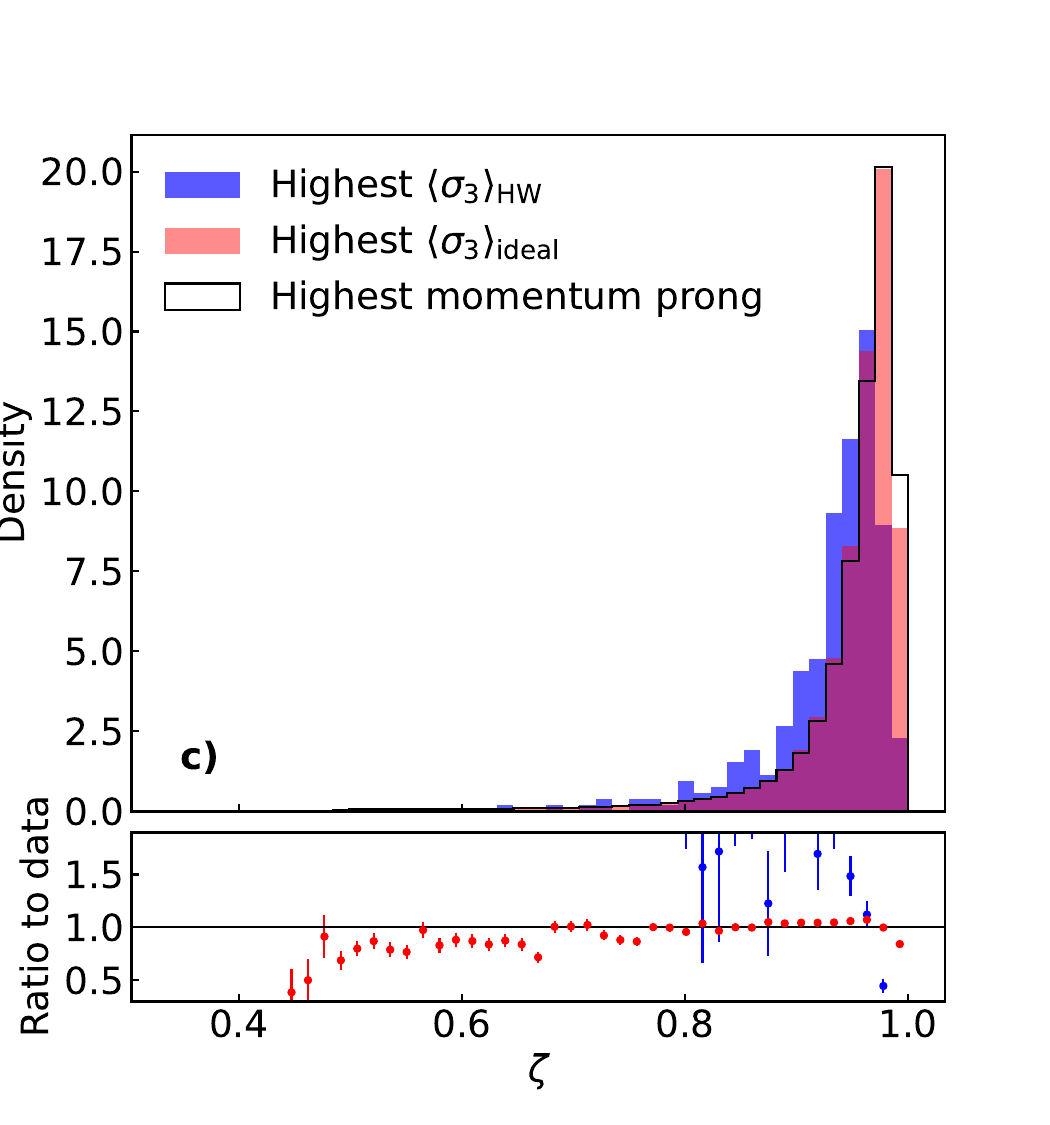}
\end{minipage}
\caption{Comparison of the momentum fraction $\zeta$ distribution of the a) lowest- b) intermediate- and c) highest-momentum prong obtained from the hardware execution of the quantum circuit on \textsc{ibm\_Marrakesh} in blue, the noiseless simulation of the quantum circuit in red, and the data from the AspenOpenJets data, shown with a black outline. The bottom panel of each plot shows the ratio of the two circuit distributions to the data in the same bins. The error bars give the statistical uncertainty of the circuit samples, while the data uncertainty is smaller than the marker size.}
\label{fig:AllProngs}
\end{figure}

After obtaining the set of parameters $\{\gamma_1, \gamma_3\}$ for the two-prong structure, we now extend the approach for more than two final particles by applying successive splitting circuits using the parameter distribution obtained above. To do so, we sample the splitting fraction at each vertex independently from the same two-prong distribution, and we always apply the next splitting to the harder of the two partons. This follows from the shower carrying no evolution variable dependence, which removes any dependence on previous splittings, and leaves nothing to select which parton branches next. The accuracy of these assumptions is tested in Appendix~\ref{app:Independence}. We first consider the application of two successive splittings with fractions $z^{(1)}$ and $z^{(2)}$, as presented in Fig.~\ref{fig:ThreeParticleCircuit},  to get the momentum fraction of three final prongs. The three prong fractions are then $\zeta = \{1-z^{(1)},\; z^{(1)}(1-z^{(2)}),\; z^{(1)}z^{(2)}\}$, sorted by momentum, and these are then compared with the data.
Simultaneously, we now reconstruct the data to obtain the three hardest prongs, obtained in an analogous way to Sec.~\ref{sec:DataPrep}, which allows us to compare them with the result from our quantum circuit. The three-prong momentum fraction distributions were obtained both with the quantum circuit running on an ideal quantum simulator and on the \textsc{ibm\_Marrakesh} hardware. The results were obtained from $600$ independent circuit executions with $1024$ shots of the quantum circuit, using independently sampled parameters $\{(\gamma_1, \gamma_3), (\gamma_1', \gamma_3')\}$ from the distribution obtained in Sec.~\ref{sec:QCParameters}. The circuits were transpiled to the device basis at optimization level one and executed on physical qubits $0$ to $3$ of \textsc{ibm\_Marrakesh}, a $156$-qubit Heron r2 processor. Each transpiled circuit contains seven two-qubit gates and has depth $33$. Error mitigation was applied at resilience level two, combining readout error mitigation, Pauli twirling of the two-qubit gates and of the measurements, and zero-noise extrapolation from noise factors $1$, $3$ and $5$, with the $1024$ shots of each execution distributed over $16$ twirling randomizations. To ensure a fair comparison between the circuit and the data, we compare the highest, intermediate, and lowest momentum fractions against the corresponding measured expected values. Furthermore, hardware execution requires additional considerations, since noise and device imperfections can cause the observed outputs to deviate from the values expected from the programmed dynamics~\cite{tuysuz2025learningresponsefunctionsanalog, Borras_2023}. We measure all four circuit outputs, not just the three equivalent to the three final prongs. This extra measurement will give us the momentum fraction of the first splitting $z^{(1)}$, but, as we have seen, this expected value will be close to one, meaning it is more resilient to shot noise than measuring $1-z^{(1)}$ in the last qubit. This occurs because $1-z^{(1)}$ should be positive in an ideal circuit, but due to hardware errors, the measured value may become negative. So, by measuring $z^{(1)}$ and $z^{(1)}z^{(2)}$ from qubits $C$ and $A$, respectively, the momentum fractions $1-z^{(1)}$ and $z^{(1)}-z^{(1)}z^{(2)}=z^{(1)}(1-z^{(2)})$ of the other particles are also obtained. Finally, we correct for the bias introduced by the device. A subset of $100$ executions, disjoint from those used in the comparison, is used for calibration. For each of the two measured qubits that output $z^{(1)}$ and $z^{(1)}z^{(2)}$, we take the difference between the mean measured value from the hardware and one from the ideal circuit, and apply that shift to the remaining $500$ executions. After this correction a fraction of the estimates still falls outside $[0,1]$, which is not physical. These estimates are excluded, corresponding to $28.8\%$ of executions. Comparisons between the ideal simulation, the hardware execution, and the real data for the three prongs are shown in Fig.~\ref{fig:AllProngs}.

\begin{figure}[H]
\begin{center}
    \includegraphics[width=0.62\textwidth]{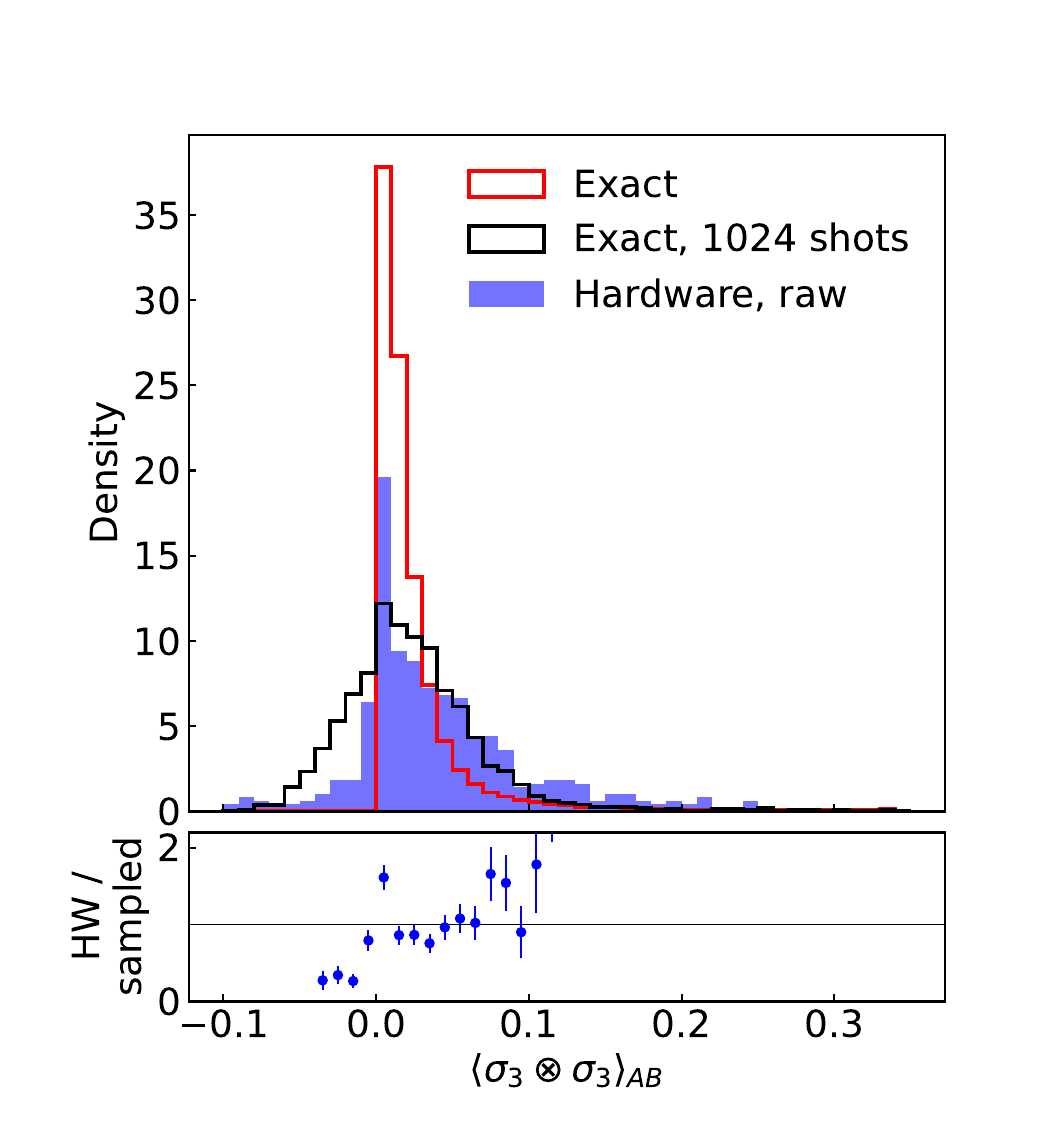}
\end{center}
    \vspace{-5mm}
    \caption{Distribution of the two-qubit correlator $\langle\sigma_3\otimes\sigma_3\rangle_{AB}$ over the calibrated parameter set. The exact circuit values (red), circuits drawn from the same parameter distribution and simulated without noise at $1024$ shots (black), and the raw hardware result on \textsc{ibm\_marrakesh} (blue). The lower panel shows the ratio of the hardware to the sampled result.}
    \label{fig:HWCorrelator}
\end{figure}

The hardware execution also gives access to a quantity sensitive to the entanglement. The two-qubit correlator $\langle\sigma_3\otimes\sigma_3\rangle_{AB}$ of qubits $A$ and $B$, the pair produced at the second splitting, is the observable introduced in Sec.~\ref{subsec:SplittingQuantumCircuit}. As mentioned before, its value is a consequence of the entanglement matching, and it is unaffected by the splittings that precede it. Figure~\ref{fig:HWCorrelator} compares the exact circuit, a noiseless simulation with $1024$ shots, and raw hardware output without postprocessing. Both measured distributions are much broader than the exact one, and comparing them separates the two effects. Firstly, the broadening is already produced by the finite number of shots, while the device adds a shift towards larger values together with a longer tail. The measured correlator remains positive on average and concentrated in the same region as the exact result. The measure is affected by device noise, and increasing the number of shots could improve the agreement. As this measure is independent of previous splittings and of the measured momentum fractions, it can be used to probe the entanglement of each splitting independently.

The corresponding plots before postprocessing (beyond the exclusion of unphysical measurements) are presented in Appendix~\ref{app:Simulation}. From the plots, we observe that after postprocessing, both the ideal simulation and the hardware reproduce the momentum fractions of the three-prong structure. This tells us that by just adapting the parameters of the two-qubit quantum circuit to replicate the two-prong structure, successive applications of this primitive generalize the momentum distribution of a higher number of prongs. The limited impact of hardware noise is attributed to the small number of qubits and the low circuit depth. Although postprocessing our measurements greatly improved our result, it remains unclear how this procedure would generalize to cases with dozens of particles, since even the highest-momentum particle could have a momentum fraction far from unity, thus being more prone to hardware noise. It is important to note that this circuit is not the only configuration that can mimic the three-prong structure. The other configuration is the one present in Fig.~\ref{fig:ThreeParticleSecondary}, where now the lower momentum particle produced in the first splitting is the one that interacts again. However, from the plots above, we see that the dominant configuration alone reproduces the data across the bulk of the range. This is expected from theory, as higher-momentum particles are much more likely to split again, whereas lower-momentum ones tend to hadronize, instead of undergoing further splitting. The circuit nevertheless falls below the data as $\zeta\to1$. Since a splitting of the softer parton leaves the leading prong undivided, this is the direction expected if such configurations retain some influence in that region. We next consider the quantum circuit in Fig.~\ref{fig:DominantFourParticle}, representing the dominant configuration for four-prong production. We run this circuit on an ideal simulation, following the same approach as in the three-prong case.

\vspace{-5mm}
\begin{figure}[H]
\begin{center}
    \begin{quantikz}
        
        \lstick{$\ket{0}$}&\gate[2]{U_S(z(\gamma_1, \gamma_3))}&&& \meter{} \\
        \lstick{$\ket{0}$}&&\ctrl{2}&& \\
        \lstick{$\ket{0}$}&&&\gate[2]{U_S(z(\gamma'_1, \gamma'_3)=z^{(2)})}&\meter{} \\
        \lstick{$\ket{0}$}&&\targ{}&&\meter{}\\
    \end{quantikz}
\end{center}
    \vspace{-5mm}
    \caption{Similar to Fig.~\ref{fig:ThreeParticleCircuit}, but instead of connecting the qubit corresponding to the highest momentum particle from the first block to the second splitting block, we connect the one corresponding to the lowest momentum, representing a secondary emission.}
    \label{fig:ThreeParticleSecondary}
\end{figure}

\begin{figure}[H]
\begin{minipage}[t]{0.49\textwidth}
    \centering
    \includegraphics[width=1.0\linewidth]{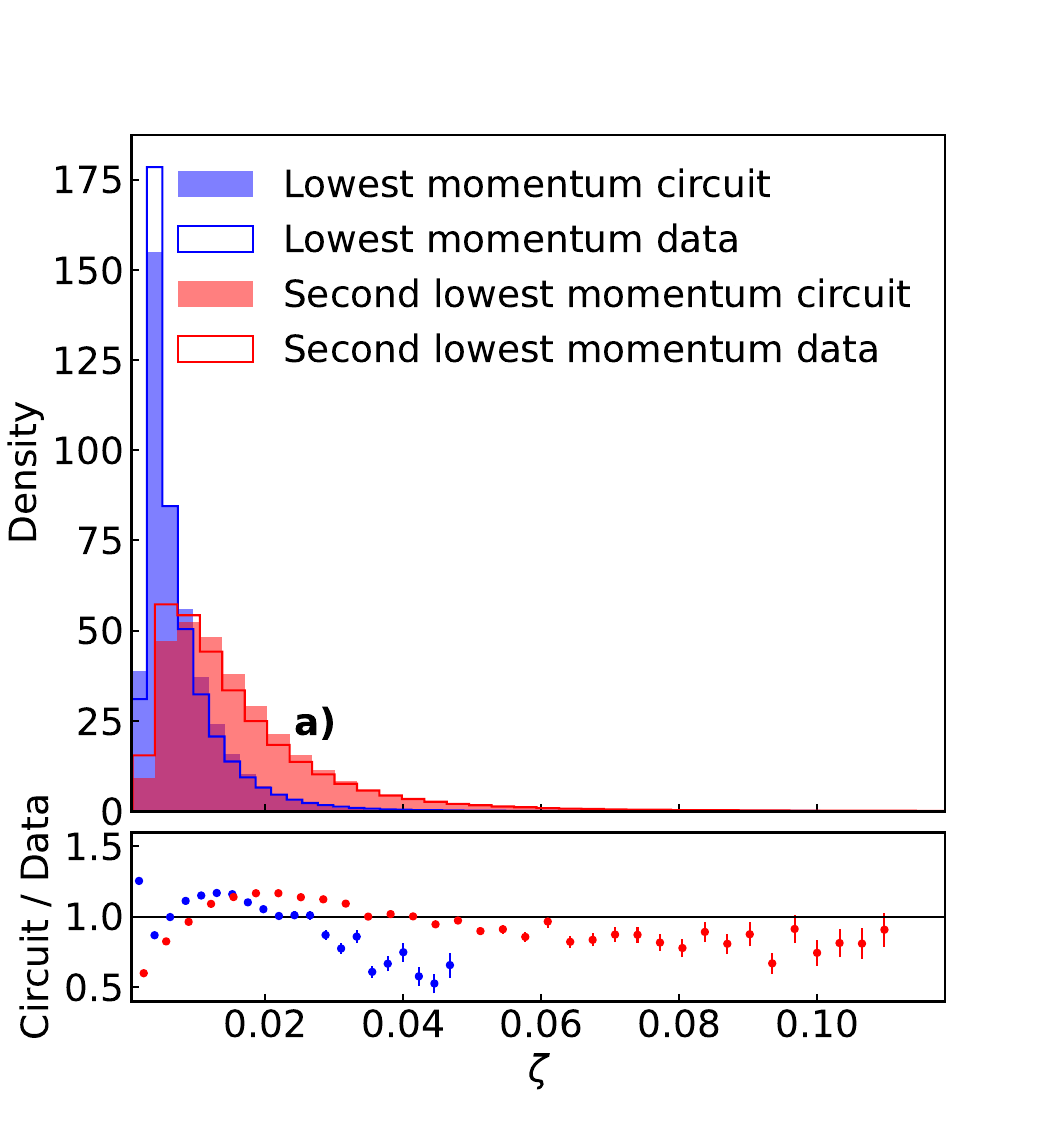}
\end{minipage}
\hfill
\begin{minipage}[t]{0.49\textwidth}
    \centering
    \includegraphics[width=1.0\linewidth]{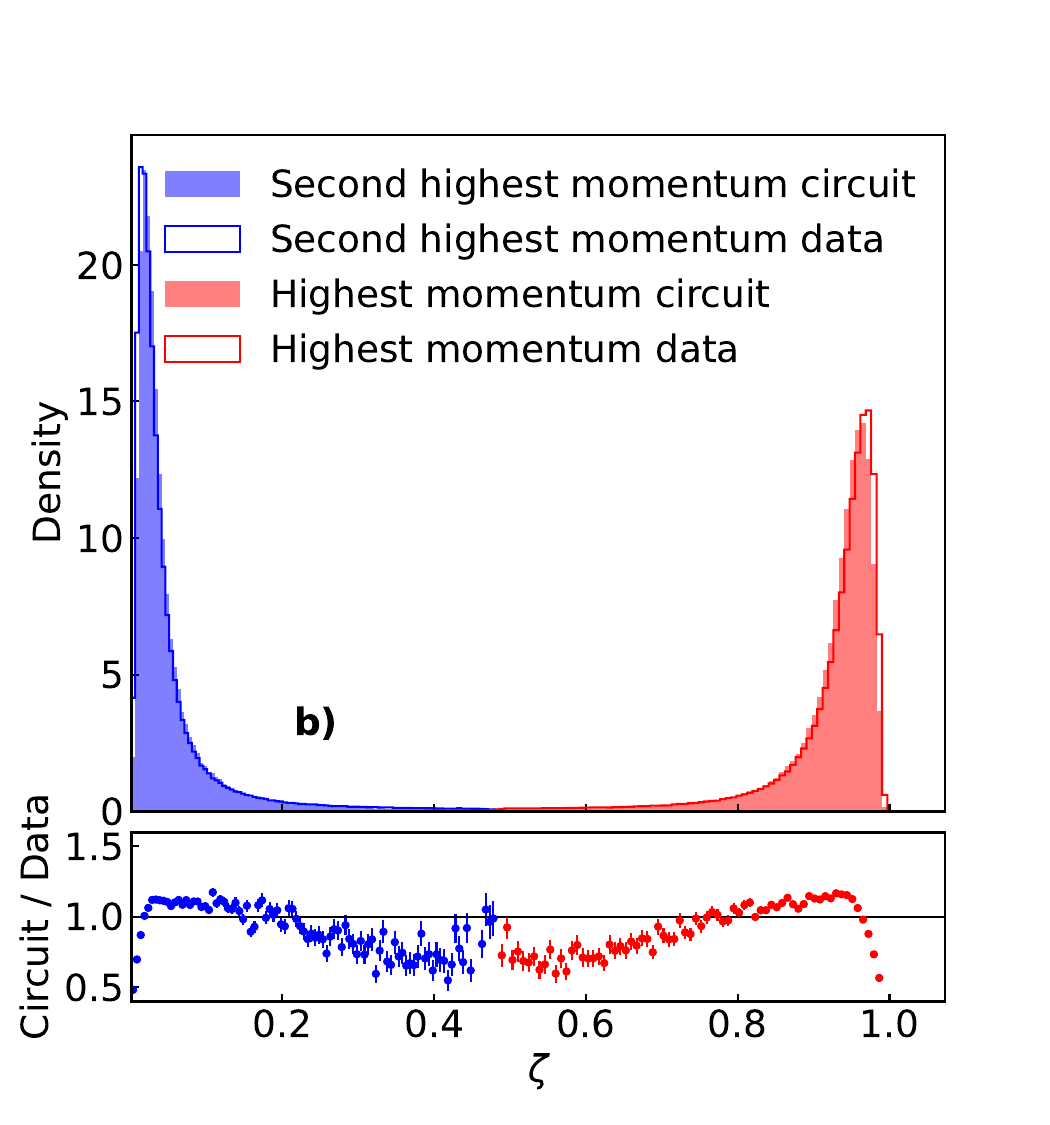}
\end{minipage}
    \vspace{-5mm}
    \caption{Comparison of the momentum fraction $\zeta$ distributions of a) the two lowest- and b) the two highest-momentum prongs. Filled histograms show the noiseless simulation of the quantum circuit, and outlines of the same colour the AspenOpenJets data. The lower panel of each figure shows the ratio of circuit to data in each bin with the statistical uncertainties.}
    \label{fig:FourProngsPlot}
\end{figure}

We obtain the momentum fraction distribution and compare it with the ones of the four main prongs extracted from the dataset. The result is shown in Fig.~\ref{fig:FourProngsPlot}. The dominant circuit, in which the highest-momentum particle undergoes successive splittings, reproduces the data across the bulk of the range. However, the circuit falls below the data when the highest momentum fraction is close to one, and consequently the remaining fractions are close to zero, as seen in the ratio plots. The comparison also becomes worse at the tails of the distributions. This follows from the assumptions considered above. In particular, we neglect splittings of the softer parton, which would leave the leading prong undivided and therefore harder. Appendix~\ref{app:Independence} quantifies how far each assumption departs from the data. To develop a complete quantum algorithm for parton showers, secondary splittings must be accounted for. The existence of these splittings is governed by the Sudakov factors~\cite{buckley2021practical}, meaning an adaptation of the procedure presented in this work that incorporates the probability of splitting is a possible future direction for this work.

We note that, since the parameters of our circuit are determined analytically to reproduce the measured two-prong distribution, applying the circuit repeatedly gives the same result as sampling the splitting fractions independently from that distribution and multiplying them. This means that the multi-prong distributions of Figs.~\ref{fig:AllProngs} and~\ref{fig:FourProngsPlot} can also be obtained classically. Nonetheless, reconstructing the entanglement structure of the splittings would require an additional computation in the classical algorithm, while our quantum algorithm automatically stores the entanglement. These comparisons show that our splitting primitive composes correctly, that momentum is conserved exactly at each step, and that the parameters fitted to the two-prong structure carry over to a higher number of prongs without refitting.

\begin{figure}[H]
\begin{center}
    \begin{quantikz}
        \lstick{$\ket{0}$}&&&&&\gate[2]{U_S(z^{(3)})}&\meter{}\\
        \lstick{$\ket{0}$}&&&&\targ{}&&\meter{}\\
        \lstick{$\ket{0}$}&&&\gate[2]{U_S(z^{(2)})}&\ctrl{-1}&& \\
        \lstick{$\ket{0}$}&&\targ{}&&&&\meter{}\\
        \lstick{$\ket{0}$}&\gate[2]{U_S(z^{(1)})}&\ctrl{-1}&&&& \\
        \lstick{$\ket{0}$}&&&&&&\meter{} \\
    \end{quantikz}
\end{center}
    \vspace{-5mm}
    \caption{Extension of Fig.~\ref{fig:ThreeParticleCircuit} to the dominant four-prong structure, where the qubit associated with the highest-momentum particle after the second splitting block is connected via a CNOT gate to a third splitting block.}
    \label{fig:DominantFourParticle}
\end{figure}

\vspace{-3mm}

\section{Summary and Conclusion} \label{sec:Section5}

In this paper, we develop a foundational building block toward a complete quantum simulation of parton showers by creating a fundamental building block for HEP simulations that incorporates quantum information aspects of QCD into its design. This tool explicitly connects a study of entanglement from the quantum information perspective of a QCD process in the perturbative regime with quantum computing. Starting from a simple, yet informative three-gluon scattering process in the perturbative regime, we first studied the entanglement generated by this process, encoding helicity states as qubit states. Next, we developed a two-qubit quantum circuit that conserves momentum and produces the momentum fraction of a scattering event. This circuit has tunable parameters that can reproduce the momentum-fraction distribution observed in a dataset while maintaining entanglement equivalent to that in quantum chromodynamics.

The circuit is designed so that successive applications are identical to successive splittings, producing the momentum fractions of the final partons. Initially, we determined the distribution of the circuit's parameters to match the two-prong momentum distribution of the provided dataset, while verifying the agreement with the entanglement produced in QCD. To validate how successive applications of the splitting primitive generalize a multi-prong structure, we applied the circuit with the parameter distribution obtained before twice to obtain predictions of the momentum fractions for three prongs. We compared the results with the dataset, observing agreement across the middle of the distribution with departures at large momentum fraction. The same behaviour is found in the four-prong scenario, where the departures are larger. In the specific case of the three-prong structure, we also ran the circuit on the \textsc{ibm\_Marrakesh} hardware and verified that, after some postprocessing of the results, there was good agreement with the experimental data, which can be attributed to the low number of qubits and low depth of the circuit.

 Despite its proof-of-concept nature, this work represents a step toward implementing a full quantum algorithm for parton showers. However, essential elements of a complete formulation remain to be developed. The parton shower process depends on using the Sudakov factor to update the virtuality of the incoming particle. This is equivalent to executing the time evolution of the shower, but a quantum algorithm that does this has yet to be developed. Moreover, one must consider the different parton species involved in a shower, an aspect that the real data we used do not account for. Similarly, the colour structure is essential for hadronization, and further work is required to develop its quantum simulation, in which the mapping of group generators to quantum gates may offer a computational advantage. Overall, this work provides a first step toward building simulations in which the role of quantum entanglement is physically motivated in the design. This may prove fundamental to the execution of future collider experiments, as quantum information becomes an integral component of particle physics. Furthermore, the circuit built could serve as the basic element for future physics-aware quantum machine learning models, with its parameters treated as trainable variables in a variational framework.

\section*{Acknowledgements}
GR acknowledges that the project that gave rise to these results received the support of a fellowship from ”la Caixa” Foundation (ID 100010434). The fellowship code is LCF/BQ/ EU23/12010091. GR also acknowledges funding by the Deutsche Forschungsgemeinschaft (DFG, German Research Foundation) under Germany's Excellence Strategy – EXC-2111 – 390814868, and the European Research Council (ERC) under the European Union's Horizon Europe research and innovation program (Grant Agreement No.~101165667)-ERC Starting Grant QuSiGauge. Y.H. is supported by the Open Quantum Institute (OQI); the OQI itself is an initiative hosted by CERN, born at GESDA, and supported by UBS. CT, MG are supported by CERN through the CERN Quantum Technology Initiative. This work is part of the Quantum Computing for High-Energy Physics (QC4HEP) working group.

\bibliographystyle{JHEP}
\bibliography{biblo}

@article{Albrecht2019Roadmap,
  title = {{A Roadmap for HEP Software and Computing R\&D for the 2020s}},
  author = {Albrecht, Johannes and Alves, Antonio Augusto and Amadio, Guilherme and Andronico, Giuseppe and Anh-Ky, Nguyen and Aphecetche, Laurent and Apostolakis, John and Asai, Makoto and Atzori, Luca and Babik, Marian and Bagliesi, Giuseppe and Bandieramonte, Marilena and Banerjee, Sunanda and Barisits, Martin and Bauerdick, Lothar A. T. and Belforte, Stefano and Benjamin, Douglas and Bernius, Catrin and Bhimji, Wahid and Bianchi, Riccardo Maria and Bird, Ian and Biscarat, Catherine and Blomer, Jakob and Bloom, Kenneth and Boccali, Tommaso and Bockelman, Brian and Bold, Tomasz and Bonacorsi, Daniele and Boveia, Antonio and Bozzi, Concezio and Bracko, Marko and Britton, David and Buckley, Andy and Buncic, Predrag and Calafiura, Paolo and Campana, Simone and Canal, Philippe and Canali, Luca and Carlino, Gianpaolo and Castro, Nuno and Cattaneo, Marco and Cerminara, Gianluca and Cervantes Villanueva, Javier and Chang, Philip and Chapman, John and Chen, Gang and Childers, Taylor and Clarke, Peter and Clemencic, Marco and Cogneras, Eric and Coles, Jeremy and Collier, Ian and Colling, David and Corti, Gloria and Cosmo, Gabriele and Costanzo, Davide and Couturier, Ben and Cranmer, Kyle and Cranshaw, Jack and Cristella, Leonardo and Crooks, David and Crépé-Renaudin, Sabine and Currie, Robert and Dallmeier-Tiessen, Sünje and De, Kaushik and De Cian, Michel and De Roeck, Albert and Delgado Peris, Antonio and Derue, Frédéric and Di Girolamo, Alessandro and Di Guida, Salvatore and Dimitrov, Gancho and Doglioni, Caterina and Dotti, Andrea and Duellmann, Dirk and Duflot, Laurent and Dykstra, Dave and Dziedziniewicz-Wojcik, Katarzyna and Dziurda, Agnieszka and Egede, Ulrik and Elmer, Peter and Elmsheuser, Johannes and Elvira, V. Daniel and Eulisse, Giulio and Farrell, Steven and Ferber, Torben and Filipcic, Andrej and Fisk, Ian and Fitzpatrick, Conor and Flix, José and Formica, Andrea and Forti, Alessandra and Franzoni, Giovanni and Frost, James and Fuess, Stu and Gaede, Frank and Ganis, Gerardo and Gardner, Robert and Garonne, Vincent and Gellrich, Andreas and Genser, Krzysztof and George, Simon and Geurts, Frank and Gheata, Andrei and Gheata, Mihaela and Giacomini, Francesco and Giagu, Stefano and Giffels, Manuel and Gingrich, Douglas and Girone, Maria and Gligorov, Vladimir V. and Glushkov, Ivan and Gohn, Wesley and Gonzalez Lopez, Jose Benito and González Caballero, Isidro and González Fernández, Juan R. and Govi, Giacomo and Grandi, Claudio and Grasland, Hadrien and Gray, Heather and Grillo, Lucia and Guan, Wen and Gutsche, Oliver and Gyurjyan, Vardan and Hanushevsky, Andrew and Hariri, Farah and Hartmann, Thomas and Harvey, John and Hauth, Thomas and Hegner, Benedikt and Heinemann, Beate and Heinrich, Lukas and Heiss, Andreas and Hernández, José M. and Hildreth, Michael and Hodgkinson, Mark and Hoeche, Stefan and Holzman, Burt and Hristov, Peter and Huang, Xingtao and Ivanchenko, Vladimir N. and Ivanov, Todor and Iven, Jan and Jashal, Brij and Jayatilaka, Bodhitha and Jones, Roger and Jouvin, Michel and Jun, Soon Yung and Kagan, Michael and Kalderon, Charles William and Kane, Meghan and Karavakis, Edward and Katz, Daniel S. and Kcira, Dorian and Keeble, Oliver and Kersevan, Borut Paul and Kirby, Michael and Klimentov, Alexei and Klute, Markus and Komarov, Ilya and Konstantinov, Dmitri and Koppenburg, Patrick and Kowalkowski, Jim and Kreczko, Luke and Kuhr, Thomas and Kutschke, Robert and Kuznetsov, Valentin and Lampl, Walter and Lancon, Eric and Lange, David and Lassnig, Mario and Laycock, Paul and Leggett, Charles and Letts, James and Lewendel, Birgit and Li, Teng and Lima, Guilherme and Linacre, Jacob and Linden, Tomas and Livny, Miron and Lo Presti, Giuseppe and Lopienski, Sebastian and Love, Peter and Lyon, Adam and Magini, Nicolò and Marshall, Zachary L. and Martelli, Edoardo and Martin-Haugh, Stewart and Mato, Pere and Mazumdar, Kajari and McCauley, Thomas and McFayden, Josh and McKee, Shawn and McNab, Andrew and Mehdiyev, Rashid and Meinhard, Helge and Menasce, Dario and Mendez Lorenzo, Patricia and Mete, Alaettin Serhan and Michelotto, Michele and Mitrevski, Jovan and Moneta, Lorenzo and Morgan, Ben and Mount, Richard and Moyse, Edward and Murray, Sean and Nairz, Armin and Neubauer, Mark S. and Norman, Andrew and Novaes, Sérgio and Novak, Mihaly and Oyanguren, Arantza and Ozturk, Nurcan and Pacheco Pages, Andres and Paganini, Michela and Pansanel, Jerome and Pascuzzi, Vincent R. and Patrick, Glenn and Pearce, Alex and Pearson, Ben and Pedro, Kevin and Perdue, Gabriel and Perez-Calero Yzquierdo, Antonio and Perrozzi, Luca and Petersen, Troels and Petric, Marko and Petzold, Andreas and Piedra, Jónatan and Piilonen, Leo and Piparo, Danilo and Pivarski, Jim and Pokorski, Witold and Polci, Francesco and Potamianos, Karolos and Psihas, Fernanda and Puig Navarro, Albert and Quast, Günter and Raven, Gerhard and Reuter, Jürgen and Ribon, Alberto and Rinaldi, Lorenzo and Ritter, Martin and Robinson, James and Rodrigues, Eduardo and Roiser, Stefan and Rousseau, David and Roy, Gareth and Rybkine, Grigori and Sailer, Andre and Sakuma, Tai and Santana, Renato and Sartirana, Andrea and Schellman, Heidi and Schovancová, Jaroslava and Schramm, Steven and Schulz, Markus and Sciabà, Andrea and Seidel, Sally and Sekmen, Sezen and Serfon, Cedric and Severini, Horst and Sexton-Kennedy, Elizabeth and Seymour, Michael and Sgalaberna, Davide and Shapoval, Illya and Shiers, Jamie and Shiu, Jing-Ge and Short, Hannah and Siroli, Gian Piero and Skipsey, Sam and Smith, Tim and Snyder, Scott and Sokoloff, Michael D. and Spentzouris, Panagiotis and Stadie, Hartmut and Stark, Giordon and Stewart, Gordon and Stewart, Graeme A. and Sánchez, Arturo and Sánchez-Hernández, Alberto and Taffard, Anyes and Tamponi, Umberto and Templon, Jeff and Tenaglia, Giacomo and Tsulaia, Vakhtang and Tunnell, Christopher and Vaandering, Eric and Valassi, Andrea and Vallecorsa, Sofia and Valsan, Liviu and Van Gemmeren, Peter and Vernet, Renaud and Viren, Brett and Vlimant, Jean-Roch and Voss, Christian and Votava, Margaret and Vuosalo, Carl and Vázquez Sierra, Carlos and Wartel, Romain and Watts, Gordon T. and Wenaus, Torre and Wenzel, Sandro and Williams, Mike and Winklmeier, Frank and Wissing, Christoph and Wuerthwein, Frank and Wynne, Benjamin and Zhang, Xiaomei and Yang, Wei and Yazgan, Efe},
  journal = {Computing and Software for Big Science},
  volume = {3},
  number = {1},
  pages = {7},
  year = {2019},
  doi = {10.1007/s41781-018-0018-8},
  url = {https://doi.org/10.1007/s41781-018-0018-8}
}

@article{PRXQuantum.5.037001,
  title = {{Quantum Computing for High-Energy Physics: State of the Art and Challenges}},
  author = {Di Meglio, Alberto and Jansen, Karl and Tavernelli, Ivano and Alexandrou, Constantia and Arunachalam, Srinivasan and Bauer, Christian W. and Borras, Kerstin and Carrazza, Stefano and Crippa, Arianna and Croft, Vincent and de Putter, Roland and Delgado, Andrea and Dunjko, Vedran and Egger, Daniel J. and Fern\'andez-Combarro, Elias and Fuchs, Elina and Funcke, Lena and Gonz\'alez-Cuadra, Daniel and Grossi, Michele and Halimeh, Jad C. and Holmes, Zo\"e and K\"uhn, Stefan and Lacroix, Denis and Lewis, Randy and Lucchesi, Donatella and Martinez, Miriam Lucio and Meloni, Federico and Mezzacapo, Antonio and Montangero, Simone and Nagano, Lento and Pascuzzi, Vincent R. and Radescu, Voica and Ortega, Enrique Rico and Roggero, Alessandro and Schuhmacher, Julian and Seixas, Joao and Silvi, Pietro and Spentzouris, Panagiotis and Tacchino, Francesco and Temme, Kristan and Terashi, Koji and Tura, Jordi and T\"uys\"uz, Cenk and Vallecorsa, Sofia and Wiese, Uwe-Jens and Yoo, Shinjae and Zhang, Jinglei},
  journal = {PRX Quantum},
  volume = {5},
  issue = {3},
  pages = {037001},
  numpages = {49},
  year = {2024},
  month = {8},
  publisher = {American Physical Society},
  doi = {10.1103/PRXQuantum.5.037001},
  url = {https://link.aps.org/doi/10.1103/PRXQuantum.5.037001}
}

@article{Bepari2022QuantumWalk,
  title = {{Quantum walk approach to simulating parton showers}},
  author = {Bepari, Khadeejah and Malik, Sarah and Spannowsky, Michael and Williams, Simon},
  journal = {Physical Review D},
  volume = {106},
  number = {5},
  pages = {056002},
  year = {2022},
  doi = {10.1103/PhysRevD.106.056002},
  url = {https://doi.org/10.1103/PhysRevD.106.056002}
}

@article{Sjostrand:2014zea,
  author       = {Torbjörn Sjöstrand and Stefan Ask and Jesper R. Christiansen and Richard Corke and Nishita Desai and Philip Ilten and Stephen Mrenna and Stefan Prestel and Christine O. Rasmussen and Peter Z. Skands},
  title        = {{An Introduction to PYTHIA 8.2}},
  journal      = {Comput. Phys. Commun.},
  volume       = {191},
  pages        = {159--177},
  year         = {2015},
  doi          = {10.1016/j.cpc.2015.01.024},
  eprint       = {1410.3012},
  archivePrefix= {arXiv},
  primaryClass = {hep-ph}
}

@article{B_hr_2008,
   title={{Herwig++ physics and manual}},
   volume={58},
   ISSN={1434-6052},
   url={http://dx.doi.org/10.1140/epjc/s10052-008-0798-9},
   DOI={10.1140/epjc/s10052-008-0798-9},
   number={4},
   journal={The European Physical Journal C},
   publisher={Springer Science and Business Media LLC},
   author={Bähr, Manuel and Gieseke, Stefan and Gigg, Martyn A. and Grellscheid, David and Hamilton, Keith and Latunde-Dada, Oluseyi and Plätzer, Simon and Richardson, Peter and Seymour, Michael H. and Sherstnev, Alexander and Webber, Bryan R.},
   year={2008},
   month=nov, pages={639–707} }

@article{Gleisberg_2004,
   title={{SHERPA 1., a proof-of-concept version}},
   volume={2004},
   ISSN={1029-8479},
   url={http://dx.doi.org/10.1088/1126-6708/2004/02/056},
   DOI={10.1088/1126-6708/2004/02/056},
   number={02},
   journal={Journal of High Energy Physics},
   publisher={Springer Science and Business Media LLC},
   author={Gleisberg, T and Hoeche, S and Krauss, F and Schaelicke, A and Schumann, S and Winter, J},
   year={2004},
   month=feb, pages={056–056} }

@article{PhysRevLett.126.062001,
  title = {{Quantum Algorithm for High Energy Physics Simulations}},
  author = {Nachman, Benjamin and Provasoli, Davide and de Jong, Wibe A. and Bauer, Christian W.},
  journal = {Phys. Rev. Lett.},
  volume = {126},
  issue = {6},
  pages = {062001},
  numpages = {6},
  year = {2021},
  month = {2},
  publisher = {American Physical Society},
  doi = {10.1103/PhysRevLett.126.062001},
  url = {https://link.aps.org/doi/10.1103/PhysRevLett.126.062001}
}

@article{Bepari2021Towards,
  title = {{Towards a Quantum Computing Algorithm for Helicity Amplitudes and Parton Showers}},
  author = {Bepari, Khadeejah and Malik, Sarah and Spannowsky, Michael and Williams, Simon},
  journal = {Physical Review D},
  volume = {103},
  number = {7},
  pages = {076020},
  year = {2021},
  doi = {10.1103/PhysRevD.103.076020},
  url = {https://doi.org/10.1103/PhysRevD.103.076020}
}

@article{Bauer2024QuantumPartonKinematics,
  title = {{Quantum Parton Shower with Kinematics}},
  author = {Bauer, Christian W. and Chigusa, So and Yamazaki, Masahito},
  journal = {Physical Review A},
  volume = {109},
  number = {3},
  pages = {032432},
  year = {2024},
  doi = {10.1103/PhysRevA.109.032432},
  url = {https://doi.org/10.1103/PhysRevA.109.032432}
}

@article{Chawdhry2023QuantumColour,
  title = {{Quantum simulation of colour in perturbative quantum chromodynamics}},
  author = {Chawdhry, Herschel A. and Pellen, Mathieu},
  journal = {SciPost Phys.},
  volume = {15},
  pages = {205},
  year = {2023},
  doi = {10.21468/SciPostPhys.15.5.205},
  url = {https://scipost.org/10.21468/SciPostPhys.15.5.205}
}

@article{Preskill_2018,
   title={{Quantum Computing in the NISQ era and beyond}},
   volume={2},
   ISSN={2521-327X},
   url={http://dx.doi.org/10.22331/q-2018-08-06-79},
   DOI={10.22331/q-2018-08-06-79},
   journal={Quantum},
   publisher={Verein zur Forderung des Open Access Publizierens in den Quantenwissenschaften},
   author={Preskill, John},
   year={2018},
   month=aug, pages={79} }

@article{Jordan_2012,
   title={{Quantum Algorithms for Quantum Field Theories}},
   volume={336},
   ISSN={1095-9203},
   url={http://dx.doi.org/10.1126/science.1217069},
   DOI={10.1126/science.1217069},
   number={6085},
   journal={Science},
   publisher={American Association for the Advancement of Science (AAAS)},
   author={Jordan, Stephen P. and Lee, Keith S. M. and Preskill, John},
   year={2012},
   month=jun, pages={1130–1133} }

@article{Garc_a_lvarez_2015,
   title={{Fermion-Fermion Scattering in Quantum Field Theory with Superconducting Circuits}},
   volume={114},
   ISSN={1079-7114},
   url={http://dx.doi.org/10.1103/PhysRevLett.114.070502},
   DOI={10.1103/physrevlett.114.070502},
   number={7},
   journal={Physical Review Letters},
   publisher={American Physical Society (APS)},
   author={García-Álvarez, L. and Casanova, J. and Mezzacapo, A. and Egusquiza, I.L. and Lamata, L. and Romero, G. and Solano, E.},
   year={2015},
   month=feb }

@article{halimehUniversalFrameworkQuantum2026,
  title = {{A Universal Framework for the Quantum Simulation of Yang-Mills Theory}},
  author = {Halimeh, Jad C. and Hanada, Masanori and Matsuura, Shunji and Nori, Franco and Rinaldi, Enrico and Sch{\"a}fer, Andreas},
  year = 2026,
  month = feb,
  journal = {Communications Physics},
  volume = {9},
  number = {1},
  pages = {67},
  publisher = {Nature Publishing Group},
  issn = {2399-3650},
  doi = {10.1038/s42005-025-02421-6},
  urldate = {2026-04-20},
  copyright = {2026 The Author(s)},
  langid = {english}
}

@article{Klco2020SU2,
  title = {{$SU(2)$ Non-Abelian Gauge Field Theory in One Dimension on Digital Quantum Computers}},
  author = {Klco, Natalie and Stryker, Jesse R. and Savage, Martin J.},
  journal = {Phys. Rev. D},
  volume = {101},
  number = {7},
  pages = {074512},
  year = {2020},
  doi = {10.1103/PhysRevD.101.074512},
  url = {https://doi.org/10.1103/PhysRevD.101.074512}
}

@article{Silvi2017FiniteDensity,
  title = {{Finite-Density Phase Diagram of a $(1+1)$D Non-Abelian Lattice Gauge Theory with Tensor Networks}},
  author = {Silvi, Pietro and Rico, Enrique and Dalmonte, Marcello and Tschirsich, Ferdinand and Montangero, Simone},
  journal = {Quantum},
  volume = {1},
  pages = {9},
  year = {2017},
  doi = {10.22331/q-2017-04-25-9},
  url = {https://doi.org/10.22331/q-2017-04-25-9}
}

@article{Tagliacozzo2013Simulation,
  title = {{Simulation of Non-Abelian Gauge Theories with Optical Lattices}},
  author = {Tagliacozzo, Luca and Celi, Alessio and Orland, Peter and Mitchell, Morgan W. and Lewenstein, Maciej},
  journal = {Nature Communications},
  volume = {4},
  pages = {2615},
  year = {2013},
  doi = {10.1038/ncomms3615},
  url = {https://doi.org/10.1038/ncomms3615}
}

@article{mildenbergerConfinement$$mathbbZ_2$$Lattice2025,
  title = {{Confinement in a $\mathbb{Z}$ Lattice Gauge Theory on a Quantum Computer}},
  author = {Mildenberger, Julius and Mruczkiewicz, Wojciech and Halimeh, Jad C. and Jiang, Zhang and Hauke, Philipp},
  year = 2025,
  month = feb,
  journal = {Nature Physics},
  volume = {21},
  number = {2},
  pages = {312--317},
  publisher = {Nature Publishing Group},
  issn = {1745-2481},
  doi = {10.1038/s41567-024-02723-6},
  urldate = {2026-04-20},
  copyright = {2025 The Author(s), under exclusive licence to Springer Nature Limited},
  langid = {english}
}

@article{Cheng_2025,
    author = "Cheng, Kun and Han, Tao and Low, Matthew",
    title = "{{Quantum tomography at colliders: With or without decays}}",
    eprint = "2410.08303",
    archivePrefix = "arXiv",
    primaryClass = "hep-ph",
    reportNumber = "PITT-PACC-2408",
    doi = "10.1016/j.physletb.2025.139675",
    journal = "Phys. Lett. B",
    volume = "868",
    pages = "139675",
    year = "2025"
}

@article{Magano2022QuantumSpeedup,
  title = {{Quantum speedup for track reconstruction in particle accelerators}},
  author = {Magano, Duarte and Kumar, Akshat and Kālis, Mārtiņš and Locāns, Andris and Glos, Adam and Pratapsi, Sagar and Quinta, Gonçalo and Dimitrijevs, Maksims and Rivošs, Aleksander and Bargassa, Pedrame and Seixas, João and Ambainis, Andris and Omar, Yasser},
  journal = {Physical Review D},
  volume = {105},
  number = {7},
  pages = {076012},
  year = {2022},
  doi = {10.1103/PhysRevD.105.076012},
  url = {https://doi.org/10.1103/PhysRevD.105.076012}
}

@article{Schuhmacher_2023,
   title={{Unravelling physics beyond the standard model with classical and quantum anomaly detection}},
   volume={4},
   ISSN={2632-2153},
   url={http://dx.doi.org/10.1088/2632-2153/ad07f7},
   DOI={10.1088/2632-2153/ad07f7},
   number={4},
   journal={Machine Learning: Science and Technology},
   publisher={IOP Publishing},
   author={Schuhmacher, Julian and Boggia, Laura and Belis, Vasilis and Puljak, Ema and Grossi, Michele and Pierini, Maurizio and Vallecorsa, Sofia and Tacchino, Francesco and Barkoutsos, Panagiotis and Tavernelli, Ivano},
   year={2023},
   month=nov, pages={045031} }

@article{Wei2020QuantumJet,
  title = {{Quantum Algorithms for Jet Clustering}},
  author = {Wei, Annie Y. and Naik, Preksha and Harrow, Aram W. and Thaler, Jesse},
  journal = {Physical Review D},
  volume = {101},
  number = {9},
  pages = {094015},
  year = {2020},
  doi = {10.1103/PhysRevD.101.094015},
  url = {https://doi.org/10.1103/PhysRevD.101.094015}
}

@article{PhysRevD.104.074014,
  title = {{Symmetry from entanglement suppression}},
  author = {Low, Ian and Mehen, Thomas},
  journal = {Phys. Rev. D},
  volume = {104},
  issue = {7},
  pages = {074014},
  numpages = {7},
  year = {2021},
  month = {10},
  publisher = {American Physical Society},
  doi = {10.1103/PhysRevD.104.074014},
  url = {https://link.aps.org/doi/10.1103/PhysRevD.104.074014}
}

@article{PhysRevLett.122.102001,
  title = {{Entanglement Suppression and Emergent Symmetries of Strong Interactions}},
  author = {Beane, Silas R. and Kaplan, David B. and Klco, Natalie and Savage, Martin J.},
  journal = {Phys. Rev. Lett.},
  volume = {122},
  issue = {10},
  pages = {102001},
  numpages = {6},
  year = {2019},
  month = {3},
  publisher = {American Physical Society},
  doi = {10.1103/PhysRevLett.122.102001},
  url = {https://link.aps.org/doi/10.1103/PhysRevLett.122.102001}
}

@article{ATLAS:2024entanglement,
  author       = {ATLAS Collaboration},
  title        = {{Observation of quantum entanglement with top quarks at the ATLAS detector}},
  journal      = {Nature},
  volume       = {633},
  number       = {8030},
  pages        = {542--547},
  year         = {2024},
  doi          = {10.1038/s41586-024-07824-z},
  url          = {https://www.nature.com/articles/s41586-024-07824-z}
}

@misc{cheng2024quantum,
  title         = {{Quantum Tomography at Colliders: With or Without Decays}},
  author        = {Cheng, Kun and Han, Tao and Low, Matthew},
  year          = {2024},
  eprint        = {2410.08303},
  archivePrefix = {arXiv},
  primaryClass  = {hep-ph},
  doi           = {10.48550/arXiv.2410.08303},
  url           = {https://arxiv.org/abs/2410.08303}
}

@article{afik2022quantum,
  title         = {{Quantum Information with Top Quarks in QCD}},
  author        = {Afik, Yoav and Mu{\~n}oz de Nova, Juan Ram{\'o}n},
  journal       = {Quantum},
  volume        = {6},
  pages         = {820},
  year          = {2022},
  doi           = {10.22331/q-2022-09-29-820},
  url           = {https://doi.org/10.22331/q-2022-09-29-820}
}

@article{Schuld_2016,
   title={{Prediction by linear regression on a quantum computer}},
   volume={94},
   ISSN={2469-9934},
   url={http://dx.doi.org/10.1103/PhysRevA.94.022342},
   DOI={10.1103/physreva.94.022342},
   number={2},
   journal={Physical Review A},
   publisher={American Physical Society (APS)},
   author={Schuld, Maria and Sinayskiy, Ilya and Petruccione, Francesco},
   year={2016},
   month=aug }

@book{schuld_supervised_2018,
  title = {{Supervised Learning with Quantum Computers}},
  author = {Schuld, Maria and Petruccione, Francesco},
  publisher = {Springer},
  year = {2018},
  series = {Quantum Science and Technology},
  volume = {17},
}

@article{lloyd_quantum_2013,
    author = "Lloyd, Seth and Mohseni, Masoud and Rebentrost, Patrick",
    title = "{{Quantum algorithms for supervised and unsupervised machine learning}}",
    eprint = "1307.0411",
    archivePrefix = "arXiv",
    primaryClass = "quant-ph",
    month = "7",
    year = "2013"
}

@article{cai_entanglement-based_2015,
  title = {{Entanglement-Based Machine Learning on a Quantum Computer}},
  author = {Cai, X.-D. and Wu, D. and Su, Z.-E. and Chen, M.-C. and Wang, X.-L. and Li, L. and Liu, N.-L. and Lu, Chao-Yang and Pan, Jian-Wei},
  journal = {Physical Review Letters},
  year = {2015},
  volume = {114},
  number = {11},
  pages = {110504},
  doi = {10.1103/PhysRevLett.114.110504},
}

@article{gibbs_dynamical_2024,
  title = {{Dynamical simulation via quantum machine learning with provable generalization}},
  author = {Gibbs, Joe and Holmes, Zo\"e and Caro, Matthias C. and Ezzell, Nicholas and Huang, Hsin-Yuan and Cincio, Lukasz and Sornborger, Andrew T. and Coles, Patrick J.},
  journal = {Physical Review Research},
  year = {2024},
  volume = {6},
  number = {1},
  pages = {013241},
  doi = {10.1103/PhysRevResearch.6.013241},
}

@article{Belis_2024,
   title={{Guided quantum compression for high dimensional data classification}},
   volume={5},
   ISSN={2632-2153},
   url={http://dx.doi.org/10.1088/2632-2153/ad5fdd},
   DOI={10.1088/2632-2153/ad5fdd},
   number={3},
   journal={Machine Learning: Science and Technology},
   publisher={IOP Publishing},
   author={Belis, Vasilis and Odagiu, Patrick and Grossi, Michele and Reiter, Florentin and Dissertori, Günther and Vallecorsa, Sofia},
   year={2024},
   month=jul, pages={035010} }

@article{baldi_searching_2014,
    author = "Guest, Dan and Cranmer, Kyle and Whiteson, Daniel",
    title = "{{Deep Learning and its Application to LHC Physics}}",
    eprint = "1806.11484",
    archivePrefix = "arXiv",
    primaryClass = "hep-ex",
    doi = "10.1146/annurev-nucl-101917-021019",
    journal = "Ann. Rev. Nucl. Part. Sci.",
    volume = "68",
    pages = "161--181",
    year = "2018"
}

@article{guest_deep_2018,
  title = {Deep Learning and its Application to LHC Physics},
  author = {Guest, Daniel and Cranmer, Kyle and Whiteson, Daniel},
  journal = {Annual Review of Nuclear and Particle Science},
  year = {2018},
  note = {arXiv:1806.11484},
  eprint = {1806.11484},
  archivePrefix = {arXiv},
  primaryClass = {hep-ex},
}

@article{deoliveira_jet-images_2016,
  title = {{Jet-Images --- Deep Learning Edition}},
  author = {{de Oliveira}, Luke and Kagan, Michael and Mackey, Lester and Nachman, Benjamin and Schwartzman, Ariel},
  year = 2016,
  month = jul,
  journal = {Journal of High Energy Physics},
  volume = {2016},
  number = {7},
  pages = {69},
  issn = {1029-8479},
  doi = {10.1007/JHEP07(2016)069},
  urldate = {2026-04-20},
  langid = {english}
}

@article{Qu_2020,
   title={{Jet tagging via particle clouds}},
   volume={101},
   ISSN={2470-0029},
   url={http://dx.doi.org/10.1103/PhysRevD.101.056019},
   DOI={10.1103/physrevd.101.056019},
   number={5},
   journal={Physical Review D},
   publisher={American Physical Society (APS)},
   author={Qu, Huilin and Gouskos, Loukas},
   year={2020},
   month=mar }

@article{komiske_energy_2019,
  title = {{Energy Flow Networks: Deep Sets for Particle Jets}},
  shorttitle = {Energy Flow Networks},
  author = {Komiske, Patrick T. and Metodiev, Eric M. and Thaler, Jesse},
  year = 2019,
  month = jan,
  journal = {Journal of High Energy Physics},
  volume = {2019},
  number = {1},
  pages = {121},
  issn = {1029-8479},
  doi = {10.1007/JHEP01(2019)121},
  urldate = {2026-04-20},
  langid = {english}
}

@article{deepshower_2018,
  title = {{Deep Learning as a Parton Shower}},
  author = {Monk, J. W.},
  year = 2018,
  month = dec,
  journal = {Journal of High Energy Physics},
  volume = {2018},
  number = {12},
  pages = {21},
  issn = {1029-8479},
  doi = {10.1007/JHEP12(2018)021},
  urldate = {2026-04-20},
  langid = {english}
}

@article{lai_explainable_2020,
    author = "Lai, Yue Shi and Neill, Duff and P{\l}osko{\'n}, Mateusz and Ringer, Felix",
    title = "{{Explainable machine learning of the underlying physics of high-energy particle collisions}}",
    eprint = "2012.06582",
    archivePrefix = "arXiv",
    primaryClass = "hep-ph",
    doi = "10.1016/j.physletb.2022.137055",
    journal = "Phys. Lett. B",
    volume = "829",
    pages = "137055",
    year = "2022"
}

@article{ghosh_towards_2022,
  title = {{Towards a deep learning model for hadronization}},
  author = {Ghosh, Aishik and Ju, Xiangyang and Nachman, Benjamin and Siodmok, Andrzej},
  journal = {Physical Review D},
  year = {2022},
  volume = {106},
  pages = {096020},
  doi = {10.1103/PhysRevD.106.096020},
}

@article{chan_fitting_2023,
  title = {{Fitting a deep generative hadronization model}},
  author = {Chan, Jay and Ju, Xiangyang and Kania, Adam and Nachman, Benjamin and Sangli, Vishnu and Siodmok, Andrzej},
  journal = {Journal of High Energy Physics},
  year = {2023},
  note = {arXiv:2305.17169},
  eprint = {2305.17169},
  archivePrefix = {arXiv},
  doi = {10.1007/JHEP09(2023)084},
}

@article{chan_flavor_2023,
  title = {{Integrating Particle Flavor into Deep Learning Models for Hadronization}},
  author = {Chan, Jay and Ju, Xiangyang and Kania, Adam and Nachman, Benjamin and Sangli, Vishnu and Siodmok, Andrzej},
  year = 2025,
  month = jun,
  journal = {Physical Review D},
  volume = {111},
  number = {11},
  pages = {116015},
  publisher = {American Physical Society},
  doi = {10.1103/hgbg-k7js},
  urldate = {2026-04-20}
}

@article{Amram:2024fjg,
    author = {Amram, Oz and Anzalone, Luca and Birk, Joschka and Faroughy, Darius A. and Hallin, Anna and Kasieczka, Gregor and Kr{\"a}mer, Michael and Pang, Ian and Reyes-Gonzalez, Humberto and Shih, David},
    title = "{{Aspen Open Jets: unlocking LHC data for foundation models in particle physics}}",
    eprint = "2412.10504",
    archivePrefix = "arXiv",
    primaryClass = "hep-ph",
    reportNumber = "FERMILAB-PUB-24-0941-AD",
    doi = "10.1088/2632-2153/ade58f",
    journal = "Mach. Learn. Sci. Tech.",
    volume = "6",
    number = "3",
    pages = "030601",
    year = "2025"
}

@article{meyer_exploiting_2023,
  title = {{Exploiting symmetry in variational quantum machine learning}},
  author = {Meyer, Johannes Jakob and Mularski, Marian and Gil-Fuster, Elies and Mele, Antonio Anna and Arzani, Francesco and Wilms, Alissa and Eisert, Jens},
  journal = {PRX Quantum},
  year = {2023},
  volume = {4},
  pages = {010328},
  doi = {10.1103/PRXQuantum.4.010328},
}

@article{park_hamiltonian_2024,
  title = {{Hamiltonian variational ansatz without barren plateaus}},
  author = {Park, Chae-Yeun and Killoran, Nathan},
  journal = {Quantum},
  year = {2024},
  volume = {8},
  pages = {1239},
  doi = {10.22331/q-2024-02-01-1239},
}

@article{west_provably_2024,
  title = {{Provably Trainable Rotationally Equivariant Quantum Machine Learning}},
  author = {West, Maxwell T. and Heredge, Jamie and Sevior, Martin and Usman, Muhammad},
  journal = {PRX Quantum},
  year = {2024},
  volume = {5},
  pages = {030320},
  doi = {10.1103/PRXQuantum.5.030320},
}

@article{marrero_expressivity_2021,
  title = {{Expressivity of quantum neural networks through entanglement}},
  author = {Marrero, Carlos O. and Kiani, Bobak and Coles, Patrick J.},
  journal = {PRX Quantum},
  volume = {2},
  pages = {040316},
  year = {2021},
  doi = {10.1103/PRXQuantum.2.040316},
}

@book{peskin1995introduction,
  title     = {{An Introduction to Quantum Field Theory}},
  author    = {Peskin, Michael E. and Schroeder, Daniel V.},
  year      = {1995},
  publisher = {Westview Press},
  address   = {Boulder, CO},
  isbn      = {978-0201503975}
}

@book{schwartz2014quantum,
  title     = {{Quantum Field Theory and the Standard Model}},
  author    = {Schwartz, Matthew D.},
  year      = {2014},
  publisher = {Cambridge University Press},
  address   = {Cambridge, UK},
  isbn      = {978-1107034730}
}

@article{Collins:1988ig,
  author    = {John C. Collins and Davison E. Soper and George Sterman},
  title     = {{Soft gluons and factorization}},
  journal   = {Nucl. Phys. B},
  volume    = {308},
  pages     = {833--856},
  year      = {1988},
  doi       = {10.1016/0550-3213(88)90130-7}
}

@book{Ellis_Stirling_Webber_1996, place={Cambridge}, series={Cambridge Monographs on Particle Physics, Nuclear Physics and Cosmology}, title={{QCD and Collider Physics}}, publisher={Cambridge University Press}, author={Ellis, R. K. and Stirling, W. J. and Webber, B. R.}, year={1996}, collection={Cambridge Monographs on Particle Physics, Nuclear Physics and Cosmology}}

@article{PhysRevLett.80.2245,
  title = {{Entanglement of Formation of an Arbitrary State of Two Qubits}},
  author = {Wootters, William K.},
  journal = {Phys. Rev. Lett.},
  volume = {80},
  issue = {10},
  pages = {2245--2248},
  numpages = {0},
  year = {1998},
  month = {3},
  publisher = {American Physical Society},
  doi = {10.1103/PhysRevLett.80.2245},
  url = {https://link.aps.org/doi/10.1103/PhysRevLett.80.2245}
}

@article{bergholm2022pennylane,
  title={{PennyLane: Automatic differentiation of hybrid quantum-classical computations}},
  author={Bergholm, Ville and Izaac, Josh and Schuld, Maria and Gogolin, Christian and Killoran, Nathan},
  journal={Computing in Science \& Engineering},
  volume={24},
  number={1},
  pages={51--61},
  year={2022},
  publisher={IEEE},
  doi={10.1109/MCSE.2021.3072906}
}

@article{Larkoski:2017Exposing,
  title = {{Exposing the QCD Splitting Function with CMS Open Data}},
  author = {Larkoski, Andrew J. and Marzani, Simone and Thaler, Jesse and Tripathee, Aashish and Xue, Wei},
  journal = {Phys. Rev. Lett.},
  volume = {119},
  pages = {132003},
  year = {2017},
  doi = {10.1103/PhysRevLett.119.132003}
}

@article{Cacciari:2008AntiKt,
    author = "Cacciari, Matteo and Salam, Gavin P. and Soyez, Gregory",
    title = "{{The anti-$k_t$ jet clustering algorithm}}",
    eprint = "0802.1189",
    archivePrefix = "arXiv",
    primaryClass = "hep-ph",
    reportNumber = "LPTHE-07-03",
    doi = "10.1088/1126-6708/2008/04/063",
    journal = "JHEP",
    volume = "04",
    pages = "063",
    year = "2008"
}

@article{Dokshitzer:1997CA,
    author = "Dokshitzer, Yuri L. and Leder, G. D. and Moretti, S. and Webber, B. R.",
    title = "{{Better jet clustering algorithms}}",
    eprint = "hep-ph/9707323",
    archivePrefix = "arXiv",
    reportNumber = "CAVENDISH-HEP-97-06",
    doi = "10.1088/1126-6708/1997/08/001",
    journal = "JHEP",
    volume = "08",
    pages = "001",
    year = "1997"
}

@article{Wobisch:1998CA,
  title = {{Hadronization Corrections to Jet Cross Sections in Deep-Inelastic Scattering}},
  author = {Wobisch, M. and Wengler, T.},
  year = 1999,
  month = jul,
  number = {arXiv:hep-ph/9907280},
  eprint = {hep-ph/9907280},
  publisher = {arXiv},
  doi = {10.48550/arXiv.hep-ph/9907280},
  urldate = {2026-04-20},
  archiveprefix = {arXiv}
}

@book{buckley2021practical,
  title        = {{Practical Collider Physics}},
  author       = {Buckley, Andy and White, Chris and White, Martin},
  year         = {2021},
  publisher    = {IOP Publishing},
  doi          = {10.1088/978-0-7503-2444-1},
  url          ={https://iopscience.iop.org/book/mono/978-0-7503-2444-1}
}

@article{alexandrouRealizingStringBreaking2025,
  title = {{Realizing String Breaking Dynamics in a $\mathbb{Z}_2$ Lattice Gauge Theory on Quantum Hardware}},
  author = {Alexandrou, Constantia and Athenodorou, Andreas and Blekos, Kostas and Polykratis, Georgios and K{\"u}hn, Stefan},
  year = 2025,
  month = dec,
  journal = {Physical Review D},
  volume = {112},
  number = {11},
  pages = {114506},
  publisher = {American Physical Society},
  doi = {10.1103/r6sr-dv13},
  urldate = {2026-04-20}
}

@article{ANDERSSON198331,
title = {{Parton fragmentation and string dynamics}},
journal = {Physics Reports},
volume = {97},
number = {2},
pages = {31-145},
year = {1983},
issn = {0370-1573},
doi = {https://doi.org/10.1016/0370-1573(83)90080-7},
url = {https://www.sciencedirect.com/science/article/pii/0370157383900807},
author = {B. Andersson and G. Gustafson and G. Ingelman and T. Sjöstrand}
}

@article{FIELD198365,
title = {{A QCD model for $e^+e^-$ annihilation}},
journal = {Nuclear Physics B},
volume = {213},
number = {1},
pages = {65-84},
year = {1983},
issn = {0550-3213},
doi = {https://doi.org/10.1016/0550-3213(83)90175-X},
url = {https://www.sciencedirect.com/science/article/pii/055032138390175X},
author = {Richard D. Field and Stephen Wolfram}
}

@article{abreuMeasurementTriplegluonVertex1993,
  title = {{Measurement of the Triple-Gluon Vertex from 4-Jet Events at LEP}},
  author = {Abreu, P. and Adam, W. and Adye, T. and Agasi, E. and Aleksan, R. and Alekseev, G. D. and Algeri, A. and Allen, P. and Almehed, S. and Alsvaag, S. J. and Amaldi, U. and Andreazza, A. and Antilogus, P. and Apel, W. -D. and Apsimon, R. J. and Arnoud, Y. and {\AA}sman, B. and Augustin, J. -E. and Augustinus, A. and Baillon, P. and Bambade, P. and Barao, F. and Barate, R. and Barbiellini, G. and Bardin, D. Y. and Barker, G. J. and Baroncelli, A. and Barring, O. and Barrio, J. A. and Bartl, W. and Bates, M. J. and Battaglia, M. and Baubillier, M. and Becks, K. -H. and Beeston, C. J. and Begalli, M. and Beilliere, P. and Belokopytov, {\relax Yu}. and Beltran, P. and Benedic, D. and Benvenuti, A. C. and Berggren, M. and Bertrand, D. and Bianchi, F. and Bilenky, M. S. and Billoir, P. and Bjarne, J. and Bloch, D. and Blyth, S. and Bocci, V. and Bogolubov, P. N. and Bolognese, T. and Bonesini, M. and Bonivento, W. and Booth, P. S. L. and Borisov, G. and Borner, H. and Bosio, C. and Bostiancic, B. and Bosworth, S. and Botner, O. and Bouquet, B. and Bourdarios, C. and Bowcock, T. J. V. and Bozzo, M. and Braibant, S. and Branchini, P. and Brand, K. D. and Brenner, R. A. and Briand, H. and Bricman, C. and Brown, R. C. A. and Brummer, N. and Brunet, J. -M. and Bugge, L. and Buran, T. and Burmeister, H. and Buytaert, J. A. M. A. and Caccia, M. and Calvi, M. and Camacho Rozas, A. J. and Campion, R. and Camporesi, T. and Canale, V. and Cao, F. and Carena, F. and Carroll, L. and Caso, C. and Castillo Gimenez, M. V. and Cattai, A. and Cavallo, F. R. and Cerrito, L. and Chabaud, V. and Chan, A. and Chapkin, M. and Charpentier, {\relax Ph}. and Chaussard, L. and Chauveau, J. and Checchia, P. and Chelkov, G. A. and Chevalier, L. and Chliapnikov, P. and Chorowicz, V. and Chrin, J. T. M. and Collins, P. and Contreras, J. L. and Contri, R. and Cortina, E. and Cosme, G. and Couchot, F. and Crawley, H. B. and Crennell, D. and Crosetti, G. and Crozon, M. and Cuevas Maestro, J. and Czellar, S. and {Dahl-Jensen}, E. and Dalmagne, B. and Dam, M. and Damgaard, G. and Darbo, G. and Daubie, E. and Daum, A. and Dauncey, P. D. and Davenport, M. and David, P. and Davies, J. and Da Silva, W. and Defoix, C. and Delpierre, P. and Demaria, N. and De Angelis, A. and De Boeck, H. and De Boer, W. and De Clerq, C. and De Fez Laso, M. D. M. and De Groot, N. and De La Vaissiere, C. and De Lotto, B. and De Min, A. and Dijkstra, H. and Di Ciaccio, L. and Dolbeau, J. and Donszelmann, M. and Doroba, K. and Dracos, M. and Drees, J. and Dris, M. and Dufour, Y. and Dupont, F. and Edsall, D. and Eek, L. -O. and Eerola, P. A. -M. and Ehret, R. and Ekelof, T. and Ekspong, G. and Elliot Peisert, A. and Engel, J. -P. and Ershaidat, N. and Fassouliotis, D. and Feindt, M. and Fernandez Alonso, M. and Ferrer, A. and Filippas, T. A. and Firestone, A. and Foeth, H. and Fokitis, E. and Fontanelli, F. and Forbes, K. A. J. and Fousset, J. -L. and Francon, S. and Franek, B. and Frenkiel, P. and Fries, D. C. and Frodesen, A. G. and Fruhwirth, R. and {Fulda-Quenzer}, F. and Furnival, K. and Furstenau, H. and Fuster, J. and Gamba, D. and Garcia, C. and Garcia, J. and Gaspar, C. and Gasparini, U. and Gavillet, {\relax Ph}. and Gazis, E. N. and Gerber, J. -P. and Giacomelli, P. and Gokieli, R. and Golob, B. and Golovatyuk, V. M. and Cadenas, J. J. Gomez Y. and Gopal, G. and Gorski, M. and Gracco, V. and Grant, A. and Grard, F. and Graziani, E. and Grosdidier, G. and Gross, E. and {Grosse-Wiesmann}, P. and Grossetete, B. and Guy, J. and Haedinger, U. and Hahn, F. and Hahn, M. and Haider, S. and Hakansson, A. and Hallgren, A. and Halpaap, M. and Hamacher, K. and De Monchenault, G. Hamel and Hao, W. and Harris, F. J. and Hedberg, V. and Henkes, T. and Hernandez, J. J. and Herquet, P. and Herr, H. and Hessing, T. L. and Hietanen, I. and Higgins, C. O. and Higon, E. and Hilke, H. J. and Hodgson, S. D. and Hofmokl, T. and Holmes, R. and Holmgren, S. -O. and Holthuizen, D. and Honore, P. F. and Hooper, J. E. and Houlden, M. and Hrubec, J. and Huet, K. and Hulth, P. O. and Hultqvist, K. and Ioannou, P. and Iversen, P. -S. and Jackson, J. N. and Jalocha, P. and Jarlskog, G. and Jarry, P. and {Jean-Marie}, B. and Johansson, E. K. and Johnson, D. and Jonker, M. and Jonsson, L. and Juillot, P. and Kalkanis, G. and Kalmus, G. and Kapusta, F. and Karlsson, M. and Karvelas, E. and Katsanevas, S. and Katsoufis, E. C. and Keranen, R. and Kesteman, J. and Khomenko, B. A. and Khovanski, N. N. and King, B. and Kjaer, N. J. and Klein, H. and Klovning, A. and Kluit, P. and {Koch-Mehrin}, A. and Koehne, J. H. and Koene, B. and Kokkinias, P. and Koratzinos, M. and Korytov, A. V. and Kostioukhine, V. and Kourkoumelis, C. and Kouznetsov, O. and Kramer, P. H. and Krolikowski, J. and Kronkvist, I. and {Kruener-Marquis}, U. and Krupinski, W. and Kulka, K. and Kurvinen, K. and Lacasta, C. and Lambropoulos, C. and Lamsa, J. W. and Lanceri, L. and Lapin, V. and Laugier, J. -P. and Lauhakangas, R. and Leder, G. and Ledroit, F. and Leitner, R. and Lemoigne, Y. and Lemonne, J. and Lenzen, G. and Lepeltier, V. and Lesiak, T. and Levy, J. M. and Lieb, E. and Liko, D. and {DELPHI Collaboration}},
  year = 1993,
  month = sep,
  journal = {Zeitschrift f\"ur Physik C Particles and Fields},
  volume = {59},
  number = {3},
  pages = {357--368},
  issn = {1431-5858},
  doi = {10.1007/BF01498617},
  urldate = {2026-03-20},
  langid = {english}
}

@article{campbellHardInteractionsQuarks2006,
  title = {{Hard Interactions of Quarks and Gluons: A Primer for LHC Physics}},
  shorttitle = {Hard Interactions of Quarks and Gluons},
  author = {Campbell, J M and Huston, J W and Stirling, W J},
  year = 2006,
  month = dec,
  journal = {Reports on Progress in Physics},
  volume = {70},
  number = {1},
  pages = {89},
  issn = {0034-4885},
  doi = {10.1088/0034-4885/70/1/R02},
  urldate = {2026-03-20},
  langid = {english}
}

@misc{cmscollaborationJetHTRun2016GUL2016_MiniAODv2v2MINIAOD2024,
  title = {/{{JetHT}}/{{Run2016G-UL2016}}\_{{MiniAODv2-v2}}/{{MINIAOD}}},
  author = {{CMS Collaboration}},
  year = 2024,
  publisher = {CERN Open Data Portal},
  doi = {10.7483/OPENDATA.CMS.1KTG.X0W4},
  urldate = {2026-03-23}
}

@misc{cmscollaborationJetHTRun2016HUL2016_MiniAODv2v2MINIAOD2024,
  title = {/{{JetHT}}/{{Run2016H-UL2016}}\_{{MiniAODv2-v2}}/{{MINIAOD}}},
  author = {{CMS Collaboration}},
  year = 2024,
  publisher = {CERN Open Data Portal},
  doi = {10.7483/OPENDATA.CMS.LT9E.T7RQ},
  urldate = {2026-03-23}
}

@article{campbellEventGeneratorsHighenergy2024,
  title = {{Event Generators for High-Energy Physics Experiments}},
  author = {Campbell, J. M. and Diefenthaler, M. and Hobbs, T. J. and H{\"o}che, Stefan and Isaacson, Joshua and Kling, Felix and Mrenna, Stephen and Reuter, J. and Alioli, S. and Andersen, Jeppe R. and Andreopoulos, C. and Ankowski, A. M. and Aschenauer, Elke Caroline and Ashkenazi, A. and Baker, M. D. and Barrow, Joshua L. and {van Beekveld}, Melissa and Bewick, Gavin and Bhattacharya, S. and Bierlich, Christian and Bothmann, Enrico and Bredt, P. and Broggio, A. and Buckley, Andy and Butter, A. and Butterworth, Jonathan Mark and Byrne, E. P. and {Carloni-Calame}, Carlo and Chakraborty, Smita and Chen, X. and Chiesa, M. and Childers, J. T. and {Cruz-Martinez}, J. and Currie, J. and Darvishi, N. and Dasgupta, Mrinal and Denner, A. and Dreyer, F. A. and Dytman, S. and {El-Menoufi}, Basem Kamal and Engel, Tim and Ferrario Ravasio, Silvia and Figueroa, Daniel and Flower, L. and Forshaw, J. R. and Frederix, Rikkert and Friedland, Alex and Frixione, Stefano and Gallagher, H. and Gallmeister, K. and Gardiner, Simon and Gauld, Rhorry and Gaunt, Jonathan and Gavardi, A. and Gehrmann, Thomas and {Gehrmann-De Ridder}, Aude and Gellersen, Leif and Giele, Walter and Gieseke, Stefan and Giuli, Francesco and Glover, Nigel and Grazzini, Massimiliano and Grohsjean, A. and G{\"u}tschow, Christian and Hamilton, Keith and Han, T. and Hatcher, R. and Heinrich, G. and Helenius, Ilkka and Hen, O. and Hirschi, Valentin and H{\"o}fer, M. and Holguin, J. and Huss, Alexander and Ilten, Philip and Jadach, Stanislaw and Jentsch, A. and Jones, Stephen and Ju, W. and Kallweit, Stefan and Karlberg, Alexander and Katori, T. and Kerner, Matthias and Kilian, C. and Kirchgae{\ss}er, M. M. and Klein, S. and Knobbe, Max and Krause, Claudius and Krauss, Frank and Lang, J. and Lang, J.-N. and Lee, G. and Li, S. W. and Lim, M. A. and Lindert, J. M. and Lombardi, D. and L{\"o}nnblad, Leif and L{\"o}schner, M. and Lurkin, N. and Ma, Y. and Machado, P. and Magerya, Vitaly and Maier, A. and Majer, I. and Maltoni, Fabio and Marcoli, M. and Marinelli, G. and Masouminia, M. R. and Mastrolia, Pierpaolo and Mattelaer, Olivier and Mazzitelli, J. and McFayden, Josh A. and Medves, Rok and Meinzinger, P. and Mo, J. and Monni, Pier Francesco and Montagna, Guido and Morgan, T. and Mosel, U. and Nachman, Benjamin and Nadolsky, Pavel and Nagar, R. and Nagy, Zoltan and Napoletano, Davide and Nason, Paolo and Neumann, Tobias and Nevay, L. J. and Nicrosini, Oreste and Niehues, J. and Niewczas, K. and Ohl, Thorsten and Ossola, Giovanni and Pandey, V. and Papadopoulou, A. and Papaefstathiou, Andreas and Paz, Gil and Pellen, Mathieu and Pelliccioli, Giovanni and Peraro, T. and Piccinini, Fluvio and Pickering, L. and Pires, J. and Placzek, Wieslaw and Pl{\"a}tzer, Simon and Plehn, Tilman and Pozzorini, Stefano and Prestel, Stefan and Preuss, Christian Tobias and Price, A. C. and Quackenbush, Seth and Re, Emanuele and Reichelt, Daniel and Reina, Laura and Reuschle, Christian and Richardson, Peter and Rocco, M. and Rocco, N. and Roda, M. and Rodriguez Garcia, A. and Roiser, Stefan and Rojo, Juan and Rottoli, Luca and Salam, Gavin P. and Sch{\"o}nherr, Marek and Schuchmann, S. and Schumann, Steffen and Sch{\"u}rmann, Robin and Scyboz, Ludovic and Seymour, Michael H. and Siegert, Frank and Signer, Adrian and Chahal, Gurpreet Singh and Si{\'o}dmok, Andrzej and Sj{\"o}strand, Torbj{\"o}rn and Skands, Peter and Smillie, Jennifer M. and Sobczyk, J. T. and Soldin, Dennis and Soper, D. E. and {Soto-Ontoso}, Alba and Soyez, Gregory and Stagnitto, Giovanni and {Tena-Vidal}, J. and Tomalak, O. and Tramontano, Francesco and Trojanowski, Sebastian and Tu, Z. and Uccirati, Sandro and Ullrich, T. and Ulrich, Yannick and Utheim, Marius and Valassi, A. and Verbytskyi, Andrii and Verheyen, Rob and Wagman, M. and Walker, D. and Webber, B. R. and Weinstein, L. and White, O. and Whitehead, James and Wiesemann, Marius and Wilkinson, C. and Williams, C. and Winterhalder, Ramon and Wret, C. and Xie, K. and Yang, T.-Z. and Yazgan, E. and Zanderighi, Giulia and Zanoli, S. and Zapp, Korinna},
  year = 2024,
  month = may,
  journal = {SciPost Physics},
  volume = {16},
  number = {5},
  pages = {130},
  issn = {2542-4653},
  doi = {10.21468/SciPostPhys.16.5.130},
  urldate = {2026-03-25},
  langid = {english}
}

@article{khalekUltimatePartonDistributions2018,
    title = {{Towards ultimate parton distributions at the high-luminosity LHC}},
    volume = {78},
    issn = {1434-6052},
    url = {https://doi.org/10.1140/epjc/s10052-018-6448-y},
    doi = {10.1140/epjc/s10052-018-6448-y},
    language = {en},
    number = {11},
    urldate = {2026-03-25},
    journal = {The European Physical Journal C},
    author = {Khalek, Rabah Abdul and Bailey, Shaun and Gao, Jun and Harland-Lang, Lucian and Rojo, Juan},
    month = nov,
    year = {2018},
    pages = {962},
}

@article{neillEntropyJet2019,
  title = {{Entropy of a Jet}},
  author = {Neill, Duff and Waalewijn, Wouter J.},
  year = 2019,
  month = sep,
  journal = {Physical Review Letters},
  volume = {123},
  number = {14},
  pages = {142001},
  publisher = {American Physical Society},
  doi = {10.1103/PhysRevLett.123.142001},
  urldate = {2026-03-25}
}

@article{vidalEfficientClassicalSimulation2003,
  title = {Efficient {{Classical Simulation}} of {{Slightly Entangled Quantum Computations}}},
  author = {Vidal, Guifr{\'e}},
  year = 2003,
  month = oct,
  journal = {Physical Review Letters},
  volume = {91},
  number = {14},
  pages = {147902},
  publisher = {American Physical Society},
  doi = {10.1103/PhysRevLett.91.147902},
  urldate = {2026-03-25}
}

@article{valassiChallengesMonteCarlo2021,
  title = {Challenges in {{Monte Carlo Event Generator Software}} for {{High-Luminosity LHC}}},
  author = {Valassi, Andrea and Yazgan, Efe and McFayden, Josh and Amoroso, Simone and Bendavid, Joshua and Buckley, Andy and Cacciari, Matteo and Childers, Taylor and Ciulli, Vitaliano and Frederix, Rikkert and Frixione, Stefano and Giuli, Francesco and Grohsjean, Alexander and G{\"u}tschow, Christian and H{\"o}che, Stefan and Hopkins, Walter and Ilten, Philip and Konstantinov, Dmitri and Krauss, Frank and Li, Qiang and L{\"o}nnblad, Leif and Maltoni, Fabio and Mangano, Michelangelo and Marshall, Zach and Mattelaer, Olivier and Fernandez Menendez, Javier and Mrenna, Stephen and Muralidharan, Servesh and Neumann, Tobias and Pl{\"a}tzer, Simon and Prestel, Stefan and Roiser, Stefan and Sch{\"o}nherr, Marek and Schulz, Holger and Schulz, Markus and {Sexton-Kennedy}, Elizabeth and Siegert, Frank and Si{\'o}dmok, Andrzej and Stewart, Graeme A. and {The HSF Physics Event Generator WG}},
  year = 2021,
  month = may,
  journal = {Computing and Software for Big Science},
  volume = {5},
  number = {1},
  pages = {12},
  issn = {2510-2044},
  doi = {10.1007/s41781-021-00055-1},
  urldate = {2026-03-25},
  langid = {english}
}

@misc{azziStandardModelPhysics2019,
  title = {Standard {{Model Physics}} at the {{HL-LHC}} and {{HE-LHC}}},
  author = {Azzi, P. and Farry, S. and Nason, P. and Tricoli, A. and Zeppenfeld, D. and Khalek, R. Abdul and Alimena, J. and Andari, N. and Bella, L. Aperio and Armbruster, A. J. and Baglio, J. and Bailey, S. and Bakos, E. and Bakshi, A. and Baldenegro, C. and Balli, F. and Barker, A. and Barter, W. and de Blas, J. and Blekman, F. and Bloch, D. and Bodek, A. and Boonekamp, M. and Boos, E. and Sola, J. D. Bossio and Cadamuro, L. and Camarda, S. and Campanario, F. and Campanelli, M. and Campbell, J. M. and Cao, Q.-H. and Cavaliere, V. and Cerri, A. and Chahal, G. S. and Chargeishvili, B. and Charlot, C. and Chen, S.-L. and Chen, T. and Cieri, L. and Ciuchini, M. and Corcella, G. and Cotogno, S. and Covarelli, R. and {Cruz-Martinez}, J. M. and Czakon, M. and Dainese, A. and Dang, N. P. and Darm{\'e}, L. and Dawson, S. and la Torre, H. De and Deile, M. and Deliot, F. and Demers, S. and Denner, A. and Derue, F. and Ciaccio, L. Di and Clemente, W. K. Di and Damiani, D. Dominguez and Dudko, L. and Durglishvili, A. and D{\"u}nser, M. and Ebadi, J. and Faria, R. B. Ferreira De and Ferrera, G. and Ferroglia, A. and Figy, T. M. and Finelli, K. D. and Fiolhais, M. C. N. and Franco, E. and Frederix, R. and Fuks, B. and Galhardo, B. and Gao, J. and Gaunt, J. R. and Gehrmann, T. and Ridder, A. Gehrmann-De and Giljanovic, D. and Giuli, F. and Glover, E. W. N. and Goodsell, M. D. and Gouveia, E. and Govoni, P. and Goy, C. and Grazzini, M. and Grohsjean, A. and {Grosse-Oetringhaus}, J. F. and Gunnellini, P. and Gwenlan, C. and {Harland-Lang}, L. A. and Harrison, P. F. and Heinrich, G. and Helsens, C. and Herndon, M. and Hindrichs, O. and Hirschi, V. and Hoang, A. and Hoepfner, K. and Hogan, J. M. and Huss, A. and Jahn, S. and Jain, Sa and Jones, S. P. and Jung, A. W. and Jung, H. and Kallweit, S. and Kar, D. and Karlberg, A. and Kasemets, T. and Kerner, M. and Khandoga, M. K. and Khanpour, H. and Khatibi, S. and Khukhunaishvili, A. and Kieseler, J. and Kretzschmar, J. and Kroll, J. and Kryshen, E. and Lang, V. S. and Lechner, L. and Lee, C. A. and Leigh, M. and Lelas, D. and Les, R. and Lewis, I. M. and Li, B. and Li, Q. and Li, Y. and Lidrych, J. and Ligeti, Z. and Lindert, J. M. and Liu, Y. and Lohwasser, K. and Long, K. and Lontkovskyi, D. and Majumder, G. and Mancini, M. and Mandrik, P. and Mangano, M. L. and Marchesini, I. and Mayer, C. and Mazumdar, K. and McFayden, J. A. and Lagarelhos, P. M. Mendes Amaral Torres and Meyer, A. B. and Mikhalcov, S. and Mishima, S. and Mitov, A. and Najafabadi, M. Mohammadi and Ll{\'a}cer, M. Moreno and Mulders, M. and Myska, M. and Narain, M. and Nisati, A. and Nitta, T. and Onofre, A. and Griso, S. Pagan and Pagani, D. and Cortezon, E. Palencia and Papanastasiou, A. and Pedro, K. and Pellen, M. and Perfilov, M. and Perrozzi, L. and Petersen, B. A. and Pierini, M. and Pires, J. and Pleier, M.-A. and Pl{\"a}tzer, S. and Potamianos, K. and Pozzorini, S. and Price, A. C. and Rauch, M. and Re, E. and Reina, L. and Reuter, J. and Robens, T. and Rojo, J. and Royon, C. and Saito, S. and Savin, A. and Sawant, S. and Schneider, B. and Schoefbeck, R. and Schoenherr, M. and {Sch{\"a}fer-Siebert}, H. and Seidel, M. and Selvaggi, M. and Shears, T. and Silvestrini, L. and Sjodahl, M. and Skovpen, K. and Smith, N. and Spitzbart, D. and Starovoitov, P. and Suster, C. J. E. and Tan, P. and Taus, R. and Teague, D. and Terashi, K. and Terron, J. and Uplap, S. and Veloso, F. and Verzetti, M. and Vesterinen, M. A. and Vladimirov, V. E. and Volkov, P. and Vorotnikov, G. and Milosavljevic, M. Vranjes and Vranjes, N. and Vryonidou, E. and Walker, D. and Wiesemann, M. and Wu, Y. and Xu, T. and Yacoob, S. and Yazgan, E. and Zahreddine, J. and Zanderighi, G. and Zaro, M. and Zenaiev, O. and Porta, G. Zevi Della and Zhang, C. and Zhang, W. and Zhu, H. L. and Zlebcik, R. and Zubair, F. N.},
  year = 2019,
  month = dec,
  number = {arXiv:1902.04070},
  eprint = {1902.04070},
  primaryclass = {hep-ph},
  publisher = {arXiv},
  doi = {10.48550/arXiv.1902.04070},
  urldate = {2026-04-20},
  archiveprefix = {arXiv}
}

@article{tuysuztracking,
  title = {{Hybrid Quantum Classical Graph Neural Networks for Particle Track Reconstruction}},
  author = {T{\"u}ys{\"u}z, Cenk and Rieger, Carla and Novotny, Kristiane and Demirk{\"o}z, Bilge and Dobos, Daniel and Potamianos, Karolos and Vallecorsa, Sofia and Vlimant, Jean-Roch and Forster, Richard},
  year = 2021,
  month = nov,
  journal = {Quantum Machine Intelligence},
  volume = {3},
  number = {2},
  pages = {29},
  issn = {2524-4914},
  doi = {10.1007/s42484-021-00055-9}
}

@article{luxetracking,
  title = {Quantum {{Algorithms}} for {{Charged Particle Track Reconstruction}} in the {{LUXE Experiment}}},
  author = {Crippa, Arianna and Funcke, Lena and Hartung, Tobias and Heinemann, Beate and Jansen, Karl and Kropf, Annabel and K{\"u}hn, Stefan and Meloni, Federico and Spataro, David and T{\"u}ys{\"u}z, Cenk and Yap, Yee Chinn},
  year = 2023,
  month = dec,
  journal = {Computing and Software for Big Science},
  volume = {7},
  number = {1},
  pages = {14},
  issn = {2510-2044},
  doi = {10.1007/s41781-023-00109-6}
}

@article{Funcke_2023,
doi = {10.1088/1742-6596/2438/1/012127},
url = {https://doi.org/10.1088/1742-6596/2438/1/012127},
year = {2023},
month = {feb},
publisher = {IOP Publishing},
volume = {2438},
number = {1},
pages = {012127},
author = {Funcke, Lena and Hartung, Tobias and Heinemann, Beate and Jansen, Karl and Kropf, Annabel and Kühn, Stefan and Meloni, Federico and Spataro, David and Tüysüz, Cenk and Chinn Yap, Yee},
title = {{Studying quantum algorithms for particle track reconstruction in the LUXE experiment}},
journal = {Journal of Physics: Conference Series}
}

@phdthesis{Tuysuz:2025mrv,
    author = {T{\"u}ys{\"u}z, Cenk},
    title = "{{Quantum machine learning with near and future term quantum computers}}",
    doi = "10.18452/33365",
    school = "Humboldt U., Berlin",
    year = "2025"
}

@misc{wang2023symmetryenhancedvariationalquantum,
      title={{Symmetry enhanced variational quantum imaginary time evolution}}, 
      author={Xiaoyang Wang and Yahui Chai and Maria Demidik and Xu Feng and Karl Jansen and Cenk Tüysüz},
      year={2023},
      eprint={2307.13598},
      archivePrefix={arXiv},
      primaryClass={quant-ph},
      url={https://arxiv.org/abs/2307.13598}, 
}

@article{tuysuz2025learningresponsefunctionsanalog,    author = {T{\"u}ys{\"u}z, Cenk and Jayakumar, Abhijith and Coffrin, Carleton and Vuffray, Marc and Lokhov, Andrey Y.},
    title = "{{Learning response functions of analog quantum computers: analysis of neutral-atom and superconducting platforms}}",
    eprint = "2503.12520",
    archivePrefix = "arXiv",
    primaryClass = "quant-ph",
    reportNumber = "LA-UR-23-31303",
    month = "3",
    year = "2025"
}

@article{Borras_2023,
doi = {10.1088/1742-6596/2438/1/012093},
url = {https://doi.org/10.1088/1742-6596/2438/1/012093},
year = {2023},
month = {feb},
publisher = {IOP Publishing},
volume = {2438},
number = {1},
pages = {012093},
author = {Borras, Kerstin and Chang, Su Yeon and Funcke, Lena and Grossi, Michele and Hartung, Tobias and Jansen, Karl and Kruecker, Dirk and Kühn, Stefan and Rehm, Florian and Tüysüz, Cenk and Vallecorsa, Sofia},
title = {{Impact of quantum noise on the training of quantum Generative Adversarial Networks}},
journal = {Journal of Physics: Conference Series}
}

@article{aoudeDecoherenceEffectsEntangled2026,
  title = {{Decoherence Effects in Entangled Fermion Pairs at Colliders}},
  author = {Aoude, Rafael and Barr, Alan J. and Maltoni, Fabio and Satrioni, Leonardo},
  year = 2026,
  month = apr,
  journal = {Physical Review D},
  volume = {113},
  number = {7},
  pages = {076007},
  issn = {2470-0010, 2470-0029},
  doi = {10.1103/yjgh-f2gw},
  urldate = {2026-04-27},
  langid = {english}
}

@article{mcginnisSymmetryEntanglementSmatrix2025,
  title = {{Symmetry, Entanglement, and the S-matrix}},
  author = {McGinnis, Navin},
  year = 2025,
  month = apr
}

@misc{mcginnisQuantumComputationalStructure2025,
  title = {{Quantum Computational Structure of $SU(N)$ Scattering}},
  author = {McGinnis, Navin},
  year = 2025,
  month = nov,
  number = {arXiv:2511.10550},
  eprint = {2511.10550},
  primaryclass = {quant-ph},
  publisher = {arXiv},
  doi = {10.48550/arXiv.2511.10550},
  urldate = {2026-04-27},
  archiveprefix = {arXiv}
}

@inproceedings{yazganMeasurementsTopQuark2026,
  title = {{Measurements of Top Quark Properties in CMS: $t\bar{t}$ Spin Density Matrix, Quantum Entanglement and Quantum Magic}},
  shorttitle = {Measurements of Top Quark Properties in {{CMS}}},
  booktitle = {Proceedings of {{The European Physical Society Conference}} on {{High Energy Physics}} --- {{PoS}}({{EPS-HEP2025}})},
  author = {Yazgan, Efe},
  year = 2026,
  month = jan,
  eprint = {2510.13743},
  primaryclass = {hep-ex},
  pages = {273},
  doi = {10.22323/1.485.0273},
  urldate = {2026-04-27},
  archiveprefix = {arXiv}
}

@article{afikQuantumInformationMeets2025,
  title = {{Quantum Information Meets High-Energy Physics: Input to the Update of the {{European}} Strategy for Particle Physics}},
  shorttitle = {Quantum Information Meets High-Energy Physics},
  author = {Afik, Yoav and Fabbri, Federica and Low, Matthew and Marzola, Luca and {Aguilar-Saavedra}, Juan Antonio and Altakach, Mohammad Mahdi and Asbah, Nedaa Alexandra and Bai, Yang and Banks, Hannah and Barr, Alan J. and Bernal, Alexander and Browder, Thomas E. and Caban, Pawe{\l} and Casas, J. Alberto and Cheng, Kun and D{\'e}liot, Fr{\'e}d{\'e}ric and Demina, Regina and Di Domenico, Antonio and Eckstein, Micha{\l} and Fabbrichesi, Marco and Fuks, Benjamin and Gabrielli, Emidio and Gon{\c c}alves, Dorival and Grabarczyk, Rados{\l}aw and Grossi, Michele and Han, Tao and Hobbs, Timothy J. and Horodecki, Pawe{\l} and Howarth, James and Hsu, Shih-Chieh and Jiggins, Stephen and Jones, Eleanor and Jung, Andreas W. and Knue, Andrea Helen and Korn, Steffen and Lagouri, Theodota and Lamba, Priyanka and Landi, Gabriel T. and Li, Haifeng and Li, Qiang and Low, Ian and Maltoni, Fabio and McFayden, Josh and McGinnis, Navin and Morales, Roberto A. and Moreno, Jes{\'u}s M. and De Nova, Juan Ram{\'o}n Mu{\~n}oz and Negro, Giulia and Pagani, Davide and Pelliccioli, Giovanni and Pinamonti, Michele and Pintucci, Laura and Ravina, Baptiste and Ruzi, Alim and Sakurai, Kazuki and Simpson, Ethan and Sioli, Maximiliano and Su, Shufang and Trifinopoulos, Sokratis and Vahsen, Sven E. and Vallecorsa, Sofia and Vicini, Alessandro and Vos, Marcel and Vryonidou, Eleni and White, Chris D. and White, Martin J. and Wildridge, Andrew J. and Wu, Tong Arthur and Zani, Laura and Zhang, Yulei and Zoch, Knut},
  year = 2025,
  month = sep,
  journal = {The European Physical Journal Plus},
  volume = {140},
  number = {9},
  pages = {855},
  issn = {2190-5444},
  doi = {10.1140/epjp/s13360-025-06752-9},
  urldate = {2026-04-27},
  langid = {english}
}

@article{florioThermalizationQuantumEntanglement2025,
  title = {Thermalization from Quantum Entanglement: {{Jet}} Simulations in the Massive {{Schwinger}} Model},
  shorttitle = {Thermalization from Quantum Entanglement},
  author = {Florio, Adrien and Frenklakh, David and Grieninger, Sebastian and Kharzeev, Dmitri E. and Palermo, Andrea and Shi, Shuzhe},
  year = 2025,
  month = nov,
  journal = {Physical Review D},
  volume = {112},
  number = {9},
  pages = {094502},
  publisher = {American Physical Society},
  doi = {10.1103/sgrx-jpp9},
  urldate = {2026-05-07}
}

@article{florioQuantumRealtimeEvolution2024,
  title = {Quantum Real-Time Evolution of Entanglement and Hadronization in Jet Production: {{Lessons}} from the Massive {{Schwinger}} Model},
  shorttitle = {Quantum Real-Time Evolution of Entanglement and Hadronization in Jet Production},
  author = {Florio, Adrien and Frenklakh, David and Ikeda, Kazuki and Kharzeev, Dmitri and Korepin, Vladimir and Shi, Shuzhe and Yu, Kwangmin},
  year = 2024,
  month = nov,
  journal = {Physical Review D},
  volume = {110},
  number = {9},
  pages = {094029},
  publisher = {American Physical Society},
  doi = {10.1103/PhysRevD.110.094029},
  urldate = {2026-05-07}
}

@article{Fromm:2024caq,
    author = "Fromm, Michael and Katschke, Lucas and Philipsen, Owe and Unger, Wolfgang",
    title = "{Quantum computational resources for lattice QCD in the strong-coupling limit}",
    eprint = "2406.18721",
    archivePrefix = "arXiv",
    primaryClass = "hep-lat",
    doi = "10.1140/epjqt/s40507-025-00395-6",
    journal = "EPJ Quant. Technol.",
    volume = "12",
    number = "1",
    pages = "92",
    year = "2025"
}

\newpage

\appendix

\section{Parton Shower Fundamentals and the Splitting Function}\label{app:MatrixElements}

In our work, we studied the splitting function, a building block of parton showers. An underlying physical principle central to their study and simulation is collinear factorization~\cite{Collins:1988ig}. It is built from the domination of soft and collinear radiation in a scattering event. Let us start by considering a parton $i$ with four-momentum $p$ that scatters into partons $l$ and $j$ with four-momenta $k$ and $q$, respectively. We call the produced partons soft radiation if one of them has, approximately, vanishing four-momentum, and we call them collinear radiation in the case where the angle between the incoming parton and the outgoing one, $\theta$, is close to zero. Let us say that the particle $j$ carries a fraction of the initial energy $z$ and $l$ carries a fraction $1-z$, so that $E_q=zE_p$ and $E_k=(1-z)E_p$. For massless particles in the collinear limit $\theta \approx 0$, the energy fraction $z$ is approximately the same as the transverse momentum fraction. To be consistent with the extracted observables from jet data, throughout this work 
$z$ denotes the transverse momentum fraction carried by the \emph{harder} daughter, $z\equiv\frac{\max(p_{T1}, p_{T2})}{p_{T1}+p_{T2}}\in [\frac{1}{2}, 1]$, so that the softer daughter carries $1-z$. Here $p_{T,1}$ and $p_{T,2}$ are the moduli of the transverse momenta of the two produced subjets and the denominator its scalar sum. Now consider that the parton $i$ is one of $n$ final-state partons produced in an event of differential cross-section $d\sigma_n$. According to the factorisation theorem, the differential cross-section for the $n+1$ final-state partons event is 
\begin{equation}
    \mathrm{d}\sigma_{n+1} = \mathrm{d}\sigma_n  \frac{\mathrm{d}t}{t}  \mathrm{d}z \frac{\alpha_s}{2\pi} P_{ji}(z),
    \label{eqn:FactorisationTheorem}
\end{equation}
where the integration measure $t\equiv p^ 2$ is the off-shell mass of the incoming parton, often called virtuality, with $t \gg q^2, k^2$, and $\alpha_s=g_s^2/4\pi$ is an energy-dependent quantity that measures the coupling strength. The variable $P_{ji}(z)$ is the spin-averaged splitting function of parton $i$ splitting into a parton $j$ with momentum fraction $z$~\cite{Ellis_Stirling_Webber_1996}.

Using factorisation, we can now model parton showers as a succession of nearly collinear emissions, with a probabilistic nature ruled by the splitting functions, which break the shower into successive splitting $a\to b+c$. To obtain a complete quantum algorithm for parton showers, we must consider all possible three-point interactions in QCD. By analyzing the gluon splitting function~\cite{Ellis_Stirling_Webber_1996}
\begin{equation}
    P_{gg}=C_A\left(\frac{z}{1-z}+\frac{1-z}{z}+z(1-z)\right),
    \label{eqn:GluonSPlittingFucntion}
\end{equation}
we see that it is enhanced in the soft limits $z \to 0$ and $z\to 1$ and shows symmetry under the exchange of $z$ and $1-z$. Studying pure gluonic processes also brings other advantages, particularly because gluons are massless. The states of massless particles can be expressed in the helicity basis~\cite{peskin1995introduction}, in which helicity is invariant under Lorentz transformations, allowing the process to be analyzed equivalently in any reference frame.

In order to obtain the splitting function in Eq.~\eqref{eqn:GluonSPlittingFucntion} and to study the interaction using quantum information tools, we need to compute the scattering amplitudes for all possible three-point gluon interactions. This interaction is governed by the following interaction term, present in the SM Lagrangian, 
\begin{equation}
   \mathcal{L} \supset   -\frac{g}{2}f^{abc}\partial_{[\mu}A_{\nu]}^aA^{b, \mu}A^{c, \nu}.
   \label{eqn:ThreePointInteraction}
\end{equation}
To obtain the non-trivial matrix elements of the S-Matrix for the diagram in Fig.~\ref{fig:ThreePointGluonDiagram}, we follow the standard QFT procedure to compute the scattering amplitudes of the process, using perturbation theory on the interacting term of the Lagrangian in Eq.~\eqref{eqn:ThreePointInteraction} and the analytical expression of the polarizations as present in~\cite{Ellis_Stirling_Webber_1996}. By labelling the left- and right-handed helicities by $L$ and $R$, respectively, we have 
\begin{equation}
\begin{gathered}
    i\mathcal{M}(L \to LL)=-i\sqrt{2}gf^{abc}(zE\theta + (1-z)E\theta)=-i\sqrt{2}gf^{abc}E\theta, \\
    i\mathcal{M}(L\to LR)=-i\sqrt{2}gf^{abc}\left(zE\theta - z(1-z)E\theta \right)=-i\sqrt{2}gf^{abc}z^2E\theta, \\
    i\mathcal{M}(L\to RL)=-i\sqrt{2}gf^{abc}(1-z)^2E\theta.
\end{gathered}
\end{equation}
The other non-zero scattering amplitudes are obtained from exchanging $L \to R$ and vice versa, which remain the same due to parity symmetry. Now, we can compute the average scattering amplitude by summing over all final helicities and colours and averaging over the initial ones, and by using that a gluons has 8 colours and that $\sum_{a,b,c}f^{abc}f^{abc}=8C_A$, with $C_A=\frac{N^2-1}{2N}$ for $SU(N)$~\cite{peskin1995introduction}, we have
\begin{equation}
\begin{gathered}
        \overline{|\mathcal{M}|^2}\equiv \frac{1}{2}\frac{1}{8}\sum_{\text{color}}\sum_{\text{helicity}}||\mathcal{M}||^2=\frac{8C_A4g^2}{16}E^2\theta^2\left( 1+ z^4 +(1-z)^4 \right)=\\
        =6g^2E^2\theta^2\left( 2(1-2z +z^2) + 2z^2 +2z^2(1-2z+ z^2) \right)=\\
        =4C_Ag^2t\left( \frac{1-z}{z} +\frac{z}{1-z} +z(1-z)\right)= 2g^2tC_A \frac{z^4 + 1 + (1-z)^4}{z(1-z)},
        \label{eqn:ScatteringAmplitudes}
\end{gathered}
\end{equation}
where we can identify the splitting function of the gluon into two gluons $P_{gg}(z)$ of Eq.~\eqref{eqn:GluonSPlittingFucntion}.

Now, using the scattering amplitudes presented in Eq.~\eqref{eqn:ScatteringAmplitudes} and the definition of the spin density matrix presented in Eq.~\eqref{eqn:RMatrixDefinition} and Eq.~\eqref{eqn:SpinDensityDefinition}, we obtain the following density matrix for the pure gluon splitting process

\begin{equation}
    \hat{\rho}_{sc} = \frac{1}{2(z^4 + (1 - z)^4 + 1)}\left(
\begin{array}{cccc}
 1 & z^2 & (1-z)^2 & 0 \\
 z^2 & z^4+(1-z)^4 & 2 (1-z)^2 z^2 & (1-z)^2 \\
 (1-z)^2 & 2 (1-z)^2 z^2 & z^4+(1-z)^4 & z^2 \\
 0 & (1-z)^2 & z^2 & 1 \\
\end{array}
\right).
    \label{eqn:ScatteringDensityMatrix}
\end{equation}

\section{Technical details for QC}\label{app:QCBasics}
Quantum circuit notation provides a graphical representation of qubit operations. Here, we represent qubits by wires that progress from left to right, and the quantum gates, which are unitary operations, by boxes placed on the wires of the qubits where they act. Finally, a final box with a meter represents a measurement of the output in the computational basis $\{\ket{0}, \ket{1}\}$.  In our work, we use both one-qubit gates and two-qubit controlled gates. An important one-qubit gate is the NOT gate, represented in matrix form by the first Pauli matrix
\begin{equation}
        X=\begin{pmatrix}
        0 & 1\\
        1 & 0
    \end{pmatrix},
\end{equation}
where $X$ clearly reproduces the negation operation, as $X\ket{0}=\ket{1}$ and $X\ket{1}=\ket{0}$. Another important one-qubit quantum gate is $R_Y(\alpha)=\exp\left(-i\frac{\alpha}{2}Y\right)$, with 
\begin{equation}
      Y=\begin{pmatrix}
        0 & -i\\
        i & 0
    \end{pmatrix},
\end{equation}
meaning that 
\begin{equation}
      R_Y(\alpha)=\begin{pmatrix}
        \cos\left(\frac{\alpha}{2}\right) & -\sin\left(\frac{\alpha}{2}\right)\\
        \sin\left(\frac{\alpha}{2}\right) & \cos\left(\frac{\alpha}{2}\right)
    \end{pmatrix},
\end{equation}
which rotates the Bloch vector an angle $\alpha$ around the y-axis. We also use the $U_3$ gate, defined as 
\begin{equation}
    U_3(\theta, \phi, \lambda) = 
\begin{pmatrix}
\cos\left(\frac{\theta}{2}\right) & -e^{i\lambda} \sin\left(\frac{\theta}{2}\right) \\
e^{i\phi} \sin\left(\frac{\theta}{2}\right) & e^{i(\phi + \lambda)} \cos\left(\frac{\theta}{2}\right)
\end{pmatrix}.
\label{eqn:U3Gare}
\end{equation}
In our work, we also use controlled, two-qubit gates. These gates have a control qubit, which, depending on its value, dictates whether a gate $U$ is applied to the target qubit. Hence, such a gate can be written as 
\begin{equation}
    \text{C-U} = \ket{0}\bra{0}\otimes\mathbb{I} + \ket{1}\bra{1}\otimes U,
\end{equation}
where, in this case, the gate $U$ is only applied to the second qubit if the first one is in the state $\ket{1}$. The specific case where $U=X$, has the name CNOT, which replicates the well-known logic operation of XOR, where $CNOT\ket{k}\ket{i}=\ket{k}\ket{k \oplus i}$, with $\oplus$ the modulo two addition. The general controlled gate C-U is represented by a black dot on the control qubit connected to a gate acting on the target qubit, as follows
\begin{center}
    \begin{quantikz}
    &\ctrl[]{1}&\\
    &\gate[1]{U}&   
    \end{quantikz}
\end{center}
in the case that the gate $U$ only acts if the control qubit is in the state $\ket{0}$, it is represented as
\begin{center}
    \begin{quantikz}
    &\ctrl[open]{1}&\\
    &\gate[1]{U}&   
    \end{quantikz}
    =
    \begin{quantikz}
    &\gate[1]{X}&\ctrl[]{1}&\gate[1]{X}&\\
    &&\gate[1]{U}&&   
    \end{quantikz}
\end{center}
while the specific case of the $CNOT$ gate is represented as
\begin{center}
    \begin{quantikz}
    &\ctrl[]{1}&\\
    &\targ{}&   
    \end{quantikz}
\end{center}

\section{Parameters and execution of the quantum circuit}\label{app:QCParameters}
From the definition of the two-qubit splitting quantum circuit
\begin{center} 
    \begin{quantikz}
    \lstick{$\ket{0}$}&\gate[1]{R_Y\left(\gamma_2\right)}&\ctrl[open]{1}&\targ{}&\meter{}\\
    \lstick{$\ket{0}$}&\gate[1]{U_3(\gamma_1,0, \pi/2)}&\gate[1]{R_Y(\gamma_3-\gamma_1)}&\ctrl[open]{-1}&\meter{}
    
    \end{quantikz}
\end{center}

We can use a numerical tool like \textit{Mathematica},  where we use the partial trace to find the state of each qubit's subsystem: subsystem $A$ for qubit $0$ and subsystem $B$ for qubit $1$, which yields
\begin{equation}
\begin{gathered}
       \hat{\rho}_A=\Tr_B(\hat{\rho}_\text{tot})=\left(\begin{array}{cc}
 \frac{1}{2} +\frac{z^{(1)}}{2}& \xi(\gamma_1, \gamma_3) \\
 \xi(\gamma_1, \gamma_3) &
   \frac{1}{2}-\frac{z^{(1)}}{2} \\
\end{array}
\right),
\end{gathered}
\label{eqn:FirstReducedState}
\end{equation}
while the reduced state of the second qubit is
\begin{equation}
\begin{gathered}
       \hat{\rho}_B=\Tr_A(\hat{\rho}_\text{tot})=\left(\begin{array}{cc}
 \frac{1}{2} +\frac{1-z^{(1)}}{2}& \xi(\gamma_1, \pi-\gamma_3) \\
 \xi(\gamma_1,\pi-\gamma_3) &
   \frac{1}{2}-\frac{1-z^{(1)}}{2} \\
\end{array}
\right), 
\end{gathered}
\label{eqn:SecondReducedState}
\end{equation}
where
\begin{equation}
    z^{(1)}\equiv z(\gamma_1, \gamma_3)=\frac{1}{2}\left(1+\cos(\gamma_3)(\sec(\gamma_1)-2)\right),
    \label{eqn:valueofz_app}
\end{equation}
and 
\begin{equation}
    \xi(\gamma_1, \gamma_3)=\frac{1}{2} \sqrt{2 \cos (\gamma_1)-1} \sec (\gamma_1) \cos
   \left(\frac{\gamma_1-\gamma_3}{2}\right).
    \label{eqn:OffDiagonalFirst}
\end{equation}

Meanwhile, for the case where we have three final particles, the reduced state of the first qubit is 

\begin{equation}
\begin{gathered}
    \hat{\rho}_A=\Tr_B(\hat{\rho}_{AB})=\\=
\begin{pmatrix}
 \frac{1}{2} \Big(1+\frac{z^{(1)}}{2}[1+(\sec (\gamma'_1)-2) \cos (\gamma'_3)] \Big) & \xi(\gamma_1', \gamma_3')  \\
\xi(\gamma_1', \gamma_3') &
    \frac{1}{2} \Big(1-\frac{z^{(1)}}{2}[1+(\sec (\gamma'_1)-2) \cos (\gamma'_3)] \Big) \\
\end{pmatrix},
\end{gathered}
\end{equation}
where $z^{(1)} \equiv z(\gamma_1, \gamma_3)$ and $z^{(2)} \equiv z(\gamma'_1, \gamma'_3)$. Similarly, the reduced state for the second qubit is

\begin{equation}
\begin{gathered}
    \hat{\rho}_B=\Tr_A(\hat{\rho}_{AB})=\\
    =
\begin{pmatrix}
 \frac{1}{2} \Big(1+\frac{z^{(1)}}{2}[1-(\sec (\gamma'_1)-2) \cos (\gamma'_3)] \Big) & \xi(\gamma_1', \pi - \gamma_3')
    \\
\xi(\gamma_1', \pi - \gamma_3')&
    \frac{1}{2} \Big(1-\frac{z^{(1)}}{2}[1-(\sec (\gamma'_1)-2) \cos (\gamma'_3)] \Big) \\
\end{pmatrix}.
\end{gathered}
\label{eqn:SecondCircuitSecondQubit}
\end{equation}

From the AspenOpenJets data set, we obtained the following parameter distribution for $\gamma_1$ and $\gamma_3$, respectively

\begin{figure}[H]
\begin{minipage}{0.48\textwidth}
    \centering
    \includegraphics[width=1.0\linewidth]{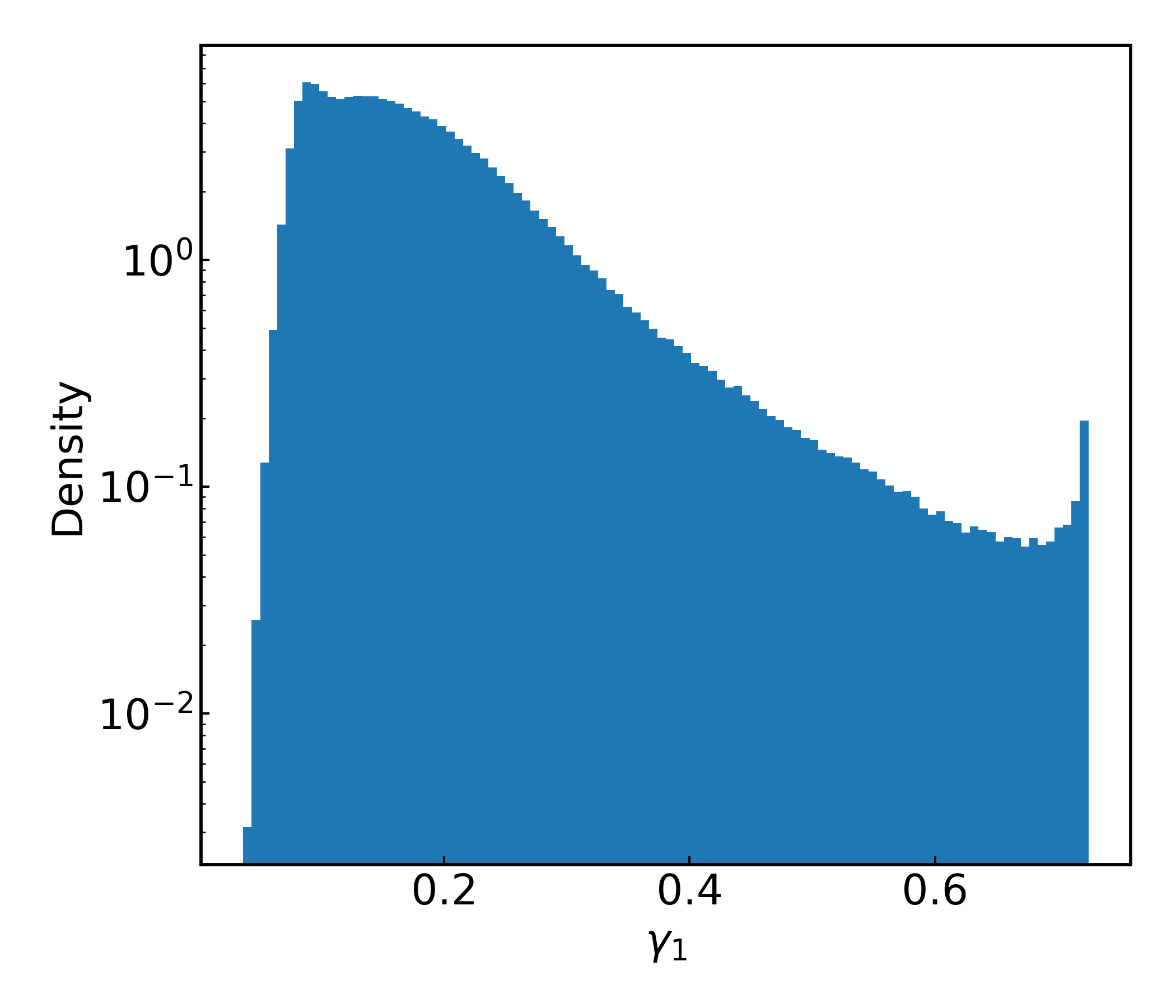}
    \vspace{-5mm}
    \caption{Analytically obtained distribution of $\gamma_1$}
    \label{fig:theta1distribution}
\end{minipage}
\hfill
\begin{minipage}{0.48\textwidth}
    \centering
    \includegraphics[width=1.0\linewidth]{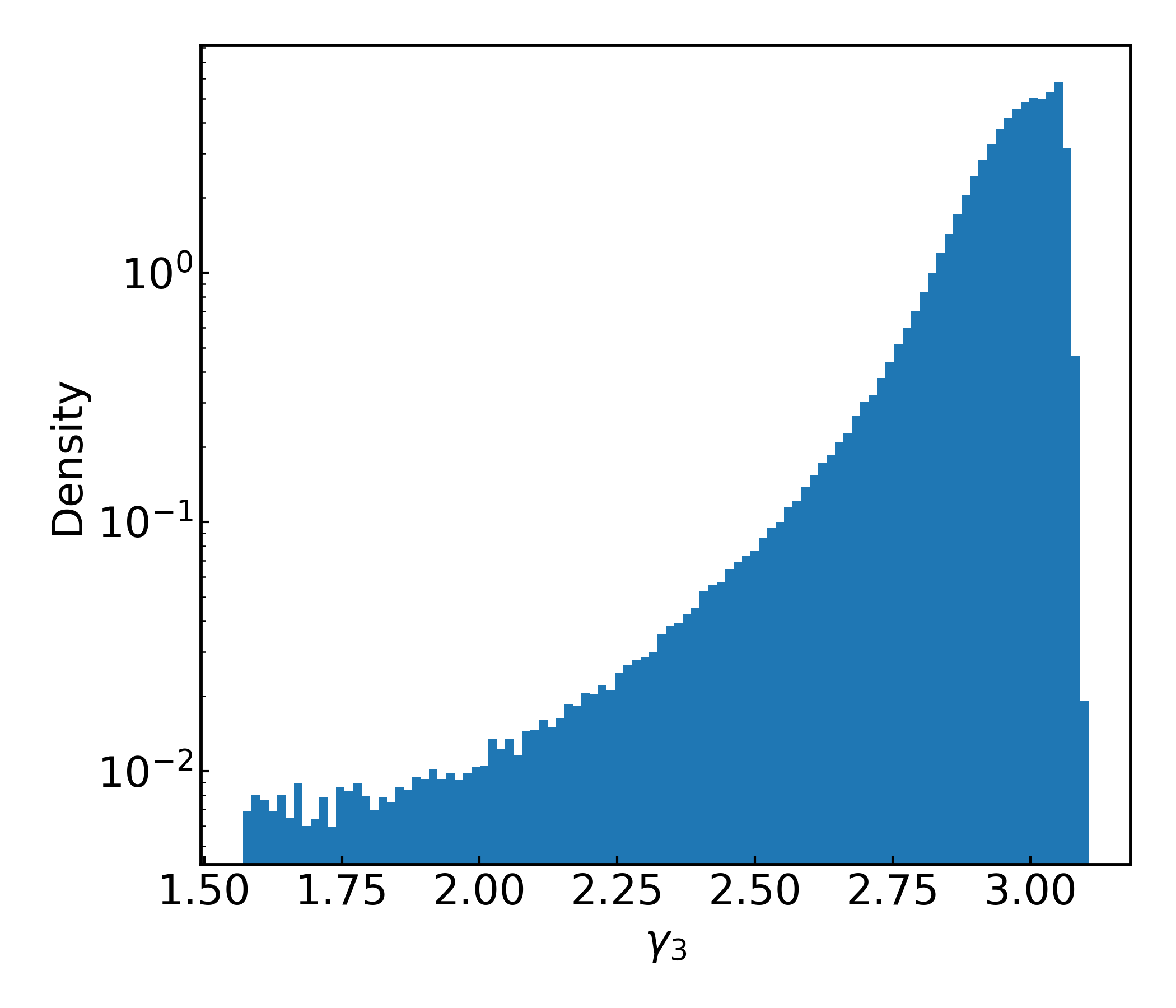}
    \vspace{-5mm}
    \caption{Analytically obtained distribution of $\gamma_3$}
    \label{fig:theta3distribution}
\end{minipage}
\end{figure}
\vspace{-5mm}

\begin{figure}[H]
\begin{minipage}{0.48\textwidth}
    \centering
    \includegraphics[width=1.0\linewidth]{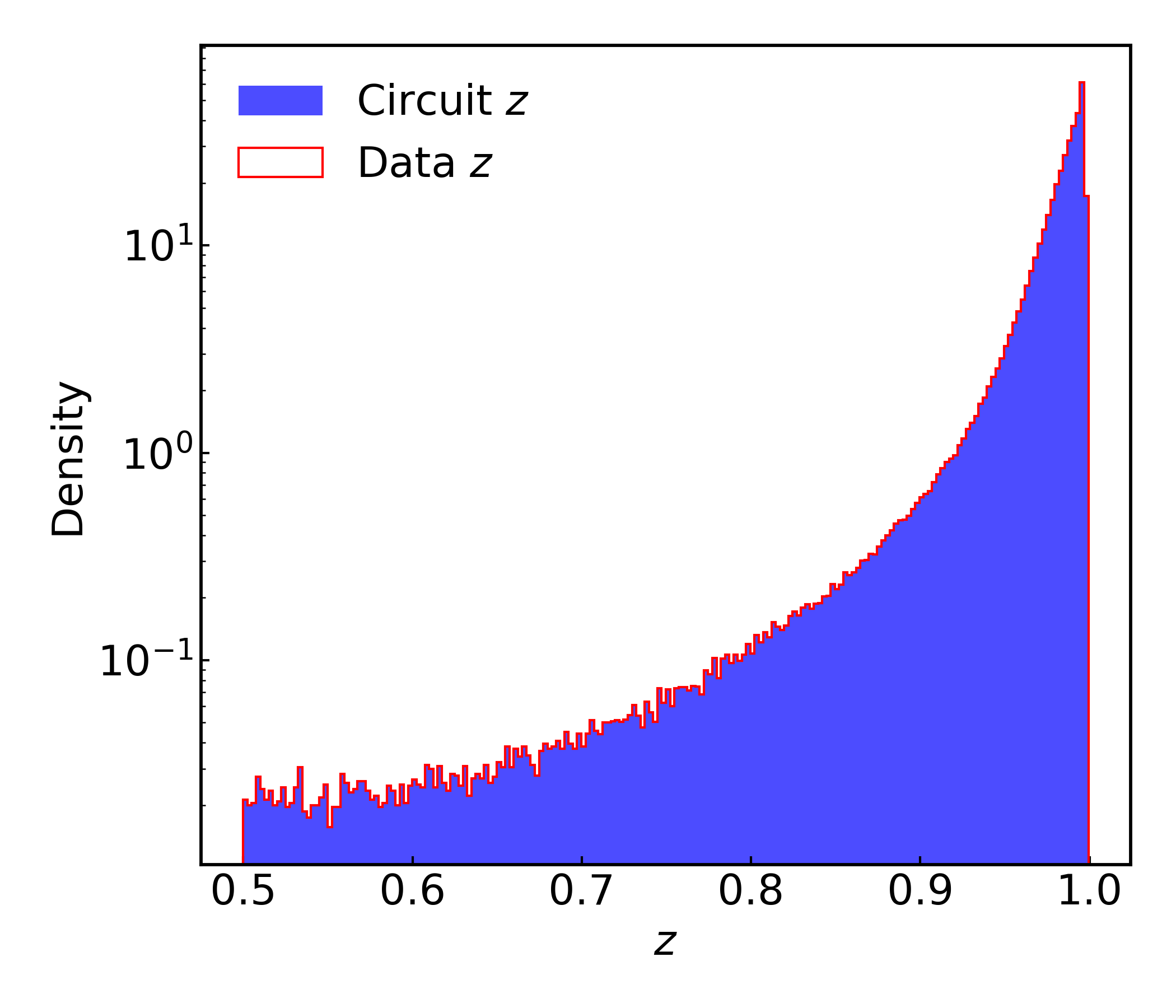}
    \caption{Comparison of the obtained distribution of $z$ from the circuit with the one from data}
    \label{fig:Comparisontwoqubit}
\end{minipage}
\hfill
\begin{minipage}{0.48\textwidth}
    \centering
    \includegraphics[width=1.0\linewidth]{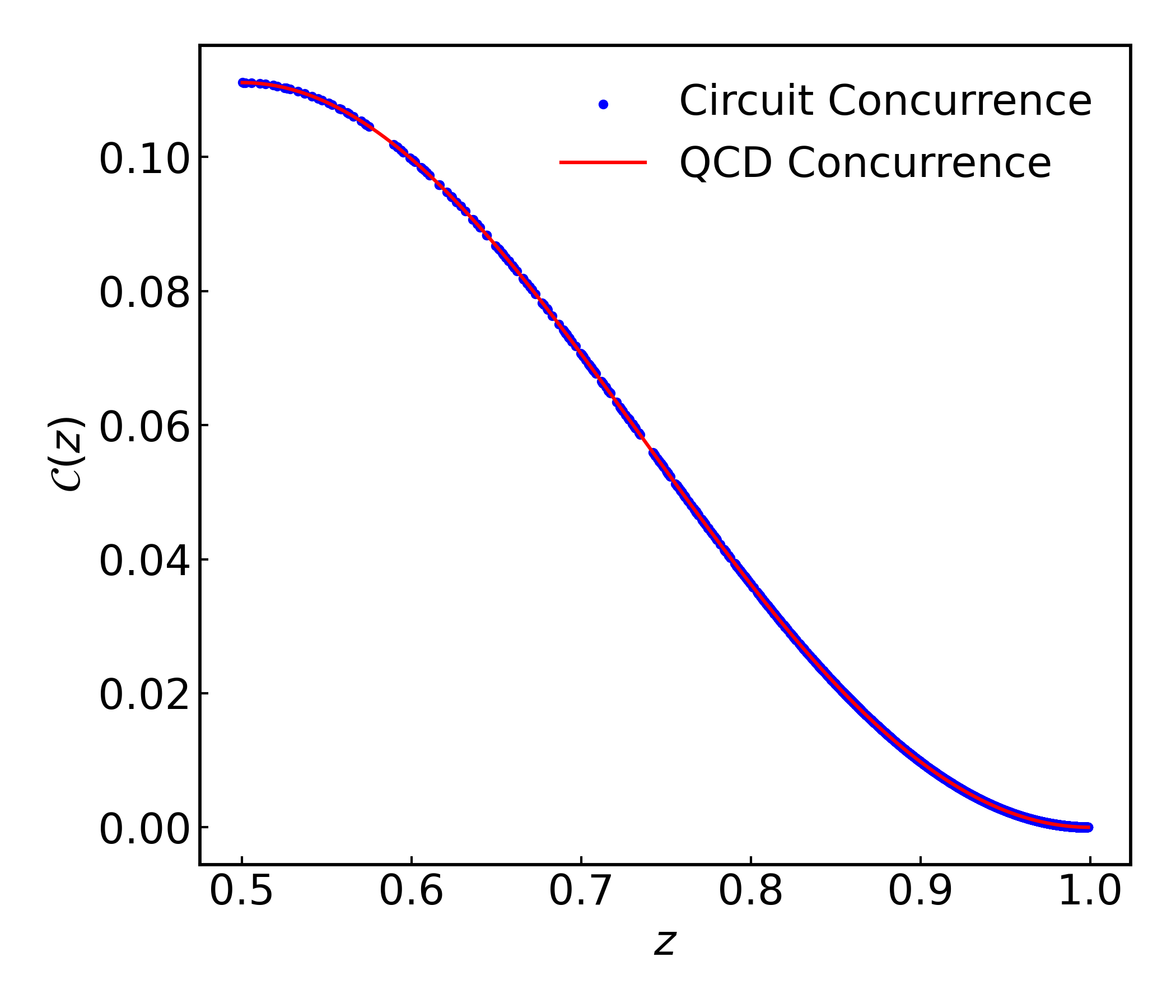}
    \caption{Close-up check of the concurrence obtained from the theory model ( Eq.~\eqref{eqn:ConcurrenceQCDProcess}) and the circuit model}
    \label{fig:TwoQubitConcurrence}
\end{minipage}
\end{figure}
By running the ideal quantum circuit using the above distribution of parameters, we can verify that they replicate the momentum fraction distribution provided by the data while replicating the entanglement. The output of the first qubit is shown in Fig.~\ref{fig:Comparisontwoqubit}, next to the distribution of the provided data. The circuit reproduces the measured two-prong distribution exactly because the parameter set is determined analytically for every jet in the sample. Similarly, in Fig.~\ref{fig:TwoQubitConcurrence} we can compare the concurrence of the sampled momentum fractions $z$ from the circuit with its dependence given by Eq.~\eqref{eqn:ConcurrenceQCDProcess}.

\section{Further discussion on the quantum circuit}\label{app:QCDiscussion}
Consider again the entanglement created by the circuit, presented in Eq.~\eqref{eqn:ConcurrenceCircuitAnal}. We first examine the symmetry under $z \to 1-z$. The first two terms inside the square root are invariant, but the third has no clear indication of being invariant. However, applying the transformation $\gamma_1 \to \gamma_1$ and $\gamma_3\to \pi -\gamma_3$, we see that $z( \gamma_1,  \pi-\gamma_3)=1-z(\gamma_1, \gamma_3)$ while $ \mathcal{C}_\text{circuit}(\gamma_1, \pi - \gamma_3)=\mathcal{C}_\text{circuit}(\gamma_1, \gamma_3)$. Furthermore, considering only positive values of $\gamma_1$ has no limiting effects on the circuit's possible outputs, as we can reach all values of $z\in[0,1]$ with $\gamma_1 \in [0, \pi/3]$. To verify that at $z=0$ the concurrence is zero as before, we note that the combination of parameters $(\gamma_1, \gamma_3)=(0, 0)$, gives $z=0$ and $\mathcal{C}_\text{circuit}(0, 0)=0$, meaning that our circuit also allows us to match the entanglement behaviour in the limits of $z\to 0$ and $z\to 1$. Finally, the symmetry of $z \to 1-z$ means that $z=1/2$ is an extremum of the function, and it can be numerically verified that it is a maximum.  
If we restrict $\gamma_3\in [\pi/2, \pi]$, the circuit covers the values of $z\in [0.5, 1]$, which is consistent with our definition of $z$ as the momentum fraction of the harder daughter. This construction imposes a consistent ordering of gluon energies across splittings. 

\section{Simulation results with no postprocessing}\label{app:Simulation}

\begin{figure}[H]
\begin{minipage}[t]{0.33\textwidth}
    \centering
    \includegraphics[width=1.0\linewidth]{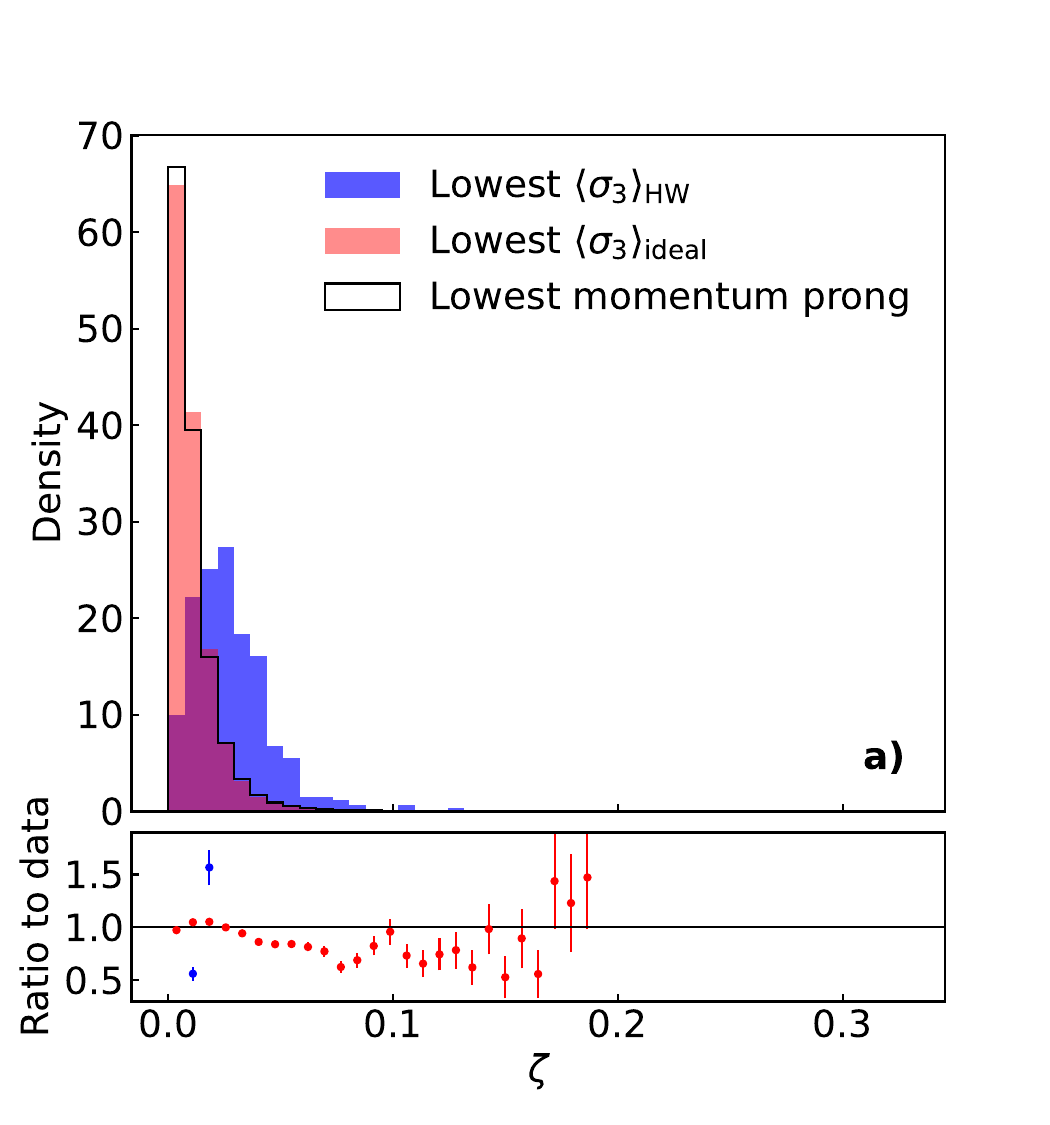}
\end{minipage}\hfill
\begin{minipage}[t]{0.33\textwidth}
    \centering
    \includegraphics[width=1.0\linewidth]{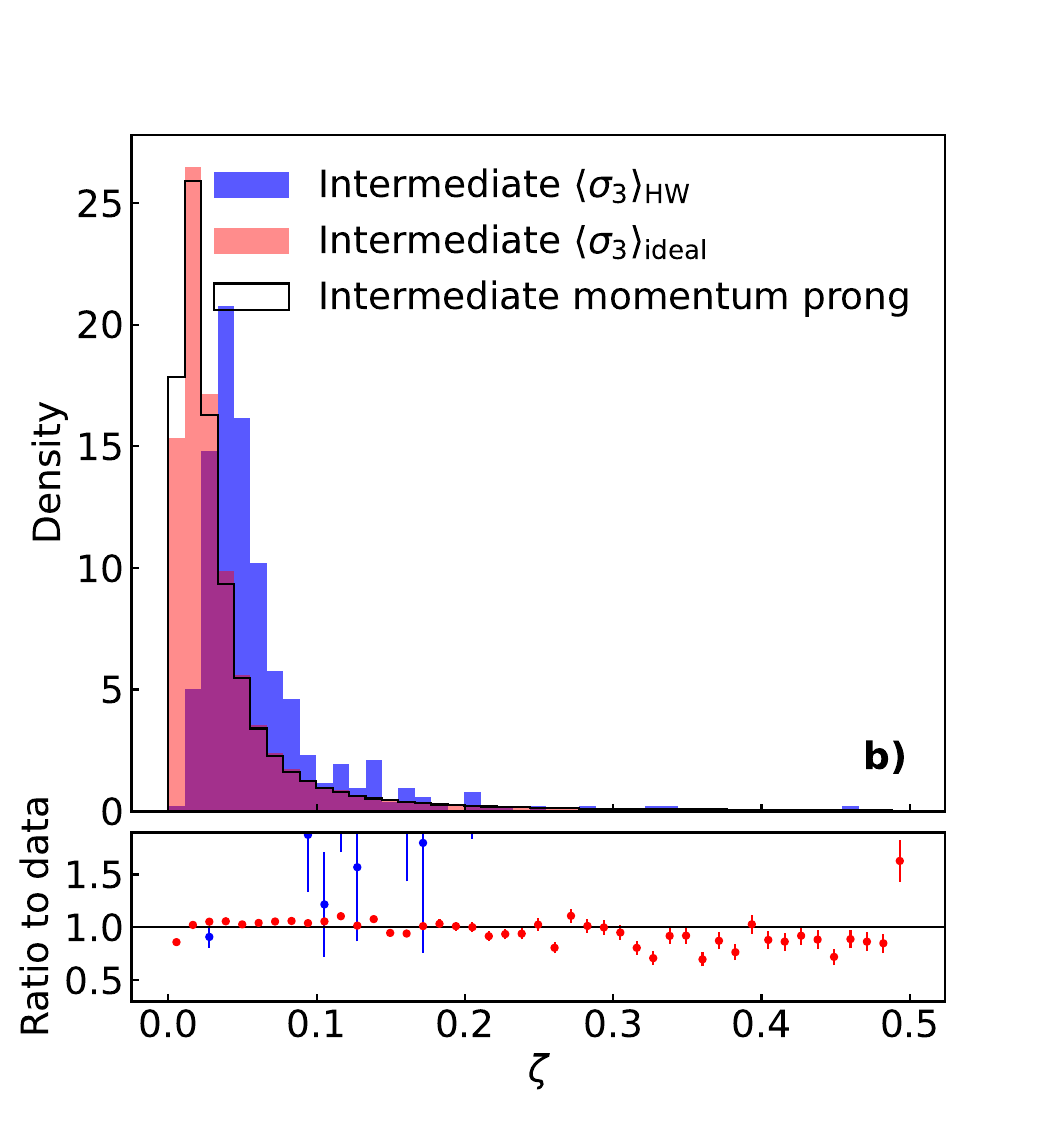}
\end{minipage}\hfill
\begin{minipage}[t]{0.33\textwidth}
    \centering
    \includegraphics[width=1.0\linewidth]{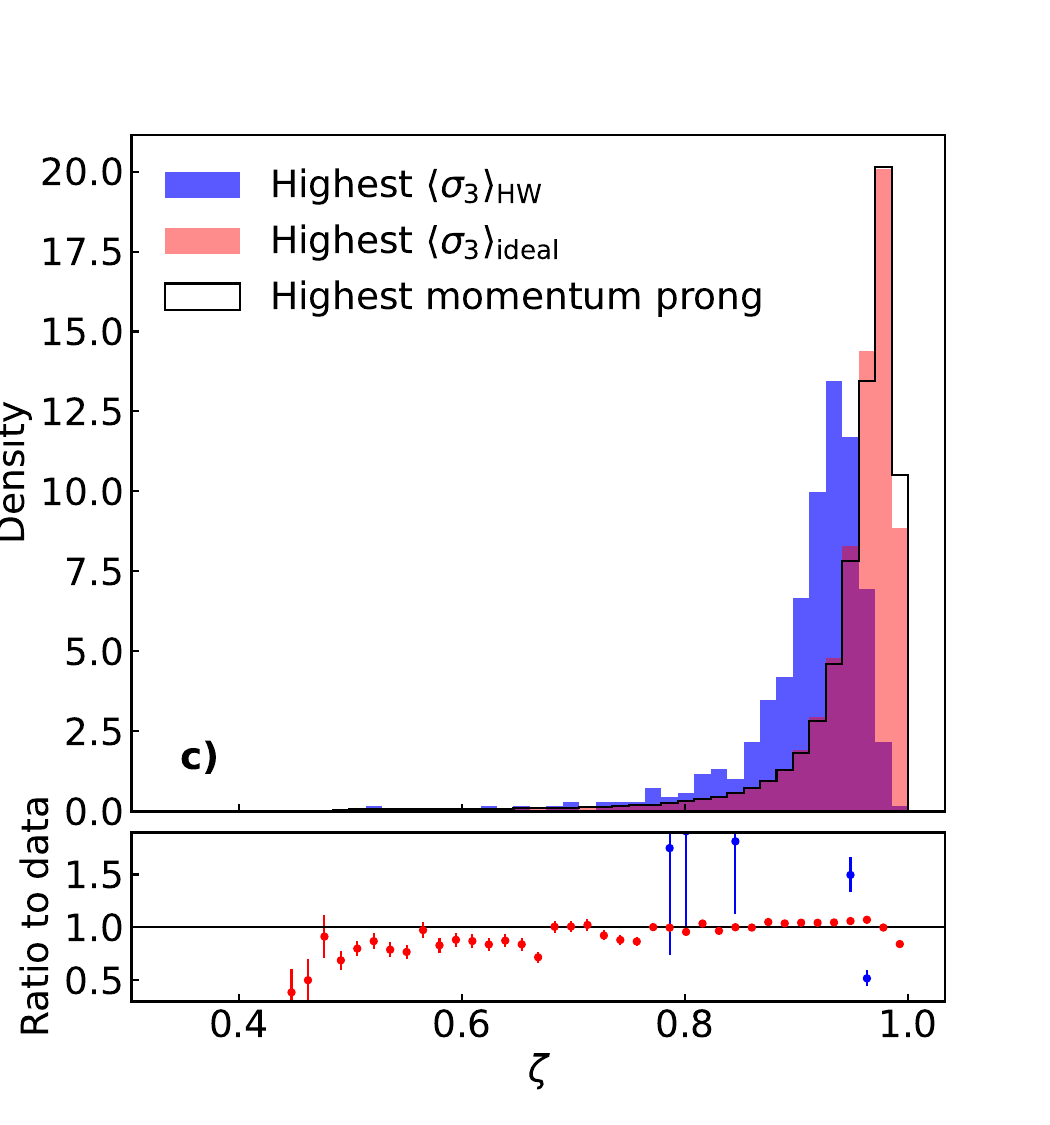}
\end{minipage}
\caption{Comparison of the momentum fraction $\zeta$ distribution of the a) lowest- b) intermediate- and c) highest-momentum prong obtained from the hardware execution of the quantum circuit on \textsc{ibm\_Marrakesh}, only excluding unphysical runs in blue, the noiseless simulation of the quantum circuit in red, and the data from the AspenOpenJets data with black outline. The error bars give the statistical uncertainty of the circuit samples, while the data uncertainty is smaller than the marker size.}
\label{fig:AllProngsNoPostprocessing}
\end{figure}

The results obtained from the hardware execution, with the bias correction removed and only the executions whose momentum fractions fall outside $[0,1]$ excluded, are presented in Fig.~\ref{fig:AllProngsNoPostprocessing}.

\section{Study of the independence assumption}\label{app:Independence}

In composing the splitting primitive we made three assumptions, stated in Sec.~\ref{sec:Section4}. We consider that the splitting distribution at each vertex is independent and identically distributed from the same two-prong distribution, and that the next splitting always acts on the harder of the two partons. All three follow from the primitive carrying no evolution variable during the shower. In this appendix we measure how far each departs from the data, and how much the departure matters for the distributions of Figs.~\ref{fig:AllProngs} and~\ref{fig:FourProngsPlot}.

Because these assumptions concern the order of splittings within a jet, they cannot be tested on the sorted prong fractions alone. We therefore follow the declustering of Sec.~\ref{sec:DataPrep} step by step for the same dataset, recording the momentum fraction $z^{(k)}$ of the harder daughter at each splitting $k$, together with which of the two partons splits next. The final prong fractions rebuilt from this sequence agree with those used in Sec.~\ref{sec:Section4}.

\begin{figure}[h]
\centering
\includegraphics[width=\linewidth]{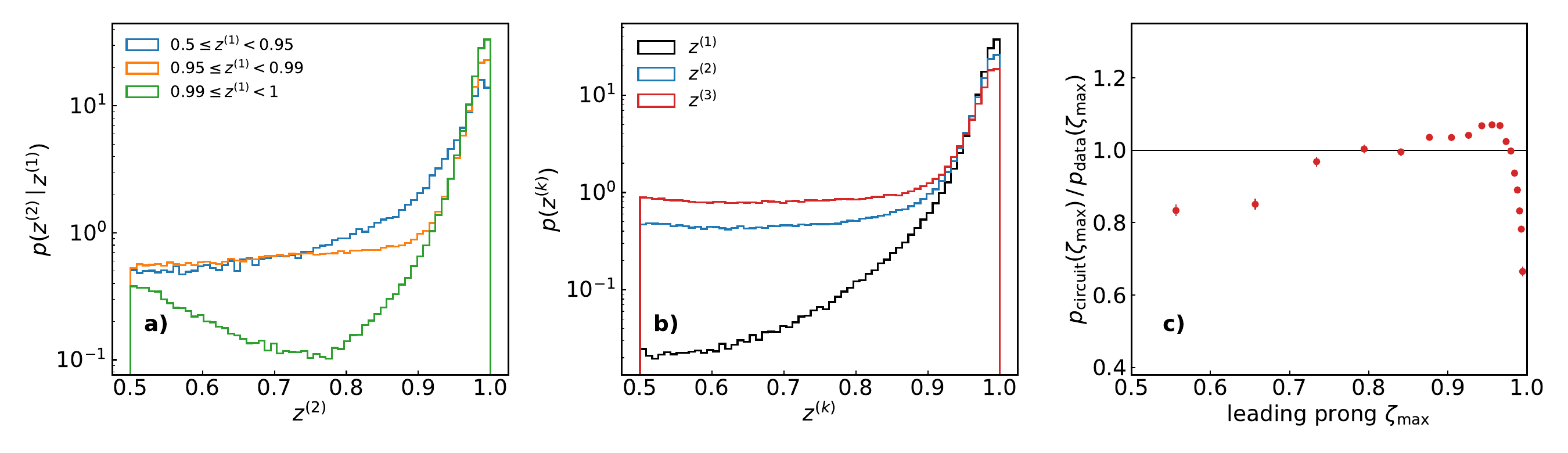}
\caption{Tests of the sampling assumptions on the declustered data. a) The distribution of the
second splitting fraction conditioned on three ranges of the first, which shows the dependence on previous splittings. b) The marginal distributions of the first three splitting
fractions, whose difference shows the different distributions. c) The ratio of
the leading-prong distribution obtained by independent sampling to the one measured in the data.}
\label{fig:Independence}
\end{figure}

None of the three assumptions holds exactly. Successive fractions are correlated as the conditional distributions $p(z^{(2)}|z^{(1)})$ in
Fig.~\ref{fig:Independence}~a) do not coincide. Following a very hard first splitting, intermediate
values of the second fraction are suppressed, and the distribution acquires a different shape
rather than merely shifting. The probability distribution of each splitting fraction also shows discrepancy, as the harder daughter takes a
progressively smaller expected value at each vertex, $\langle z^{(k)}\rangle = 0.974,\,0.925,\,0.883$, as
Fig.~\ref{fig:Independence}~b) shows. This also confirms the different splitting fractions are not identically distributed.

Despite this the multi-prong distributions are well reproduced over most of their range, it is
worth seeing why the deviation concentrates where it does. The momentum fraction of the leading
prong is a product of the individual splitting fractions. A typical value of that product
can be obtained in many ways, one hard splitting compensating a softer one, so the centre of the
distribution is insensitive to how the individual fractions are related to one another. Its
extreme, by contrast, is reached only when every splitting along the branch is hard at the same
time, and is therefore controlled by their joint distribution rather than by the marginals alone. A
correlation too weak to be visible in the bulk becomes the dominant effect in the tail. 

This is what we observe. For the leading prong of the three-prong sample we find $P(\zeta>0.95)=0.694$
from the circuit against $0.703$ in the data, while $P(\zeta>0.99)=0.047$ against $0.061$. The ratio of
the two distributions, Fig.~\ref{fig:Independence}~c), is flat to within a few percent across the
bulk and falls away at the largest fractions. The approximation is therefore accurate where the great
majority of the jets lie, and deteriorates in the region most sensitive to the correlations it discards. The ratio also falls below unity at small leading fractions, a region reached only when the harder parton splits almost evenly. There the assumptions act together rather than separately, and relaxing any one of them alone moves the agreement further from the data.

Relaxing these assumptions would require a different construction. All three originate from the lack of an evolution variable that makes the
splitting probability depend on the scale, correlates successive vertices, and decides which parton branches next. Supplying it means introducing the Sudakov factor and the associated ordering, which is the extension discussed in Sec.~\ref{sec:Section5}. We have kept the independent-sampling approximation because our aim is to demonstrate that the splitting primitive composes correctly and can be calibrated to data, not to present a complete parton shower.

\end{document}